\documentclass[aps,graphicx,reprint,amsmath,amssymb,groupedaddress,rp]{revtex4-1}
\usepackage{color}
\usepackage{bbm}
\usepackage{mathrsfs}
\usepackage{amsfonts}
\usepackage{booktabs}
\usepackage{graphicx}
\usepackage{amscd}
\begin{document}
\title{The topological properties in topological insulators and superconductors}
\author{Chunbo Zhao}
\email[]{cbzhao@semi.ac.cn}
\affiliation{State Key Laboratory of Superlattices and Microstructures, Institute of Semiconductors\\Chinese Academy of Sciences, P.O.Box 912,Beijing 100083, People's Republic of China}
\date{\today}
\begin{abstract}
We give a self-contained and enriched review about topology properties in the rapidly growing field of topological states of matter (TSM). This review is mainly focus on the beautiful interplay of topology mathematics and condensed matter physics that issuing TSM. Fiber bundle theory is a powerful concept to describe the non-trivial topology properties underlying the physical system. So we briefly present some motivation of fiber bundle theory and following that several effective topological methods have been introduced to judge whether a fiber bundle is trivial or not. Next, we give some topological invariants that characterizes the non-trivial TSM in the non-interacting systems in all dimensions, which is called topological band theory. Following that, we review and generalize the topological response using topological field theory called Chern-Simons effective theory.
 Finally, the classification of free-fermion systems have been studied by loop space and K-theory.

\end{abstract}
\maketitle
\textcolor[rgb]{0.00,0.00,1.00}{\tableofcontents}
\section{Introduction}
Since the discovery of inter quantum Hall effect (IQHE) by von Kiltzing , el.\cite{PhysRevLett.45.494} and the geometry phase in quantum systems by Berry\cite{berry1984quantal} during 1980s, the mathematic of differential geometry and topology was introduced to condensed matter physics. Later, Thouless, el \cite{thouless1982quantized} found that the topological invariant (TKNN number) which characteristic the valance band bundle over the magnetic Brillouin zone (BZ) contribute to the robust quantization Hall conductance in IQHE. In this paper, we want to address the issue why and how physicists apply topology to study condensed matter physics.
\par
The  first motivation may be that physicists want to generalize 2d IQHE and try to find  more non-trivial states  from a viewpoint of topology.  Among the process of exploration, Zhang and Hu\cite{zhang2001four} successfully establish a 4d version quantum Hall effect in 2001, not like 2d IQHE , the base manifold is 4d sphere and replace the U(1) fiber with SU(2) fiber. Based on this model, Marakami,el \cite{murakami2003dissipationless} predict dissipationless quantum spin current in hole-doped semiconductors, which eventually leads to the find of spin Hall effect\cite{kato2004observation}.  From that, physicists began to attention spin-orbit coupling system and developed the topological band theory to study the topological property of energy band over the entire Brillouin zone (BZ).  For example, physicists build some subtle and practical models with non-trivial topology property in time-reversal invariant system, such as quantum spin Hall (QSH) state\cite{kane2005quantum,bernevig2006quantum}, 3d topological insulators\cite{fu2007topological1,zhang2009topological}. Fortunately, the helical edge states of QSH  are observed in 2d HgTe quantum well experimentally in 2007\cite{konig2007quantum}. Soon after that, the non-trivial Dirac fermion surface states of 3d topological insulator such as $Bi_{x}Sb_{1-x}$ alloy\cite{hsieh2008topological} and $Bi_2Se_3$ class materials\cite{xia2009observation} are also been observed from experiments. Tremendous interests have been recently attracted by the study of topological state of matter (TSM), readers can reference the reviews\cite{hasan2010colloquium,qi2011topological}. Here I want to emphasize that these experiments discoveries purely from the topological property  of free fermions Hamiltonian. And it's the quest for beauty of topology in physics that leads to the discoveries of these novel states.
\par
However, real physical systems are actually interacting. Then, how to interpret the robust quantization Hall conductance such as in IQHE if it's interacting?  Actually, Niu, el\cite{niu1985quantized} solved this problem by proposing a topological invariant via Green function with twisted boundary. For interacting topological insulators in 3d and QSHE in 2d, Wang, el \cite{wang2010topological} propose the general topological order parameters expressed with Green function.  The topological property extracted from Green function also are used by Vovlovik \cite{volovik2003universe}in Helium liquid to describe the non-trivial topology property in momentum space.  Using Green function, Gurarie, el\cite{essin2011bulk} propose a general bulk-boundary correspondence which related the topological invariants in the bulk and the one its boundary in all space-time dimensions. However, these Green function description is very difficult for practice application. Recently, Wang propose a simplified topological invariant only from the zero-frequency Green function\cite{wang2012simplified,wang2012topological}. Wang's approach to this problem can greatly help the search for realistic materials . In addition, these simplified topological invariants may be served as the topological order parameters for interacting TSM including recent proposed topological anderson insulators\cite{li2009topological} and topological mott insulators\cite{raghu2008topological}.

\par
Another reason for application of topology concept in condensed matter physics is the topological field theory (TFT)\cite{qi2008topological,cho2011topological} description of physical responses of these non-trivial systems. TFT is powerful due to the physical responses such as electromagnetic and gravitational responses \cite{ryu2012electromagnetic} can be read out directly from the effective theory with the coefficients expressed using topological invariants. The axion topological field theory predicts the non-trivial topological magnetoelectric effect\cite{qi2008topological} in topological insulators and the presence of magnetic monopole\cite{qi2009inducing}. This field theory recently has been successfully used to describe time-reversal invariant topological superconductors\cite{qi2013axion}, and some novel observable effects have been predicted. In physics, these non-trivial TFTs are stem from the anomalies in quantum field theory\cite{nakahara2003geometry,ryu2012electromagnetic}. The bulk-boundary correspondence in TSM can be easily understood through the anomalies polynomial and descent realtion\cite{nakahara2003geometry}, which is related to a deep mathematic fact called Atiyah-Singer index theorem \cite{atiyah1968index1}.
\par
The last motivation of using topology in condensed matter physics is the classification of states of matter beyond the symmetry breaking principle. How many TSM can exist in nature?  This problem has been solved partly by Schnyder, el\cite{schnyder2008classification} in three spatial dimensions for free-fermion systems. Qi, el \cite{qi2008topological} proposed the classification for interacting systems can  be classified using Chern-Simons field theory by imposing on the discrete symmetry. And later, Kitaev\cite{kitaev2009periodic} gives a complete classification periodic table or classification using K theory for free-fermion systems.  This beautiful periodic table is an interesting mathematic result called Bott periodicity and there are only ten symmetric space that discovered by \'{E}lie. Cartan in 1926. It tells us that there are five distinct classes of TSM in every spatial dimension and only ten symmetry classes. With this periodic table, physicists can search more non-trivial TSM purposefully. For interacting boson and fermion systems, Wen, el \cite{chen2012symmetry,gu2012symmetry} propose a general framework to classify the symmetry protected topological states using group-cohomoloy or supercohomology. Wen's approach is powerful because it describes a large class of new topological states in general dimensions.

\par
This paper is mainly focus on the interplay of topology and condensed matter physics, it can be used as a self-contained introduction of topology and their application in condensed matter physics. The rest parts of the paper are organized as follows. First, we want to give an intuitive introduction of topology mathematics as possible as we can, readers who are familiar this knowledge can skip this subsection. In the second part, we show how to apply the mathematic tool to investigate the non-trivial topology property in solid state band theory and how getting the topological observable effect from TFT. During this process, we present the topological invariants expressed with Berry phase connection or curvature for non-interacting systems and corresponding topological order parameter for interacting systems expressed using Green functions.
  Finally, we review the tenfold way and dimensional hierarchy of topological states of matter using fiber bundle and K theory. This paper is mainly for readers who are interested in the topology application in condensed matter physics or who like to apply topological tool to investigate the physic systems.

\section{Topology for Physics}
In this section, we briefly review some important mathematic tools in this field. Readers who pay close attention to the mathematics details can consult this subsection.
\subsection{Fiber Bundles}
Before introducing the abstract definition of fiber bundle, we show  the appearance of it is very natural and justifiability. As we all know, the famous Bloch theorem tells us the wave function in condensed matter is nothing but the irreducible representation of translation invariant of reciprocal lattice $\vec{k}$. Hence, corresponding Hamiltonian can be interpreted as a map from the first Brillouin zone (which is equivalent to a torus $T^2$ in two dimension) to the group of Hamiltonian.  It means, for a fixed point $\vec{k}$ in Brillouin zone, there are Hamiltonians belonging to the symmetry group required in physics. This object is called fibre bundles in mathematics.
\par
  A fiber bundle consists of the following: There are a manifold $F^k$ (called the fiber), a manifold $E$ (the bundle space) and a manifold $M^n$ (the base space) together with a map
  \begin{equation}\label{1}
    \pi: E \rightarrow M^n\nonumber
  \end{equation}
  $E$ onto $M$, where $k$ and $n$ are the dimension of fiber and base space, respectively. We demand that $E$ is locally a product space in the following sense:
  There is a covering of $M^n$ by open sets, $U, V,$ . . . , such that $\pi^{-1}(U)$ is diffeomorphic to $U\times F$; there is a diffeomorphism
 $$ \phi_{U}:U \times F \rightarrow \pi^{-1}(U)$$. In an overlap $U\cap V$ a point $e \in
\pi^{-1}(U\cap V)$ will have two representations $$ e = \pi_{U}(p, y_{U}) = \pi_{V}(p, y_{V})$$ and we demand that $$ y_{V}= c_{VU}(p)y_{U}$$ where $c_{VU}: F\rightarrow F$ is a diffeomorphism of the fiber, while $p$ is a point of base space. In the case of vector bundle, $F= \mathbb{C}^{K}, \mathbb{R}^{K}$, each $c_{VU}(p): \mathbb{R}^{K}\rightarrow\mathbb{R}^{K}$ was a linear transformation called transition function is a subgroup of the structural group $G$ of the fibre bundle. So we still demand that $$c_{VU}=c_{UV}^{-1}$$ and $$c_{UW}\circ c_{WV}\circ c_{VU}=I (U\cap V\cap W)$$ where $I$ is identity. In summary, a fiber bundle $\mathscr{E}= \{E, M,\pi, G, F, \phi\}$ consists total space $E$, base space $M$, typical fiber $F$, projection $\pi$, structural group $G$ and local trivialization $\phi$. In special case, if the fiber $F$ is the same as the group $G$, and if the transition functions $c_{UV}(x)$ act on $F = G$ by left translations as $g_{U}(p)= c_{UV}(p)g_{V}(p)$, the fiber bundle is called principle fiber bundle. In addition, if the total space $E$ globally equate $M\times F$, the corresponding fiber is trivial fibre bundle. Finally, a local section is again a map $s : U\rightarrow E$ such that
$\pi\circ s$= identity. A section $s$ is simply a collection of maps $\{s_U : U\rightarrow F\}$ such that in $U\cap V$ we have $s_U(p)=c_{UV}(p)s_V(p)$.
 \par
  Now it's time to describe the classification question in a mathematical way. The mathematical structural of a non-interacting insulator of spatial dimension $d$ is that of a vector bundle or principle fiber bundle $E \underrightarrow{\pi} M$. The base manifold $M$ of $E$ is the d dimensional $k$-space of the system $T^{d}$, and the fiber over a point $k$ given by the space of occupied states $P(k)$ (also called valence band). If no further symmetry conditions are imposed (like IQHE in class A), the structural group of the bundle is $U(n)$, where $n$ is the dimension of $P$. Hence, the fiber bundle of class A is $\{E, T^{d}(k), \pi, U(n),P(k) \}$, and the question of a insulator is a topological non-trivial or not is tantamount to asking the total space of fiber bundle is non-trivial or not. In other words, we should investigate the difference between  the total space $E$ and multiple space $T^{d}(k)\times P(k)$. For the local trivialization property of fiber bundle, we know that if a base manifold can be covered with one open set, the fiber will be trivial. However, when the base space is a torus or sphere that cannot be covered by one open set, the total space cannot be constructed in a multiplied way. So we obtained that the obstruction to defined a global multiplied space is  the transition function  in the overlap region is not identity.

\subsection{Chern Character}
  Characteristic classes are subsets of cohomology of classes of the base space and measure the non-triviality or twisting of a bundle. In this sense, they are obstructions which prevent a bundle from a trivial bundle. Chern character is one of significant characteristic classes given by de Rahm cohomology classes. It is used extensively in physics and mathematics.
  \par
  We give here a brief introduction of the de Rahm cohomology group. Let $M$ be an $m$-dimensional manifold. An $r$-form $\omega\in\Omega^r(M)$ is closed if d$\omega$=0 and exact if $\omega$=d$\eta$ for some $\eta\in\Omega^{r-1}(M)$. The set of closed $r$-forms is denoted by $Z^{r}(M)$ and the set of exact $r$-forms by $B^{r}(M)$. Since d$^2$=0, it follows that $Z^{r}(M)\supset B^{r}(M)$. We define the $r$-th de Rahm cohomology group $H^r(M)$ by
  $$H^{r}(M)\equiv Z^r(M)/B^r(M).$$
  Actually, $H^r(M)$ is a dual space of the homology group $H_r(M)$. The duality is provided by Stokes' theorem, it can take a compact form:
  \begin{equation}\label{stokes}
    (c,d\omega)=(\partial c,\omega)
  \end{equation}
  where $c$ is the chain group in homology and $\partial$ is the boundary operator. In this sense, the exterior derivative d is the adjoint of the boundary operator $\partial$ and vice versa.
  \par
    Let $P(M,G)$ be a principle bundle. A connection on $P$ is a unique separation of the tangent space $T_uP$ into the vertical subspace $V_uP$ and the horizontal subspace $H_uP$ such that
    \\\textbf{(i)} $T_uP= H_uP \bigoplus V_uP.$
    \\\textbf{(ii)} A smooth vector field $X$ on $P$ is separated into smooth vector fields $X^H\in H_uP$ and $X^V\in V_uP$ as $X=X^H+X^V.$
    \\\textbf{(iii)}$H_{ug}P=R_{g\ast }H_uP$ for arbitrary $u\in P$ and $g\in G.$\\
    The condition (iii) states the horizontal subspaces $H_uP$ and $H_{ug}P$ on the same fiber are related by a linear map $R_{g\ast }$ induced by the right action.
    In practical computations, we need separate $T_uP$ into $V_uP$ and $H_uP$ in a systematic way. This can be achieved by introducing a Lie-algebra-valued one-form $\omega\in\mathfrak{g}\bigotimes\Omega^1(P)$ called the connection one-form, where $\mathfrak{g}$ denotes the Lie algebra of group $G$ and $\Omega^1(P)$ is the one-form of fibre bundle. It is a projection of $T_uP$ onto the vertical component $V_uP\simeq\mathfrak{g}$. Another important concept is curvature two-form $\Omega$ which is defined by the covariant exterior derivative of the one-form $\omega$, $\Omega\equiv$ D$\omega\in\Omega^2(P)\bigotimes\mathfrak{g}$. Based on the above definitions, an important property about curvature two-form called Bianchi identity can be given as D$\Omega$ = 0.
    \par
    Until now, the Berry phase gauge connection and Berry phase field strength in CMP have not expressed with a mathematic way. The relationship will be clear after introducing the local connection and local curvature. Let $\{U_i\}$ be an open covering of $M$ and let $\sigma_i$ be a local section defined on each $U_i$. It's convenient to introduce a Lie-algebra-valued 1-form $\mathcal{A}_i$ on $U_i$ via the pullback map of $\sigma_{i}$ such that $$\mathcal{A}_i\equiv\sigma^{*}_i\omega\in\mathfrak{g}\otimes\Omega^1(U_i),$$
    where $\sigma_{i}: U_i\rightarrow\pi^{-1}(U_i)$. The local form $\mathcal {F}_i$ of curvature $\Omega$ is defined by $$\mathcal{F}_i\equiv\sigma^{*}_i\Omega\in\mathfrak{g}\otimes\Omega^2(U_i)$$
    In physics, $\mathcal{A}_i$ is identified with Berry phase gauge potential, $\mathcal {F}_i$ is corresponding to the Berry field strength. Take local sections $\sigma_i$ and $\sigma_j$ over $U\cap V$, such that $\sigma_j(p)=\sigma_i(p)g(p)$. Then the corresponding local form $\mathcal{A}_i$ and $\mathcal{A}_j$ are related by the compatibility condition as
    \begin{equation}\label{connection}
     \mathcal{A}_j=g^{-1}\mathcal{A}_ig +g^{-1}\textrm{d}g.
    \end{equation}
    It's nothing but the gauge transformation in Berry gauge theory. $\mathcal{F}$ can be expressed in terms of gauge potential $\mathcal{A}$ as
    $$\mathcal{F}=\textrm{d}\mathcal{A}+\mathcal{A}\wedge\mathcal{A}$$
    where d is exterior derivative on $M$ and $\wedge$ is wedge product. The most important property of $\mathcal{F}$ is the following covariability
    \begin{equation}\label{1}
        \mathcal{F}_j=g^{-1}\mathcal{F}_ig
    \end{equation}
    it's also called the compatibility condition for local form $\mathcal{F}$. Similarly, the Bianchi identity of local form $\mathcal{F}$ is following:
    \begin{equation}\label{2}
       \textrm{D}\mathcal{F}=\textrm{d}\mathcal{F}+\mathcal{A}\wedge\mathcal{F}-\mathcal{F}\wedge\mathcal{A}=0
    \end{equation}
     Can we get the topology information of fibre bundle from these local geometrical quantities such as $\mathcal{F}$ and $\mathcal{A}$? The answer is yes, due to the famous work by S. S. Chern and Weil, one can obtain de Rahm cohomology group by the properties \ref{1} and \ref{2} via the invariant polynomial of $\mathcal{F}$. An invariant polynomial is defined as
     $$P(g^{-1}Ag)=P(A)~~  A\in\mathfrak{g}, g\in G.$$
     Chern and Weil found that if the Lie algebra $A$ in invariant polynomial $P$ is replaced by curvature $\mathcal{F}$, the $P(\mathcal{F})$ is characteristic classes called Chern classes. The Chern-Weil theorem is following:
     \\$\textbf{(a)} \textrm{d}P(\mathcal{F})=0$
     \\$\textbf{(b)} P(\mathcal{F})-P(\mathcal{F'})$ is exact, where $\mathcal{F}$ and $\mathcal{F'}$ are curvature two forms corresponding to different connections $\mathcal{A}$ and $\mathcal{A'}$.
    \par
    Now, it's easy to write Chern classes and Chern character according to the Chern-Weil theorem. For example, let $E\underrightarrow\pi M$ be a class A topological insulator, whose fiber is complex vector bundle $\mathbb{C}^{k}$, and the structure group is U(n).
    From the invariant polynomial  such as determinant and trace, one can get Chern classes and Chern character with Berry curvature $\mathcal{F}$, respectively.
    Explicitly, the total Chern class $c$ can be expressed as
    $$c(\mathcal{F})\equiv\textrm{det}(I+\frac{i\mathcal{F}}{2\pi}).$$
    Since Berry curvature $\mathcal{F}$ is a two-form, $c(\mathcal{F})$ is a direct sum of forms of degree of even
    $$c(\mathcal{F})=1 + c_1(\mathcal{F})+c_2(\mathcal{F})+\cdots$$
    where $c_j(\mathcal{F})\in \Omega^{2j}(M)$ is called $j$th Chern class. Obviously, $c_j$ can only be non-vanishing for 2$j\leq d$, where $d$ is the dimension of the base manifold $M$, i.e., the physical system of IQHE $d$ equal to 2. The Chern character can be written as
    $$\textrm{ch}(\mathcal{F})\equiv \textrm{tr}\exp(\frac{i\mathcal{F}}{2\pi})=\sum_{j=1}\frac{1}{j!}\textrm{tr}(\frac{i\mathcal{F}}{2\pi})^j.$$
    The $j$th Chern character $\textrm{ch}_j(\mathcal{F})$ defined as
    $$ \textrm{ch}_j(\mathcal{F})\equiv\frac{1}{j!}\textrm{tr}(\frac{i\mathcal{F}}{2\pi})^j.$$
    The integral of the Chern character in $d = 2n$ dimensions is an integer, the $n$-th Chern number,

    \begin{eqnarray}\label{chern number}
      \textrm{Ch}_n[\mathcal{F}] &=& \int_{\textrm{BZ}^{d=2n}}\textrm{ch}_n(\mathcal{F})\nonumber \\
       &=&  \int_{\textrm{BZ}^{d=2n}}\frac{1}{n!}\textrm{tr}(\frac{i\mathcal{F}}{2\pi})^n\in\mathbb{Z}.
    \end{eqnarray}
    Here, $\int_{\textrm{BZ}^{d}}$ denotes the integration over $d$-dimensional $k$-space. For lattice models, it can be taken as a Wigner-Seitz cell in the reciprocal lattice space. On the other hand, for continuum models, it can be taken as $\mathbb{R}^d$. Assuming that the asymptotic behavior of the Bloch wavefunctions approaches a $k$-independent value as $|k|\rightarrow\infty$, the domain of the integration can be regarded as $S^d$.
    When $d$=2, $\textrm{Ch}_1[\mathcal{F}]$ is the TKNN integer,
    \begin{equation}\label{ch1}
       \textrm{Ch}_1[\mathcal{F}]=\frac{i}{2\pi}\int_{\textrm{BZ}^{d=2}}\textrm{tr}(\mathcal{F}),
    \end{equation}
    which is nothing but the quantized Hall conductance $\sigma_{xy}$ in units of $(e^2/h)$.
    \par
    Unfortunately, the Chern character is only suitable for use in even-dimensional manifold. Deriving from the Chern form, one can obtain a characteristic class called Chern-Simons forms for odd-dimensional manifold. Since $\textrm{ch}_n(\mathcal{F})$ is closed, it can be written locally as an exact form like following,
    \begin{equation}\label{chern-simons}
        \textrm{ch}_n(\mathcal{F})=\textrm{d}Q_{2n-1}(\mathcal{A},\mathcal{F}),
    \end{equation}
    where $Q_{2n-1}(\mathcal{A},\mathcal{F})\in\mathfrak{g}\otimes\Omega^{2n-1}(M)$, $M$ is the base manifold. Here, the Chern-Simons form is defined as
    \begin{equation}\label{chern-simons defenition}
        Q_{2n-1}(\mathcal{A},\mathcal{F}):=\frac{1}{(n-1)!}\left(\frac{i}{2\pi}\right)^{n}\int_0^1\textrm{d}
        t\ \textrm{tr}(\mathcal{A}\mathcal{F}_t^{n-1})
    \end{equation}
    where
    \begin{equation}\label{Ft}
        \mathcal{F}_t=t\textrm{d}\mathcal{A}+t^2\mathcal{A}^2=t\mathcal{F}+(t^2-t)\mathcal(A)^2.
    \end{equation}
    Similarly, integrating the Chern-Simons form over BZ, we can obtain a number called Chern-Simons invariant
    \begin{equation}\label{csn}
        \textrm{CS}_{2n-1}[\mathcal{A},\mathcal{F}]:=\int_{\textrm{BZ}^{2n-1}}Q_{2n-1}(\mathcal{A},\mathcal{F}).
    \end{equation}

  For U(1) Abelian group, due to $f_t=t\textrm{d}a$ ($f$ is the curvature form and $a$ is connection form of U(1) group), the Chern-Simons can be simply written as:
  \begin{equation}\label{abelian}
    Q_{2n-1}(A)=\frac{1}{n!}\left(\frac{i}{2\pi}\right)^n
    [A(\textrm{d}A)^{n-1}].
  \end{equation}
  The topology field theory for the linear responses of a topology insulator in symmetry class A or AII and in $d$ spatial dimensions is then given by the Chern-Simons action in $D=d+1$ spacetime dimensions,

  \begin{eqnarray}\label{linear response}
     S_{\textrm{CS}}^{D=2n+1}&=& 2\pi \cdot\textrm{Ch}_n\int_{\mathbb{R}^{D=2n+1}}Q_{2n+1}(A)\nonumber \\
     &=&2\pi i\cdot\textrm{Ch}_n\cdot\textrm{CS}_{2n+1}[A].
  \end{eqnarray}

  Here, the physical response of the topological insulator in even-dimensional spatial space $d$ can be obtained by coupling an external space-time dependent U(1) Abelian gauge group. While the Chern-Character number $\textrm{Ch}_n$ computed from the the Berry curvature form in $d=2n$ space dimensions. For the gauge transformation (\ref{connection}) of $\mathcal{A}$ is not covariant, the Chern-Simons character not unique defined. For two different choices of gauge $\mathcal{A}$ and $\mathcal{A'}$, which are connected by a gauge transformation like equation (\ref{connection}) and (\ref{1}), we have
  \begin{equation}\label{chern-simons different}
    Q_{2n-1}(\mathcal{A'},\mathcal{F'})-Q_{2n-1}(\mathcal{A},\mathcal{F})
    =Q_{2n-1}(g^{-1}\textrm{d}g,0)+\textrm{d}\alpha_{2n},
  \end{equation}
  where $\alpha_{2n}$ is some $2n$ form\cite{ryu2010topological,nakahara2003geometry}. Implementing exterior $d$ to the equation (\ref{chern-simons different}) two sides and in the light of the relation with Chern characteristic (\ref{chern-simons}), one can prove the equation (\ref{chern-simons different}) is correct (Note that $Q_{2n-1}(g^{-1}\mathcal{A}g,g^{-1}\mathcal{F}g)=Q_{2n-1}(\mathcal{A},\mathcal{F})$).
  The first term of right hand side of (\ref{chern-simons different}) is a very important invariant called winding number density, it's explicit formula can be obtained according to (\ref{chern-simons defenition})
 \begin{eqnarray}
    &&Q_{2n-1}(g^{-1}\textrm{d}g,0)\nonumber \\
    &=& \frac{1}{(n-1)!}\left(\frac{i}{2\pi}\right)^{n}\int_0^1
     \textrm{d}t\ \textrm{tr}(\mathcal{A}(t^2-t)^{n-1}\mathcal{A}^{2n-2})\nonumber \\
    &=&\frac{1}{(n-1)!}\left(\frac{i}{2\pi}\right)^{n}\textrm{tr}[(g^{-1}\textrm{d}g)^{2n-1}]
      \int_0^1\textrm{d}t\ (t^2-t)^{n-1}\nonumber  \\
    &=& \frac{(-1)^{n-1}}{(n-1)!}\left(\frac{i}{2\pi}\right)^{n}\textrm{tr}
     [(g^{-1}\textrm{d}g)^{2n-1}]\textrm{B}(n,n)\nonumber \\
    &=& \frac{(-1)^{n-1}(n-1)!}{(2n-1)!}\left(\frac{i}{2\pi}\right)^{n}
     \textrm{tr}[(g^{-1}\textrm{d}g)^{2n-1}]\label{winding number density} \\
    &=&\omega_{2n-1}[g]
 \end{eqnarray}

where $\textrm{B}(n,n)$ is Beta function. The integral of $\omega_{2n-1}[g]$ is winding number
\begin{equation}\label{winding number}
    \textrm{W}_{2n-1}[g]:=\int_{\textrm{BZ}^{2n-1}}Q_{2n-1}(g^{-1}\textrm{d}g,0)
    =\int_{\textrm{BZ}^{2n-1}}\omega_{2n-1}[g]
\end{equation}
it's a integer, which counts the nontrivial winding of the map $g(k): \textrm{BZ}^{2n-1}\rightarrow \textrm{U(N)}$. Since it's very important for use in physics, we outline a prove for base manifold $S^{2n}$.
\par Consider a principle fibre bundle $(P, S^{2n}, \pi, G)$ with the connection one-form $\omega$ on $P$. Since the base manifold $S^{2n}$ is non-trivial, it is not possible to define gauge potential globally.  We thus need to split the $S^{2n}$ into two parts to cover .  We define the gauge potential $A_{\pm}=\sigma_{\pm}^{*}\omega$ where $\sigma_{\pm}$ are sections associated with the trivialization $(H_{\pm},\phi_{\pm})$ over the slightly enlarged hemispheres of $S^{2n}$ as the local coverings, and with the overlap region $S^{2n-1}\subset H_{+}\cap H_{-}$. Define a local curvature two-form $F_{\pm}=\textrm{d}A_{\pm}+A_{\pm}\wedge A_{\pm}$. The transformation function is $g\in G$. Then, we have a $2n-1$-form on the overlap region $H_{+}\cap H_{-}\supset S^{2n-1}$ according to equation (\ref{chern-simons different}):
\begin{equation}\label{}
    \omega_{2n-1}[g]=Q_{2n-1}(A_{+},F_{+})-Q_{2n-1}(A_{-},F_{-})+\textrm{d}\alpha.
\end{equation}
 Based on Stokes' theorem (\ref{stokes}) then we have

\begin{eqnarray}\label{wd}
   &&(S^{2n-1}, \omega_{2n-1}[g])\nonumber \\
   &=&(S^{2n-1},Q_{2n-1}(A_{+},F_{+})-Q_{2n-1}(A_{-},F_{-})+\textrm{d}\alpha)\nonumber  \\
   &=&(S^{2n-1},Q_{2n-1}(A_{+},F_{+})-Q_{2n-1}(A_{-},F_{-})) \nonumber \\
   &=&(\partial H_{+},Q_{2n-1}(A_{+},F_{+}))-(\partial H_{-},Q_{2n-1}(A_{-},F_{-}))\nonumber  \\
   &=&(H_{+},\textrm{d}Q_{2n-1}(A_{+},F_{+}))-(H_{-},\textrm{d}Q_{2n-1}(A_{-},F_{-}))\nonumber \\
   &=& (H_{+},\textrm{ch}_n(F_{+}))-(H_{-},\textrm{ch}_n(F_{-}))\nonumber \\
   &=& (S^{2n},\textrm{ch}_n(F))\nonumber  \\
   &=& \textrm{Ch}_n[F]=\textrm{W}_{2n-1}[g]\in\mathbb{Z}\label{chern number and winding n},
\end{eqnarray}
where we have used the $(S^{2n-1},\textrm{d}\alpha)=(\partial S^{2n-1}, \alpha )=0$ due to the boundless of sphere.
\par
The Chern-Simons effective action of topology field theroy with interacting can be written in terms of the imagine-time single particle Green function $G\equiv G(\mathbf{k},\omega) $\cite{qi2011topological},($\mathbf{k}$ is the reciprocal vector, $\omega$ is frequency), the prefactor of this form in $2n+1$ D can be expressed as\cite{wang2010topological,volovik2003universe}
\begin{equation}\label{N}
    N_{2n+1}[G]\equiv -\int_{BZ\times\mathbb{R}^{\omega}}\omega_{2n+1}[G^{-1}]=-\textrm{W}_{2n+1}[G^{-1}],
\end{equation}
it is an integer, where $G$ is the imagine-time Green function. Therefore, $N_{2n+1}[G]$ is a inter valued topological invariant. Finally, the interacting Chern-Simons effective action can be written in a more compact way,
\begin{eqnarray}\label{interacting chern-simons}
  S_{\textrm{CSi}}^{D=2n+1}&=& 2\pi \cdot N_{2n+1}[G]\cdot\textrm{CS}_{2n+1}[A]\nonumber \\
   &=& 2\pi \cdot\textrm{W}_{2n+1}[G^{-1}]\cdot\textrm{CS}_{2n+1}[A],
\end{eqnarray}
this beautiful mathematical structure of topology field theory is amazing, from which one can obtain various of physical quantities, such as quantization of hall conductance, topological magnetoelectrical effect, and so on\cite{qi2008topological}. We will give a systematic discussion in later of our paper.
\par
In summary, Chern character is an very important characteristic classes denoting the non-triviality of complex principle fiber bundles. It sets up the relationship between differential geometry and topology, one can find the beautiful mathematical topology structure in CMP through the Berry phase gauge potential. Another important mathematical tool in CMP to study topological phenomenons is homotopy theory, which investigating the map from $S^{n}$ to the manifold being studied. This will be discussed in the following subsection.
\subsection{Homotopy theory}\label{Homotopy}
 Ever since 1970s\cite{mermin1979topological}, the language of homotopy group has been used to classify defects in ordered media in CMP\cite{volovik2003universe}. For example, the point defects in 3D space are determined by classes of mapping $S^2\rightarrow R$ of a closed surface $S^2$ embracing the defect point into the vacuum manifold. These topological classes form the second homotopy group $\pi_2(R)$. Another example is the winding number defined in the last subsection, it actually indicates different mapping from $S^{2n-1}\rightarrow G(\mathbb{C})$ see the equation of (\ref{winding number}) and (\ref{wd}). In homotopy theory, it's written as $\pi_{2n-1}(G(\mathbb{C}))$. It indicates whether the manifold of $G(\mathbb{C})$ has the $2n$ dimensional "hole" or not. So we can also get fire bundle topology properties by investigating how many different mapping from base manifold to structure group that can't continue deformation into each other. In fact, the homotopy theory tells us that $\pi_{2n-1}(G(\mathbb{C}))=\mathbbm{Z}$, and $\pi_{2n}(G(\mathbb{C}))$= 0 for large enough rank matrix. Hence, it's a rough classification principle, but one can immediately know whether the fiber bundle is non-trivial or not although don't know the specific value of $\mathbb{Z}$.
 \par
In homotopy theory, we are interested in continuous deformation of maps of one to another. Let X and Y be topological spaces and let F be the set of continuous maps, from X to Y. We introduce an equivalence relation, called 'homotopic to', in F by which two maps $f,g \in$ F
are identified if the image $f(X)$ is continuously deformed to $g(X)$ in Y.
\par
Let $X,Y$ are two $m$ dimensional manifolds. $f,g: X\rightarrow Y$ are continuous maps, if there is a map
\begin{equation}\label{homotopy}
    F: X\times[0,1]\rightarrow Y
\end{equation}
such that $F(x,0)=f(x), F(x,1)=g(x), x \in X$, then the map $F$ is called  a homotopy between $f$ and $g$, denoting as $f\stackrel{F}{\simeq} g$. Homotopy is an equivalence relation, the topology property of $X$ and $Y$ can be compared according to the equivalence class. For example, if the manifold of $X$ is the Hamiltonian, then one can classify them by homotopy discussion as Qi\cite{qi2008topological} for non-chiral topological insulator and Ryu\cite{ryu2010topological} for chiral topological insulator.
\par
The most important thing of this article is to classify fibre bundles. How many different principle bundle $P(M,G)$ can be constructed for a given base manifold $M$ and structural group $G$ with homotopy language? For the n-sphere base manifold $S^n$, it can be covered with two open sets because the non-triviality of sphere. The principle fibre bundle $P$ can be written as a trivial direct product restricted in every open set,
\begin{equation}\label{sphere open}
    P|_{x\in U_\alpha}=U_\alpha\times G.
\end{equation}
However, we need a gauge transformation to connect in the overlap region of coverings (homotopically to $S^{n-1}$), it's a map from overlap region to Lie group $G$,
\begin{equation}\label{overlap of sphere}
    \psi_{\alpha\beta}:U_\alpha\cap U_\beta\rightarrow G.
\end{equation}
It is homotopically to the map form $S^{n-1}$ to $G$. Consequently, the classification of $P(S^n, G)$ is tantamount to calculating $\pi_{n-1}(G)$.
\par For the classification of a general fiber bundle $P(M,G)$, we need classification theorem of $G$-bundle in mathematic\cite{nakahara2003geometry}. That's to say for one principle fiber bundle there is a corresponding universal bundle is defined as
\[\begin{CD}
G  @>>>\xi(G)\\
@.        @VVV \\
       @. B_G,
\end{CD}\]
with the property
\begin{equation}\label{universal bundle}
    \pi_{k}(\xi(G))=0,~~~~ k\geq1.
\end{equation}
One can classify the fiber bundles homotopically because the universal bundles has the following properties:
  If there is a continuous map from manifold $M$ to $B_G$
\begin{equation}\label{Classifying space}
    f: M\longrightarrow B_G,
\end{equation}
and the associated bundle map $f^{*}$
\begin{equation}\label{fibre map}
    f^{*}: \xi(G)\longrightarrow P(M,G)
\end{equation}
we will get that the given principle bundle is isomorphism to the pull back of the universal bundle
\begin{equation}\label{isomorphism}
    P(M,G)\simeq f^{*} \xi(G).
\end{equation}
In addition, if $f\simeq g$, then $f^{*} \xi(G)\simeq g^{*} \xi(G)$. The base manifold of universal bundle is called classifying space.
 That is to say there is a one to one corresponding between the equivalence of principle fiber bundle $P(M,G)$
and the set of continuous mapping homotopy class $[M,B_G]$. The key point is now that homotopically different bundles are distinguished by homotopically distinct maps.
Because of the following long fiber bundle exact sequence\cite{hou1997differential},
$$\cdots\rightarrow\pi_k(\xi(G))\rightarrow\pi_k(B_G)\rightarrow\pi_{k-1}(G)\rightarrow\pi_{k-1}(\xi(G))\rightarrow\cdots$$
one can get the set $[S^n,B_G]=\pi_n(B_G)=\pi_{n-1}(G)$ due to equation (\ref{universal bundle}), which agrees well with the results discussed above.
\par
For example, for unitary fibre bundles, the classifying space is Grassmanian $G_{N,n}(\mathbb{C})=U(N)/(U(n)\times U(N-n))$ for sufficiently large $N$. Thus, the set of different TSM in class A is the set $[M,G_n(C^{\infty})]=[M,BU]$, where $G_n(C^{\infty})=\textrm{lim}_{N\rightarrow\infty}G_{N,n}(\mathbb{C})\equiv BU$. We thus found for the set $Vect_n(M,\mathbb{C})$ of inequivalent $U(n)$ bundles over $M$ the expression
\begin{equation}\label{class A Homotopy}
    Vect_n(M,\mathbb{C})=[M,BU],
\end{equation}
which is known for some rather simple base manifolds. In particular for spheres $S^d$ (which can be demonstrated the continuum model of class A topological matter ), the different state can be expressed as
\begin{equation}\label{class a}
    Vect_n(S^d,\mathbb{C})=[S^d,BU]=\pi_d(BU)=\pi_{d-1}(U),
\end{equation}
so the classifying space for class A is $BU$, which is denoted as $C_0=(U(k+m)/U(k)\times U(m))\times \mathbb{Z}$ in ref\cite{kitaev2009periodic}.

\par
One more thing we need to introduced is Bott periodicity, it's an very important theorem both in homotopy and topology K theory (will be discussed later). It's firstly proved by Bott using Morse theory, one can reference the book of Milnor\cite{milnor1963morse}.
 \par
 Next, we introduce the 2-periodicity theorem for large rank unitary group. Firstly, we show that the homotopy group will be stable with the rank increasing. Geometrically, the unitary matrices satisfying $U^\dag U=I$ are just unit vector in $C^N$, the shape is tantamount to sphere $S^{2N-1}$. The action of $U(N)$ is transitive and the stable group of $x\in S^{2N-1}$ is $U(N-1)$, so sphere can be seen as quotient group like $S^{2N-1}\simeq U(N)/U(N-1)$.
For the quotient group structure of sphere, there will be a short exact sequence
\begin{equation}\label{sphere exact sequence}
    0\longrightarrow U(N-1)\stackrel{i}{\longrightarrow }U(N)\stackrel{\pi}{\longrightarrow} S^{2N-1}\longrightarrow0,
\end{equation}
where $i$ and $\pi$ are inclusion map and projection map, respectively. Since the simplicity of low rank homotopy group of simply-connected sphere, one can get the long exact sequence
\begin{eqnarray}\label{long exact sequence of unitary}
   &&\rightarrow\pi_{k+1}(S^{2N-1})\stackrel{\partial}\rightarrow\pi_k(U(N-1))\nonumber  \\
   &&\stackrel{i_*}\hookrightarrow\pi_k(U(N))
    \stackrel{\pi_*}\rightarrow\pi_k(S^{2N-1})\stackrel{\partial}\rightarrow
\end{eqnarray}
where $\partial$ is the boundary map, $i_*$ and $\pi_*$ are the push forward map of $i$ and $\pi$.
Applying the long exact sequence above (\ref{long exact sequence of unitary}) and the well know triviality of low rank homotopy group of sphere, we get the high rank of unitary group is stable as following
\begin{equation}\label{stable U}
    \pi_k(U(N))=\pi_k(U(N+1))=\cdots=\pi_k(U_{\infty}), ~~~~ k<2N.
\end{equation}
The homotopy group of $U(N)$ will be stable with enough big rank $N$, now one may ask the question: How does the homotopy group for stable range $U(N)$ varies with the $k$ increasing? In order to answer this question, a few definitions should be introduced.
For a given connected manifold $M$, a map $\gamma(t)~(0\leq t\leq1)$ denotes a path from point $p=\gamma(0)$ to point $q=\gamma(1)$ on $M$, the space of all of paths is called path space, denoting as $\Omega(M,p,q)$. In the special case, $p=q$, the space is called loop space of manifold, denoting as $\Omega(M,p)=\Omega(M)$. One can get the beautiful formula from the long exact sequence of fiber bundle
\begin{equation}\label{loop space}
    \pi_i(\Omega(M))=\pi_{i+1}(M), ~~~~i\geq1.
\end{equation}
Combining above formula and equation (\ref{class a}) we obtain
\begin{equation}\label{loop space of class a}
    \Omega (BU)=U,
\end{equation}
that is to say, $U$ is the loop space of $BU$. Now we can describe the Bott periodicity theorem for unitary group with Morse theory\cite{milnor1963morse,hou1997differential,freedman2011projective}. We approximation the loop space of $U(2N)$ by minimal geodesics from $\mathbb{I}$ to $-\mathbb{I}$ : $d(\lambda)=e^{i\lambda A}$, where $A$ is a Hermitian matrix satisfying $A^2=1$ so that $d(0)=\mathbb{I}$ and $d(\pi)=-\mathbb{I}$ (one can prove this is a minimal geodesic). Any such matrix  $A$ divides $\mathbb{C}^{2N}$ into +1 and -1 eigenspaces or, in other words, it can be written in the form
\begin{equation}\label{Unitary A}
    A=U^{\dag}\left(
                   \begin{array}{cc}
                     \mathbb{I}_N & 0 \\
                     0 & -\mathbb{I}_N \\
                   \end{array}
                 \right)U,
\end{equation}
where $\mathbb{I}_N$ is the $N\times N$ matrix. Therefore $A$ can be viewed as the $N$ dimensional plane in $\mathbb{C}^{2N}$, then $A\in G_{N,2N}\cong U(2N)/U(N)\times U(N)$. Thus the manifold $\Omega^d$ of the set minimal geodesics from $\mathbb{I}$ to $-\mathbb{I}$ on $U(2N)$ homotopiclly equivalence to $G_{N,2N}$, denoting as $\Omega^d\simeq G_{N,2N}$. If denoting the path space of $U(2N)$ as $\Omega (U(2N))$, the $\Omega^d$ and $\Omega$ will have the homotopy relationship according to the foundation theorem of Morse,
\begin{equation}\label{homotopy of path space}
    \pi_i(\Omega^d)=\pi_i(\Omega),~~~~~~i<2N+2.
\end{equation}
Due to the equation (\ref{loop space}), we have the following
\begin{equation}\label{Grassminan and U}
    \pi_i(G_{N,2N})=\pi_{i+1}(U(2N)),~~~~~i<2N.
\end{equation}
In other hand, Grassmann maifold $G_{N,2N}$ is the classifying space of $U(N)$ fiber bundle, one can write the fibration sequence like followings
$$U(N)\rightarrow V_{N,2N}\rightarrow G_{N,2N}$$,
where $V_{N,2N}$ is the Stiefel manifold, with the property $\pi_i(V_{N,2N})=0$ for $i<2N$.
Therefore, we get the relationship following:
\begin{equation}\label{u and grassmann}
    \pi_i(G_{N,2N})=\pi_{i-1}(U(2N)),~~~~~i<2N.
\end{equation}
Combining the eqution (\ref{Grassminan and U}) and above we get the famous Bott 2-periodicity theorem for stable range unitary group,
\begin{equation}\label{bott periodicity for U}
    \pi_{i-1}(U(2N))=\pi_{i+1}(U(2N)),~~~~~1<i<2N.
\end{equation}
For the stable homotopy group we have
\begin{equation}\label{bott periodicity for stable u}
    \pi_{i+2}(U)=\pi_i(U),~~\pi_{2k}(U)=0,~~\pi_{2k+1}(U)=\mathbb{Z}.
\end{equation}
\par
Until now, we can write the classification question in class A TSM in homotopy theory elegant , the distinct states can be denoted with the homotopy group element $\pi_{d-1}(U)=\pi_d(\Omega U)=\pi_d(BU)$, so the classifying space of calss A is just the loop space of the group of fiber bundle, the next element in the Bott clock\cite{stone2010symmetries}. For the class AIII, the group is Grassmann, so the classifying space is the loop space of $\Omega U$, denoting as $\Omega^2 U=U$. From the Bott periodicity theorem, we know there are non-trivial TSM in even dimensional space for class A, while TSM can be found in odd space in class AIII chiral symmetry unitary. For example, the QIHE can be found in 2\cite{PhysRevLett.45.494} or 4-dimension\cite{zhang2001four} space but not 3-dimension space, the reasons are the Bott periodicity theorem (\ref{bott periodicity for stable u}).
\par
For systems which obey anti-unitary symmetries such as time reversal symmetry $T$ and charge conjugation $C$, the structure group will be orthogonal group $O(n)$ at the  symmetry invariant points of Brillion zone\cite{teo2010topological}. The real analogue of Bott periodicity with period 8 have another version of modular of Clifford algebra. Firstly, we show the modular space with 8 period. Assuming $n$ is a multiple of 16, so we consider the action of $O(16n)$ on a vector space $V$ over $\mathbb{R}$ of dimension 16$n$. Let the set $\{J_i: i=1,\cdots ,k-1\}$ be the anti-commuting orthogonal complex structures, which means they obey
\begin{equation}\label{complex structure}
    J_iJ_j+J_jJ_i=-2\delta_{ij}\mathbb{I}.
\end{equation}
Now let $\Omega_{k}(16n)$ be the space of complex structure $J$ set that anti-commuting with fixed $\{J_i\}_1^{k-1}$. Hence, it is the subspace of $\Omega_{k-1}(16n)$. The sequence of seeking for the homogenous space of $\Omega_i$ can be viewed as a sequence symmetry breaking process\cite{stone2010symmetries},
$$\Omega_{k}(16n)\subset\Omega_{k-1}(16n)\subset\cdots\subset\Omega_1(16n)\subset O(16n)$$
This process can be physically realized in chiral symmetry field theory\cite{dejonghe2012bott}.
\par
 The subgroup of $O(16n)$ that commutes with $J_1$ is $U(8n)$ and $O(16n)$ action on the space is transitivity. So the space is isomorphism to the homogenous space
\begin{equation}\label{omega1}
    \Omega_1(16n)\simeq O(16n)/U(8n).
\end{equation}
\par In a similar manner, the subgroup of $U(8n)$ that commutes with both $J_1$ and $J_2$ is the symplectic group $Sp(4n)$, so the space of $\Omega_2(16n)$ is homogenous space as
\begin{equation}\label{omega2}
    \Omega_2(16n)\simeq U(8n)/Sp(4n).
\end{equation}
\par After introducing $J_3$ we consider the operator $\alpha=J_1J_2J_3\in O(16n)$, satisfying $\alpha^2=\mathbb{I}$. Therefore it possesses two eigenspaces $V_{\pm}$, in which it takes the value $\pm 1$ respectively. Because of  $[\alpha, J_1]=0$ and $[\alpha,J_2]=0$, $V_{\pm}$ doesn't change under the action of $J_1$ or $J_2$. The subgroup of $Sp(4n)$ commuting with $J_1, J_2, J_3$, and preserving this structure is $Sp(2n)\times Sp(2n)$. Finally, the homogenous space is the quaternionic Grassmann manifold
\begin{equation}\label{omega3}
    \Omega_3(16n)\simeq Sp(4n)/(Sp(2n)\times Sp(2n)).
\end{equation}
\par Next introduce $J_4$ and let $\beta=J_3J_4$. We have $[\beta,J_1]=[\beta,J_2]=0$, $\alpha\beta+\beta\alpha=0$ and $\beta^2=-\mathbb{I}$. $\beta$ therefore preserves the quaternionic structure and is a quanternionic isometry from $V_+$ to $V_-$. It is an element of $Sp(2n)$. So the group preserving this structure is the diagonal subgroup of $Sp(2n)\times Sp(2n)$. Finally, the space of $\Omega_4(16n)$ is the homogenous space as following
\begin{equation}\label{omega4}
    \Omega_4(16n)\simeq (Sp(2n)\times Sp(2n))/Sp(2n)=Sp(2n).
\end{equation}
\par Let introduce $J_5$ and construct $\gamma=J_1J_4J_5$ which has $\gamma^2=\mathbb{I}$ and commutes with $\alpha$ and $J_1$ but anti-commutes with $J_2$. So $\gamma$ therefore acts within the $V_+ (V_-)$ eigenspace of $\alpha$ and divides it into two mutually orthogonal eigenspaces $W_{\pm}$ with $W_-= J_2 W_+$. Conversely such a decomposition uniquely determines $J_5$. $\gamma$ is transitivity on the action of $Sp(2n)$, and $U(2n)$ is preserving $W$ complex structure $J_1$. So the space of $\Omega_5$ is
\begin{equation}\label{omega5}
    \Omega_5(16n)\simeq Sp(2n)/U(2n).
\end{equation}
\par
Next introduce $J_6$ and set $\delta=J_2J_4J_6$ which commutes with $\alpha$ and $\gamma$ but anti-commutes with $J_1$. Therefore, it will splits again in the $W_{\pm}$, denoting the mutually orthogonal eigenspaces $X_{\pm}$. The eigenspaces satisfies $X_- = J_1 X_+$. Since $J_1$ is the complex structure, $U(2n)$ is transitivity , the subgroup preserving this structure is $O(2n)$. So the range of choices for $J_6$ is the homogenous space
\begin{equation}\label{omega6}
    \Omega_6(16n)\simeq U(2n)/O(2n).
\end{equation}
Now introduce $J_7$ and set $\eta=J_1J_6J_7$ which commutes with $\alpha, \gamma, \delta$ and splits $X_{\pm}$ into $\pm 1$ subspaces $Y_{\pm}$. The group preserving this decomposition is therefore $O(n)\times O(n)$. So it's easy written the space of $\Omega_7(16n)$, just the orthogonal Grassmann manifold like following
\begin{equation}\label{omega7}
    \Omega_7(16n)\simeq O(2n)/(O(n)\times O(n)).
\end{equation}
Finally, we introduce $J_8$. Now the orthogonal transformation $\epsilon=J_7J_8$ commutes with $\alpha, \gamma, \delta$ but anticommutes with $\eta$. It is therefore an isometry mapping $ Y_+$ to $Y_-$. So the group preserving this structure is the diagonal group of $O(n)\times O(n)$. The space of $\Omega_8(16n)$ is following
\begin{equation}\label{omega8}
    \Omega_8(16n)\simeq (O(n)\times O(n))/O(n)=O(n).
\end{equation}
\par
From above tedious enumeration, we obtain the modular space of Clifford algebra has the period 8. This process closely follow \cite{milnor1963morse} and \cite{hou1997differential}, readers who are interesting in the details may reference the books. Morse theory have established the relationship between the loop space $\Omega (O(16n))$ and the minimal geodesic space of $\Omega^d(O(16n))$. In addition, for the homogenous space like above $\Omega_i(16n)$, one can prove the space of minimal geodesic set from $J_i$ to $-J_i$ is homeomorphism to $\Omega_{i+1}(16n)$. In summary, the link can be set up using minimal geodesic space as following
\begin{equation}\label{omega^d}
    \Omega(\Omega_i(16n))\simeq\Omega^d(\Omega_{i}(16n))\simeq \Omega_{i+1}(16n)
\end{equation}
Combing this relationship and equation (\ref{loop space}) we obtain
$$\pi_{i-1}(\Omega_i(16n))=\pi_i(\Omega_{i+1}(16n))$$
where $n$ is big enough.
With the $n$ limit to infinity, we have the following homotopy equivalence
\begin{equation}\label{Bott O1}
    \pi_kO=\pi_{k-1}\Omega_1=\cdots=\pi_1\Omega_{k-1}.
\end{equation}
Orthogonal group will be in the stable range as the unitary group with the enough big rank, we can write the explicit relationship as following (n is large enough)
\begin{eqnarray}\label{BottO1}
   \pi_k(O(16n))&=& \pi_{k-1}(O(16n)/U(8n))\nonumber \\
   &=&\pi_{k-2}(U(8n)/Sp(4n)) \nonumber \\
   &=&\pi_{k-3}(Sp(4n)/(Sp(2n)\times Sp(2n)))\nonumber \\
   &=&\pi_{k-4}(Sp(2n))\nonumber  \\
   &=&\pi_{k-5}(Sp(2n)/U(2n))\nonumber  \\
   &=&\pi_{k-6}(U(2n)/O(2n))\nonumber \\
   &=&\pi_{k-7}( O(2n)/(O(n)\times O(n)))\nonumber \\
   &=&\pi_{k-8}(O(n)).
\end{eqnarray}
It is straightforward to calculate the zero rank homotopy group of these homogenous space
\begin{eqnarray}\label{BottO2}
  \pi_0(O(n)) &=&\mathbb{Z}_2,\nonumber \\
  \pi_0(O(2n)/U(n)) &=& \mathbb{Z}_2,\nonumber \\
  \pi_0(U(2n)/Sp(n)) &=& 0,\nonumber \\
  \pi_0 (Sp(2n)/(Sp(n)\times Sp(n)))&=&\mathbb{Z},\nonumber  \\
  \pi_0(Sp(n)) &=& 0,\nonumber \\
  \pi_0(Sp(n)/U(n)) &=& 0,\nonumber \\
  \pi_0(U(n)/O(n)) &=& 0,\nonumber\\
  \pi_0 (O(2n)/(O(n)\times O(n)))&=& \mathbb{Z}.
\end{eqnarray}
Combining (\ref{BottO1}) and (\ref{BottO2}), we can compute any of the stable homotopy groups.
\par  Until now, we have introduced the well known significant Bott periodicity theorem of unitary group and orthogonal group in the stable limit range. We can study the fiber bundle topology through the homotopy classification $G$-bundle theorem. For the structure group like homogenous space $\Omega_i$, one can easily know the corresponding classifying space is just the loop space of it as $\Omega(\Omega_i)=\Omega_{i+1}$. Consequently, we can homotopically classify all structural group like $\Omega_i$ principle fiber bundles. Actually, the eight real classes and two complex classes homogenous spaces are all symmetry spaces in mathematics, called Carten ten symmetry spaces.
\par
In physics, the complex structure  space $\Omega_i(O)$ is the same of imposition a symmetry such as time reversal to a system classified by the space of $\Omega_{i-1}(O)$. So when applying a symmetry on the homogenous spaces of a physical system, the resulting space is tantamount the loop space of before.
\par The homotopy theory is a roughly classification principle compared with the homology characteristic classes, but it is very convenient to
classify the non-trivial TSM quickly. For example, the time-reversal invariant topological insulators in class AII, the group of fiber bundles can be identified with $U(8n)/Sp(4n)$, the classifying space will be its loop space $\Omega_3$. From homotopy theory discussed in this subsection, we know in 4 dimension space (because time reversal change k to -k, so we should compute -4 rank homotopy group\cite{kitaev2009periodic}) $\pi_{-4}(\Omega_3)=\pi_{4}(\Omega_3)=\pi_0(\Omega_7)=\mathbb{Z}$, this indicates there are non-trivial TSM, which is the 4D quantum Hall effect. As the same reason, in 3 dimension and 2 dimension space, we can obtain $\pi_{-3}(\Omega_3)=\pi_0(\Omega_8)=\mathbb{Z}_2$ and $\pi_{-2}(\Omega_3)=\pi_0(\Omega_1)=\mathbb{Z}_2$, respectively. These topological insulators correspond to quantum spin Hall effect and 3 dimension strong topological insulator, respectively.
\par
In addition, we can classify all non-trivial  TSMs  in the special dimension $d=0$ until now, because in this case, we don't need to think about the problem of involution of reciprocal $k$ space\cite{kitaev2009periodic,freedman2011projective,wen2012symmetry}. Following the convention of Ref.\cite{kitaev2009periodic}, we rename the classify spaces by $R_q, q =0, 1,\cdots, 7.$  for real case and $C_q, q=0, 1$ for complex case, respectively. We obtain five non-trivial TSMs in 0-dimensional space, which is denoted in the table $\textrm{\ref{table1}}$.
\begin{table}
  \centering
  \caption{Table of classifying space $R_q, C_q$. The first column denotes the symmetric spaces. The second column 'Cartan label' is the name given to the corresponding symmetric space in $\acute{\textrm{E}}$lie Cartan's classification scheme. The last column is the corresponding zero homotopy group.}\label{table1}
\begin{tabular}{c|c|c}
  \hline\hline
  Classifying Space & Cartan label & $\pi_0$ \\\hline
       $ R_0= \{O(2n)/(O(n)\times O(n))\}\times \mathbb{Z}$ & AI & $\mathbb{Z}$\\
       $R_1=O(16n) $& BDI &$\mathbb{Z}_2$\\
       $R_2=O(16n)/U(8n) $& D&$\mathbb{Z}_2$ \\
       $R_3=U(8n)/Sp(4n) $& DIII &0\\
       $R_4=\{Sp(4n)/(Sp(2n)\times Sp(2n))\}\times\mathbb{Z} $& AII& $\mathbb{Z}$\\
       $R_5=Sp(2n) $& CII& 0\\
       $R_6=Sp(2n)/U(2n)  $& C&0\\
       $R_7=U(2n)/O(2n) $& CI&0\\\hline
       $C_0={U(2n)/(U(n)\times U(n))}\times \mathbb{Z}$& A&$\mathbb{Z}$\\
       $C_1= U(n)$& AIII &0\\ \hline\hline
\end{tabular}
\end{table}
\par  Bott periodicity predict the pattern of topology state of matter have 8-period for real case and 2-period for complex case in the space dimension. In fact, the homogenous space $\Omega_iO$ is called symmetry class in mathematic,  $\acute{\textrm{E}}$lie  Cartan observed that there are precisely ten families of compact symmetric spaces in the 1920s, it contains two complex and eight real symmetric spaces, respectively. In physics, Altland-Zirnbauer \cite{PhysRevB.55.1142} (AZ) have found the one to one correspondence between single-particle Hamiltonians and the set of symmetric spaces.
\subsection{Map Degree}
 In this subsection we briefly introduce some differential topology contents. In order to investigate the topology of manifold, we can study the Characteristic class or homotopy group like above subsections. Another tool to obtain the topology properties is more subtle. We mainly list the important Brouwer degree\cite{wang2010equivalent}, which is used to prove the equivalence of topological invariants discrete formula used by Kane\cite{kane2005z_} and integration formula used by Qi\cite{qi2008topological}. A reference book\cite{milnor1997topology} of Milnor is recommend to read, which is a very classic and many important theorems are contained.
 \par
 Firstly, we introduce some definitions used in differential topology. Let $f: M_1\rightarrow M_2$ be a smooth map, denoting $m_1, m_2$ as the dimension of corresponding manifold. If (rank$f$)$_p < m_2, p\in M_1$, then the point of $p$ is called a critical point of $f$, the image $f(x)$ is called a critical value. While for (rank$f$)$_p = m_2, p\in M_1$, it's named a regular point of $f$ and the image $f(x)$ is called a regular value, where (rank$f$)$_p$ is the usual Jacobin determinant at point $p$. According to the famous theorem of Morse-Sard\cite{milnor1997topology}, we know the measure of the set of critical value is zero, so nearly all points in $M_2$ are in regular value set. Secondly, a manifold is called orientable, if the sign of Jacobian determinant is preserved during the overlap of any open coverings.
 \par
 Now we can introduce the degree of map called Brouwer degree. Let $M_1, M_2$ are $m$ dimension orientable manifolds, $f: M_1\rightarrow M_2$ is a smooth map, $p\in M_1$ is the regular point of $f$, $q=f(p)$. Now we can consider the differential isomorphism associated the smooth map $f$, $T_pf = (df)_p :T_pM_1\rightarrow T_qM_2$, if the orientable is preserved, denoting deg$_pf$=1; while  it's reversal, denoting deg$_pf$=-1. In summary, deg$_pf$ is called the degree of $f$ at point $p\in M_1$. Assuming the manifold $M_1$ is compact, $q$ is the regular point of $f$, then
 \begin{equation}\label{degree of f}
    \textrm{deg}(f,q)=\sum_{p\in f^{-1}(q)}\textrm{ deg}_pf
 \end{equation}

 is called the Brouwer degree of $f$ about regular value $q$, where $p\in f^{-1}(q)$ are the regular points of $f$. It's evidence Brouwer degree is an integer. The geometrical meaning of Brouwer degree is very easy. Because of the inverse function theorem, there is a neighborhood open $V\subset M_2$ set of $q$ and every point $p\in f^{-1}(q)$ has an open neighborhood set $U(p)\subset M_1$, such that $f: U(p)\rightarrow V$ is a diffemorphism, it will preserve or reversal orientation according to the type of $p$. Hence, $\textrm{deg}(f,q)$ is the algebra sum of covering $V$ under the map of $f$. This is the  discret formula in differential topology.
 \par
 Another formula is an integrate form. Generally, for a map $f: M_1\rightarrow M_2$, where the two manifold have the same dimension and both are orientable. We can define the degree of $f$ as
 \begin{equation}\label{degree f}
    \textrm{deg}(f)=\int_{M_1}f^*(\omega),
 \end{equation}
 where $\omega$ is the volume form on $M_2$ satisfying $\int_{M_1}\omega=1$, and $f^*(\omega)$ is the pullback of $\omega$ on $M_1$ under $f$. One can prove the two formula is identity.
 \par
 In order to prove this, assuming every $U(p)$ is a open set of $f^{-1}(q)$. Let volume form $\omega$ on $M_2$ is $\int_{M_2}\omega\neq0$ and supp$(\omega)\subset V$, then supp$(f^*\omega)\subset\cup U(p)$ (suup stands for suuport set, it means the volume on suuport set is not zero, otherwise it is zero),
\begin{eqnarray}\label{degree f1}
   \int_{M_1}f^*(\omega)&=&\sum_p\int_{U(p)}f^*(\omega) \nonumber \\
    &=& \sum_p \textrm{deg}_pf\int_V\omega  \nonumber\\
    &=& \sum_p \textrm{deg}_pf= \textrm{deg}(f,q),
 \end{eqnarray}
 where we have used the volume form transformation formula, it is just the Jacobian determination, and the normalization condition of the volume form on $M_2$.

 \par
 In physics, Wang\cite{wang2010equivalent} have proved  the equivalent topological invariants between FKM descret\cite{fu2007topological} formula and QHZ\cite{qi2008topological} integrate formula with the Brouwer degree. He investigate the manifold $T^3$ and $SU(2)$, which are all 3-dimension, so we can study the map degree. Duan's\cite{duan2000topological} tensor current theory is a beautiful topology tool using the $\delta$ function and Brouwer degree, he established some beautiful relationship between the zeros of a vector field and the topology property, giving an inner structure of the Gauss-Binet-Chern theorem\cite{duan1993topological}. The readers who are interested can reference his articles\cite{duan2002decomposition,duan2003many}.
\subsection{Topology K-theory}\label{Topoloy K-theory}
Topology K theory firstly was constructed by Grothendieck, then was developed primarily by Atiyah and Hizebruch in the early 1960s, and for a general reference, the reader is referred to Atiyah's classic text\cite{atiyah1968index}. The topology K theory mainly intent to construct a Abelian group for vector bundles. As we all know, the direct sum of two vector bundles $E\bigoplus F$ is the direct sum of their fibers over each point. But this addition has only a semi-group structure, since $E\bigoplus G=F\bigoplus G\nRightarrow E\simeq F$. From the minimal counterexample we can get some clues, given a vector bundles of sphere by $E= TS^2$, $F=S^2\times\mathbb{R}^2$, $F$ is clearly trivial, whereas $E$, the tangent bundle of $S^2$ is well known to be non-trivial $(\pi_1(SO(2))=\pi_1(U(1))=\mathbb{Z})$. However, adding $NS^2$, the bundle of normal vectors to $S^2$, to both bundles $E, F$ we obtain the same trivial bundle $S^2\times\mathbb{R}^3$. This motivates the concept of stable equivalence,
\begin{equation}\label{stable equivalence}
    E\stackrel{s}{\simeq}F\Longleftrightarrow E\oplus Z^m\simeq F\oplus Z^n,
\end{equation}
where $Z^n=M\times K^n, K=\mathbb{R}, \mathbb{C}$ is the trivial bundle over the fixed base manifold $M$, which plays the role of an additive zero as far as stable equivalence is concerned. Let Vec$_K(M)$ be the isomorphism classes of $K-$vector bundles over $M$. Note that stably equivalent bundles can have different fiber dimension, as $m\neq n$ in general in equation (\ref{stable equivalence}). Because of a theorem called triviality of vector bundles in mathematics,  for any compact base manifold, every bundle can be augmented to a trivial bundle, i.e.,
\begin{equation}\label{triviality theorem}
    \forall G,~ \exists H, l~~ such~~that~~ G\oplus H=Z^l,
\end{equation}
this theorem has a deep connection with Whitney embedding theorem.
Combing (\ref{stable equivalence}) and (\ref{triviality theorem}), we can loose the stable equivalence condition, such that,
\begin{equation}\label{stable equivalence1}
    E\oplus G\simeq F\oplus G\Longrightarrow E\stackrel{s}{\simeq}F.
\end{equation}
Now let us construct the K group about $M$ by virtue of the Grothendieck construction: Consider the pairs $(E_1, E_2)\in\textrm{ Vec}_K(M)\times\textrm{Vec}_K(M)$. We define K($M$) to be all pairs formal difference $E - F$ of vector bundles, modulo the equivalence relation
\begin{equation}\label{k group}
    E_1-E_2=F_1-F_2\Longleftrightarrow\exists G~,E_1\oplus F_2\oplus G\simeq E_2\oplus F_1\oplus G
\end{equation}
The resulting set K($M$) is then an Abelian group with negatives given by $-(E-F)=(F-E)$.
\par
Given a smooth map $f:M\rightarrow N$ and a vector bundle $F$ over $N$, we can construct the induced bundle $f^*F$ over $M$ satisfying $(f*F)_x=F_{f(x)}$ for all $x\in M$. In particular, if the map $i:pt\rightarrow M$ is the inclusion of a point in $M$, then we obtain a map $i^*: \textrm{K}(M)\rightarrow \textrm{K}(pt)\cong\mathbb{Z}$. A vector bundle over a point is a vector space, which is given up to ismorphism by its dimension. Defining the reduced K-group by $\tilde{\textrm{K}}(M)=Ker i^*$, we obtain the splitting $\textrm{K}(M)\cong\tilde{\textrm{K}}(M)\oplus \textrm{K}(pt)$. Viewing $\tilde{\textrm{K}}(M)$ as $\textrm{K}(M)$ modulo the trivial bundles, $\tilde{\textrm{K}}(M)$ then consists of classes of stable vector bundles.
\par
The key idea in K-theory is that of stable equivalence, which is also our main physical motivation to study it.  Augmenting them by some bundles doesn't affect our physical classification, because we can adding an extra flat band on an insulator. The final twist is that K-theory deals with differences between objects rather than objects themselves.  The stable equivalence means $\tilde{\textrm{K}}(M)$ have a homotopy demonstration. Actually, K group of a base manifold is identity to the class of homotopy equivalence from base manifold to classifying space discussed in the last subsection\cite{hatcher2003vector},
\begin{equation}\label{k group u}
   \textrm{ Vec}_K(M)= [M,BG]=\tilde{\textrm{K}}(M).
\end{equation}
The isomorphism between  $[M,BG]$ and $\tilde{\textrm{K}}(M)$ can be proved explicitly by the homotopically classification principle-G bundles theorem talked in subsection homotopy. Let $\xi(G)$ is the universal fiber bundle, $f_1\simeq f_2$ and $g_1\simeq g_2$ are the maps from base manifold $M$ to classifying space $BG$. Define $f_{1,2}^*\xi(G)$ and $g_{1,2}^*\xi(G)$ are the pull back fire bundles on the base manifold $M$. We can construct the following isomorphism map according to the maps of homotopically equivalence have the isomorphism pull back bundles,
\begin{equation}\label{isomorphism}
    [f_1^*\xi(G)]-[g_1^*\xi(G)]= [f_2^*\xi(G)]-[g_2^*\xi(G)]\leftrightarrow\tilde{\textrm{K}}(M)
\end{equation}
 where $[~]$ stands for equivalence fire bundles,  and the equation is correct because of ($f_1^*\xi(G)\simeq f_2^*\xi(G)$) and ($g_1^*\xi(G)\simeq g_2^*\xi(G)$). Hence, the zero element in $\tilde{\textrm{K}}(M)$ just the pull back bundles with the homotopically equivalence maps. Consequently, we have the equation (\ref{k group u}).
 As discussed in the homotopy subsection, we know the 2-period and 8-period for complex and real stable equivalence bundles. This immediately implies in the language of K-theory for the base manifold of sphere $S^d$
 \begin{eqnarray}
    \pi_{d+2}(C_0)=\widetilde{\textrm{KU}}(S^{d+2})&=& \widetilde{\textrm{KU}}(S^d) =\pi_d(C_0)\label{K Theory}\\
   \pi_{d+8}(R_0)= \widetilde{\textrm{KO}}(S^{d+8})&=&  \widetilde{\textrm{KO}}(S^d)=\pi_d(R_0),\label{K Theory1}
 \end{eqnarray}
 where we have called $\tilde{\textrm{KU}}$ for complex K-theory and $\tilde{\textrm{KO}}$ for real K-theory, with 2-period and 8-period, respectively.
 We define
\begin{equation}\label{suspension}
    \widetilde{\textrm{KU}}^{-d}(M)=\widetilde{\textrm{KU}}(S^dM),
    ~~\widetilde{\textrm{KO}}^{-d}(M)=\widetilde{\textrm{KO}}(S^dM),\nonumber
\end{equation}
where $S$ is the reduced suspension\cite{hatcher2003vector} which for a sphere $S^k$ indeed satisfies $SS^k=S^{k+1}$. The stronger version of the Bott periodicity in K-theory now reads\cite{atiyah1968index,hatcher2003vector}
\begin{eqnarray}
  \widetilde{\textrm{KU}}^{-d-2}(M)&=& \widetilde{\textrm{KU}}^{-d}(M), \label{bott of ku}\\
 \widetilde{\textrm{KO}}^{-d-8}(M)&=&\widetilde{\textrm{KO}}^{-d}(M) ,
\end{eqnarray}
which only for $M=S^d$ trivially follows from equation (\ref{K Theory}) and (\ref{K Theory1}). Using this periodicity, the definition of $\widetilde{\textrm{KU}}^{-d}$ or $\widetilde{\textrm{KO}}^{-d}(M)$ can be formally extended to $d\in\mathbb{Z}$. From the loop space definition discussed in the subsection homotopy, we easily know the loop operator is the adjoint topological operator with suspension,
\begin{equation}\label{loop-suspention}
    Map_{*}(SX, Y)=Map_{*}(X,\Omega Y)
\end{equation}
which is the natural identification\cite{crossley2011essential} between basepoint-preserving maps from the reduced suspension of $X$ to $Y$ and from $X$ to the space of based loops in $Y$.
From this we can get the following relationship
\begin{equation}\label{KO}
    \widetilde{\textrm{KO}}^{-q}(M)=[S^qM, R_0]=[M,\Omega^qR_0]=[M,R_q],
\end{equation}
\begin{equation}\label{KU}
    \widetilde{\textrm{KU}}^{-q}(M)=[S^qM, C_0]=[M,\Omega^qC_0]=[M,C_q].
\end{equation}
Hence, one can get all topology information about the stable equivalence bundles for arbitrary base manifold $M$ by K-theory.
\par
The classification of all TSMs is more subtle because of the interdependence of the two wave vectors $k$ and $-k$ as the time reversal symmetry in physics. Which makes the real Bott clock tick counter clockwise. This requires a more abstract K-theory, called KR theory\cite{atiyah1988michael}. It's a real vector bundles structure over a space $(\mathbb{R}^d, \tau)$, where the involution $\tau$ is given by $k\mapsto -k$. On one point compactification, this real space becomes a sphere $\bar{S}^d=(S^d, \tau)$ with the same involution. Whereas the ordinary sphere $S^d$ can be viewed as a reduced suspention $S$ of $S^d$ over the real axis, $\bar{S}^d$ of $\bar{S}^{d-1}$ over the imaginary axis. This picture is the ordinary complex conjugation which, restricted to the imaginary axis of the complex plane is of the same form\cite{atiyah1988michael,budich2012adiabatic}. Atiyah defined the KR-theory, $\widetilde{\textrm{KR}}^{p,q}(M)$ for $p$ imaginary axis suspention $\bar{S}^p$ and $q$ real axis suspention $S^q$, it will be the usual $\widetilde{\textrm{KO}}$-theory when $p$ is zero like following
\begin{equation}\label{KR}
    \widetilde{\textrm{KR}}^{0,q}(M)= \widetilde{\textrm{KO}}^{-q}(M),
\end{equation}
he have proved the (1,1) periodicity theorem like
\begin{equation}\label{KR periodicity}
    \widetilde{\textrm{KR}}^{p,q}(M)=\widetilde{\textrm{KR}}^{p+1,q+1}(M).
\end{equation}
Combing equation (\ref{KR}) and (\ref{KR periodicity}) we obtain
\begin{equation}\label{kr relation ko}
    \widetilde{\textrm{KR}}^{p,q}(M)=\widetilde{\textrm{KO}}^{p-q}(M)=\widetilde{\textrm{KO}}^{r}(M)
\end{equation}
where $r=p-q$ (mod 8).
\par
Now we can compute all $ \widetilde{\textrm{KR}}^{p,q}$ for continuum topology model, the base manifold is identify sphere $S^d$ in $d$ space dimension. Now we can use equation $(\ref{kr relation ko})$ and (\ref{K Theory1}) to compute the symmetry class $R_q$ as
\begin{equation}\label{Sphere model}
    \widetilde{\textrm{KR}}^{d,q}(pt)=\widetilde{\textrm{KO}}^{d-q}(pt)=\pi_0(R_{q-d}),
\end{equation}
from the table $\ref{table1}$ we can obtain all symmetry class non-trivial TSMs in arbitrary space dimension for sphere model denoting as table $\ref{2}$, which is consistent with ref\cite{kitaev2009periodic}. From the equation of (\ref{Sphere model}), we know there are both 8-period in space dimensions and symmetry classes. However, the Bott clock is clockwise with the symmetry classes increasing, while it will reversal the direction as counterclockwise with the space dimension increasing.
\par
Through discussion above, we can classify all TSMs in all symmetry classes defined in the homotopy subsection by compute the   $\widetilde{\textrm{KO}}$ and $\widetilde{\textrm{KU}}$. For the classification of the one point compactificaion continuum model with base manifold $\bar{S}^d$, the problem can be solved by zero homotopy group of the classifying spaces due to the Bott periodicity. The  most interesting problem in CMP is the classification of fiber bundles with base manifold $\bar{T^d}=(T^d,\tau)$, this is a more complicated mathematical problem. Here we only review the general result calculated in ref(\cite{kitaev2009periodic})
\begin{equation}\label{k group of T}
    \widetilde{\textrm{KO}}^{-q}(\bar{T^d})\cong \bigoplus_{s=0}^{d-1}C_{d}^{s} \widetilde{\textrm{KO}}^{-q}(\bar{S^s})=\bigoplus_{s=0}^{d-1}C_{d}^{s}\pi_0(R_{q-s})
\end{equation}
where $C_{d}^{s}$ is the binomial factor.  Such a result can be interpreted by stacking the lower dimensional topological phases to obtain a higher dimensional ones\cite{wen2012symmetry}. For 3-dimension time-reversal symmetry topological insulator, i.e., $d=3, q=4$, we get:
\begin{equation}\label{k group of T^3}
   \widetilde{\textrm{KO}}^{-4}(\bar{T^3})\cong \pi_0(R_4)\oplus3\pi_0(R_3)\oplus3\pi_0(R_2)=\mathbb{Z}\oplus3\mathbb{Z}_{2}.
\end{equation}
The $\mathbb{Z}$ is the number of valence bands, whereas $3\mathbb{Z}_2$ pertains to "weak topological insulator" in physics. For strong topological insulator we can only computer the K group of base manifold $\bar{S^3}$ with (\ref{Sphere model}) like following
\begin{equation}\label{strong S3}
    \widetilde{\textrm{KO}}^{-4}(\bar{S^3})\cong\pi_0(R_{4-3})=\pi_0(R_1)=\mathbb{Z}_2.
\end{equation}
The most familiar and famous topological state of QIHE in 2-dimension is describe by the following with equation (\ref{KU})
\begin{equation}\label{ku qihe}
    \widetilde{\textrm{KU}}^{-0}(S^2)=\pi_2(C_0)=\pi_0(C_0)=\mathbb{Z},
\end{equation}
where the term $\mathbb{Z}$ is just the TKNN number\cite{kohmoto1985topological}.
\par
Topology K-theory is a deep and useful mathematic tool in investigating the topology properties of fire bundles both in physics and mathematics. For the classification of all TSMs which we are interested in physics all can be obtained by the K theory or the simple zero homotopy group of each symmetry classes with the Bott periodicity. This is a very deep and amazing relationship between physics and mathematics.
\begin{table}
\caption{Classification of topological insulators and superconductors as a function of spatial dimension $d$ and symmetry class, indicated by 'Cartan label' (first column) The symmetry classes are grouped into two separate lists, the complex and real cases, depending on whether the Hamiltonian is complex or whether one (or more) reality conditions (arising from time-reversal or charge-conjugation symmetries) are imposed on it. Chiral classes are denoted by bold letters. The symbols $\mathbb{Z}$ and $\mathbb{Z}_2$ indicate that the topologically distinct phases within a given symmetry class of topological insulators (superconductors) are characterized by an integer invariant ($\mathbb{Z}$) or a ($\mathbb{Z}_2$) quantity, respectively.}\label{table2}
\begin{tabular}{c|cccccccc}
  \hline\hline
  Cartan$\setminus$dimension & 0 & 1 & 2 & 3 & 4 & 5 & 6 & 7 \\\hline
  $Complex ~case:$ &  &  &  &  &  & &  &  \\
  A  & $\mathbb{Z}$ & 0 & $\mathbb{Z}$ & 0 & $\mathbb{Z}$ & 0 & $\mathbb{Z}$ & 0 \\
  AIII& 0 & $\mathbb{Z}$ & 0 & $\mathbb{Z}$ & 0 & $\mathbb{Z}$ & 0 &$\mathbb{Z} $\\
  $Real~ case:$ &  &  &  &  &  & &  &  \\
  AI & $\mathbb{Z} $& 0 & 0 & 0 & $\mathbb{Z}$ & 0 & $\mathbb{Z}_2$ &$\mathbb{Z}_2$\\
 \textbf{ BDI}& $\mathbb{Z}_2$&$\mathbb{Z}$ & 0 & 0 & 0 & $\mathbb{Z} $& 0 &$ \mathbb{Z}_2$\\
  D & $ \mathbb{Z}_2$&$\mathbb{Z}_2$&$\mathbb{Z}$ & 0 & 0 & 0 & $\mathbb{Z}$ & 0\\
  \textbf{DIII} & 0&$\mathbb{Z}_2$&$\mathbb{Z}_2$&$\mathbb{Z}$ & 0 & 0 & 0 & $\mathbb{Z}$\\
  AII &  $\mathbb{Z}$& 0&$\mathbb{Z}_2$&$\mathbb{Z}_2$&$\mathbb{Z}$ & 0 & 0 & 0\\
 \textbf{ CII} &0 &  $\mathbb{Z}$& 0&$\mathbb{Z}_2$&$\mathbb{Z}_2$&$\mathbb{Z} $& 0 & 0 \\
  C & 0 &0 &  $\mathbb{Z}$& 0&$\mathbb{Z}_2$&$\mathbb{Z}_2$&$\mathbb{Z}$& 0 \\
  \textbf{CI} & 0 &0 &0 &  $\mathbb{Z}$& 0&$\mathbb{Z}_2$&$\mathbb{Z}_2$&$\mathbb{Z} $\\
  \hline\hline
\end{tabular}
\end{table}

\subsection{Index Theory}
  Atiyah-Singer (AS) index theorem\cite{atiyah1968index} is the most deepest mathematic theorems in the last century. It bridges a relationship between analysis, topology and differential geometry in mathematics, which contains a series formulas as its special case, such as Gauss-Bionet-Chern theorem. In physics, it is the source of some important quantum anomalies, such as axial anomaly and Abelian gauge anomaly\cite{ryu2012electromagnetic,nakahara2003geometry,hou1997differential}. There is an intuitive phenomenon which can let readers know what this theorem to say. As we all know, the wave fluctuation  on the drumheads can be demonstrated with a differential equation in physics, this theorem will tell you can listen out how many "holes" on the drumheads. In other words, one can obtain some robust topology information from analysis. AS theorem tells us that the analytic-index of differential equation like Dirac equation can be derived from some cohomology classes discussed in subsection Chern character. The prove of this theorem is very complicated, Atiyah proved the general formula with K theory in the series texts\cite{atiyah1968index}.
 \par
Let $D: \Gamma E\rightarrow\Gamma F$  be an elliptic operator like Dirac operator in physics on a compact space $X$ , where $\Gamma E$ and $\Gamma F$ are the sections of  fiber bundle. Then for each $x\in X$ and $\xi\in T^*X_x$, the symbol $\sigma(x,\xi)$ \cite{nakahara2003geometry}of $D$ is a homomorphism $E_x\rightarrow F_x$ varying smoothly with $x$ and $\xi$. It thus defines a homogeneous complex
$$0\longrightarrow\pi^*E\longrightarrow\pi^*F\longrightarrow0$$
over the cotangent bundle $T^*X$. Since $D$ is elliptic, we know that $\sigma(x,\xi)$ is invertible for $\xi\neq0$, and so this complex is exact outside the zero section. So it defines an element of $K(T^*X)$, called the symbol of $D$ and denoted by $\sigma(D)$. The kernel of $D$ is the set of null eigenvectors
$$\textrm{ker}D=\{s\in \Gamma E|Ds=0\}$$
Suppose $E$ and $F$ are endowed with fiber metrics, which will be denoted $\langle,~\rangle_E$ and $\langle,~\rangle_F$, respectively. The adjoint $D^{\dag}:\Gamma F\rightarrow\Gamma E$ of $D$ defined by
$$\langle s^{'}, Ds\rangle=\langle D^{\dag}s^{'}, s\rangle$$
where $s^{'}\in \Gamma E$ and $s\in \Gamma F$. Now we can define the cokernel of $D$ by
$$\textrm{coker}(D)=\Gamma F/\textrm{im}D$$
where $\textrm{im}$ is the image set of $D$. The elliptic operators whose kernel and conkernel are both finite dimensional are called Fredholm operator. The analytic index
\begin{equation}\label{anylitic index}
    \textrm{ind}D\equiv\textrm{dim~ker}D-\textrm{dim~coker}D.
\end{equation}
It is known from the general theory of operators that elliptic operators on a compact manifold are Fredholm operators. It also can be expressed as
 \begin{equation}\label{anylitic index1}
    \textrm{ind}D=\textrm{dim~ker}D-\textrm{dim~ker}D^{\dag}.
\end{equation}
This is the analytic definition of index of elliptic operator, now there is a corresponding topological definition of it using the symbol $\sigma(D)$ of operator.
If $X$ is a compact manifold, choose an embedding $i: X\rightarrow R^m$ for large enough $m$, and let $j: P\rightarrow R^m$ be the inclusion of the origin. In fact, $j!: K(TP)\rightarrow K(TR^m)$ is simply the Thom isomorphism $\varphi: K(pt)\rightarrow K(\mathbb{C}^m)$, and so we can define the topological index to be the composition
$$\textrm{t-ind}: K(TX)\stackrel{i!}\rightarrow K(TR^m)\stackrel{j!^{-1}}\rightarrow K(TP)\cong\mathbb{Z}$$
\par
Now it is time to introduce the AS theorem. If $D$ is an elliptic operator on a compact manifold then
\begin{equation}\label{AS theorem}
    \textrm{ind}D=\textrm{t-ind}\sigma(D).
\end{equation}
AS theorem states that the index of elliptic operators can be computed with topological invariants. In other words, the zero-mode solutions of elliptic operators elucidate the topological properties of the manifold which the operators act on.

 \par
 In the following, we briefly introduce some important theorems required to write the topological index in terms of cohomology, which is very useful for direct computation. Firstly, for cohomology and K-theory, there is a profound and significant theorem called Thom isomorphism theorem, respectively. They establish a relationship between the cohomology group or K-group of base space and fire bundle space, respectively\cite{landweber2005k}. Suppose that $E$ is an oriented real $n$-plane bundle over $X$, there exists a unique generator $u\in H^n(E,E_0)=H^n_c(E)$, where $H^*_c()$ is cohomology with compact supports and $E_0$ is the deleted space obtained by removing the zero section of $E$, called Thom class of $E$. Then the following homomorphism
 \begin{equation}\label{Thom isomorphism of H}
    \psi: H^k(X)\longrightarrow H_c^{k+n}(E),
 \end{equation}
 given by $\psi: x\mapsto\pi^*x\cup u$, where $\pi:E\rightarrow X$ is the projection, and $\cup$ is the cup product in cohomology\cite{hatcher2003vector},  is an isomorphism.
  \par
  In a similar sense, there is a corresponding Thom isomorphism theorem in K-theory.  Let $E$ is a complex vector bundle over a compact space $X$, then the following homomorphism is isomorphism
 \begin{equation}\label{Thom isomorphism of K}
    \varphi: K(X)\longrightarrow K(E)=\tilde{K}(E^+),
 \end{equation}
 where $ \varphi: x\mapsto\pi^*(x)\cdot \lambda_E$, $\tilde{K}(E^+)$ is the one point compactification of the total space, and $\lambda_E\in K(E)$ is the class of Thom\cite{landweber2005k}.
 \par
 Another important principle is the splitting principle. It states that every complex vector bundle $E$ over $X$, there exists a space $F(E)$ and a map $\pi: F(E)\rightarrow X$ such that $\pi^*: H^*(X)\rightarrow H^*(F(E))$ is injective and $\pi^*E$ splits as the direct sum $E=L_1\oplus\cdots\oplus L_n$ of complex line bundles, where $H^*()$ is the set of cohomology group of $X$. Based on this principle, one can prove the following identity of Chern character for  any complex vector bundles defined in subsection Chern character satisfying,

 \begin{eqnarray}\label{chern property}
   \textrm{ch}(E\oplus F)&=&\textrm{ch}(E)+\textrm{ch}(F), \nonumber\\
    \textrm{ch}(E\otimes F)&=& \textrm{ch}(E)\cdot\textrm{ch}(F),
 \end{eqnarray}
 where $\oplus$ and $\otimes$ stand for the direct sum  and tensor product operator in vector space. So the Chern character sets up a ring homomorphism between K-group and cohomology group, like following:
 \begin{equation}\label{ch map}
    \textrm{ch}:K(X)\rightarrow H^*(X).
 \end{equation}

 \par
 Using the Chern character, we can compare the Thom isomorphisms of K-theory and cohomology. Let $E$ be a complex $n-$plane bundle over a compact space $X$, we obtain the following diagram:
 \[\begin{CD}\label{relation}
 K(X)                                       @>\varphi>>K(E) \\
 @V\textrm{ch}VV                            @V\textrm{ch}VV \\
 H*(X)                                      @>\psi>>H^*_c(E)
 \end{CD}\]
where the vertical maps are the corresponding Chern characters and the horizontal maps are the Thom isomorphisms $\varphi: x\mapsto\pi^*x\cdot\lambda_E$ and $\psi:\pi^*x\cup u$. Since this diagram does not commute, it is necessary to introduce a correction factor. Precisely, we set
\begin{equation}\label{mu class}
    \psi(\mu(E)\cup\textrm{ch}(x))=\textrm{ch}(\varphi(x)),
\end{equation}
where
\begin{equation}\label{mu}
    \mu(E)=\psi^{-1}\circ\textrm{ch}\circ\varphi(1)=\psi^{-1}\circ\textrm{ch}(\lambda_E)\in H^*(X)
\end{equation}
Letting $i: X\rightarrow E$ be the zero section, according to the pullback of it, $i^*\lambda_E=\sum_i(-1)^i\wedge^i(E)$, and $i^*\psi(x)=x\cup e$ for $x\in H^*(X)$, where $\wedge$ is the wedge product and $e$ is the Euler class\cite{landweber2005k}.
 In particular, taking $x=\mu(E)$, it follows by the naturality of Chern classes that
\begin{eqnarray}
   \mu(E)\cup e&=& i^*\psi(\mu(E))= i^*\textrm{ch}(\lambda_E)\nonumber\\
   &=& \textrm{ch}(i^*\lambda_E) =\textrm{ch}\left(\sum_{0}^{n}(-1)^i\wedge^i(E)\right)
\end{eqnarray}
Then, invoking the splitting principle, we obtain
\begin{equation}\label{mu dot euler}
     \mu(E)\cup e=\textrm{ch}(\prod_{i=1}^{n}(1-L_i))=\prod_{i=1}^{n}(1-e^{x_i})
\end{equation}
where $x_i =c_1(L_i)$ is the Chern class. Since $e=c_n(E)=\prod_ix_i$ we conclude
$$\mu(E)=\prod_{i=1}^{n}\frac{1-e^x_i}{x_i}.$$
To be simply in the formula of index theorem, we defined another class called Todd class\cite{nakahara2003geometry} for a complex vector bundle $E$, like following
\begin{equation}\label{Todd class}
    \textrm{td}(E)=\prod_{i=1}^{n}\frac{x_i}{1-e^{-x_i}}
\end{equation}
If we denote by $\bar{E}$ the conjugate bundle of a complex vector bundle $E$, so we have $x_i=-x_i$ then the correction factor is given by
$$\mu(E)=(-1)^n\textrm{td}(\bar{E})^{-1}.$$
The Todd class satisfies $\textrm{td}(E\oplus F)=\textrm{td}(E)\cdot\textrm{td}(F)$ for any two vector bundles $E$ and $F$, and $\textrm{td}(n)=1$ for trivial complex $n$-plane bundle.
\par
Now we can reformulate the AS theorem in terms of characteristic classes and cohomology. To write the topological index in terms of cohomology, let $X$ be a compact $n$-dimensional manifold embedded in $\mathbb{R}^k$ and consider the following non-commutative diagram\cite{landweber2005k}:

\[\begin{CD}\label{relation}
 K(TX) @>\varphi>>K(TN)@>h>>K(T\mathbb{R}^k)@<\varphi^{'}<<K(TP)\\
@VVV           @VVV                 @VVV          @VVV\\
H^*_c(TX) @>\psi>> H^*_c(TN)@>k>>H^*_c(T\mathbb{R}^k)@<\psi^{'}<<H^*_c(TP)
\end{CD}\]
where $\varphi, \psi, \varphi^{'}, $ and $\psi^{'}$ are various Thom isomorphisms, $h$ and $k$ are the extension homomorphisms, and vertical map is the Chern character $\textrm{ch}$, $T^*X$ is replaced by $TX$ because they are isomorphism.
As discussed above, this diagram does not commute, but they are related by (\ref{mu class}) as
\begin{equation}\label{mu and todd}
    \psi^{-1}\circ\textrm{ch}\circ\varphi: x\mapsto (-1)^n \textrm{td}(\bar{E})^{-1}\textrm{ch}(x).
\end{equation}
In our case, we have $\textrm{td}(\overline{T\mathbb{R}}^k)$=1 since $\overline{T\mathbb{R}}^k$ is clearly a trivial bundle over the point $TP$. Since $TN\cong\pi^*N\oplus\pi^*N\cong(\pi^*N)\otimes_\mathbb{R}\mathbb{C}$, (one can impose on $TN$ the structure of a complex vector bundle) we see that $TN\cong\overline{TN}$. Noting that $T(TX)\oplus TN=T(T\mathbb{R}^k)|_{TX}$ is a trivial bundle over $TX$ and that $T(TX)\cong\pi^*TX\oplus\pi^*TX=\pi^*TX\otimes_{\mathbb{R}}\mathbb{C}$, we obtain
\begin{equation}\label{tdN}
    \textrm{td}(\overline{TN})^{-1}=\textrm{td}(TN)^{-1}=\textrm{td}(T(TX))
    =\pi^*\textrm{td}(TX\otimes_{\mathbb{R}}\mathbb{C})\nonumber.
\end{equation}
If we define the index class $\mathfrak{T}(X)=\textrm{td}(TX\otimes_{\mathbb{R}}\mathbb{C})$ of a manifold $X$,
then $\textrm{td}(\overline{TN})^{-1}=\pi^*\mathfrak{T}(X)$ and it's independent of the embedding. We thus have the following

\begin{eqnarray}
   \psi^{'-1}\circ\textrm{ch}\circ\varphi^{'}:x&\mapsto& (-1)^k \textrm{td}(\overline{T\mathbb{R}}^k)^{-1}\textrm{ch}(x)=(-1)^k \textrm{ch}(x)\nonumber \\
   \psi^{-1}\circ\textrm{ch}\circ\varphi:x&\mapsto&  (-1)^{n-k}\textrm{td}(\overline{TN})^{-1}\textrm{ch}(x)\nonumber\\
   &=&
   (-1)^{n-k}\mathfrak{T}(X)\cup\textrm{ch}(x)
\end{eqnarray}
Recalling that $\textrm{t-ind}=\varphi^{'-1}\circ h\circ\varphi$, by an extended diagram chase we obtain
$$\textrm{ch}\circ h\circ \varphi(x)=(-1)^{k}(\textrm{t-ind}x)\psi^{'}(1).$$
where $\psi^{'}(1)$ is the canonical generator of $H^*_c(T\mathbb{R})^k$.
Then letting $[T\mathbb{R}]$ be the fundamental homology class of $T\mathbb{R}$,(pairing with which can be interpreted integrating over the manifold) we have $\psi^{'}(1)[T\mathbb{R}]=1$, and thus we obtain\cite{atiyah1968index1,atiyah1968index}

\begin{eqnarray}
   \textrm{t-ind}x &=&(\textrm{t-ind}x\cup\psi^{'}(1))[T\mathbb{R}^n]
   =(-1)^k(\textrm{ch}\circ h\circ\varphi(x))[T\mathbb{R}^n]\nonumber \\
 &=&(-1)^k(\textrm{ch}\circ\varphi(x))[TN]=(-1)^n(\psi(\textrm{ch}
 \cup\mathfrak{T}(X)))[TN] \nonumber\\
   &=&(-1)^n(\textrm{ch}(x)\cup\mathfrak{T}(X))[TX].
\end{eqnarray}
\par
So the AS index theorem can be expressed with cohomology characteristics.
If $D$ is an elliptic operator on a compact $n$-dimensional manifold $X$ then
$$\textrm{ind}(D)=(-1)^n(\textrm{ch}(\sigma(D))\cup\mathfrak{T}(X))[TX],$$
where $\mathfrak{T}(X)=\textrm{td}(TX\otimes_{\mathbb{R}}\mathbb{C}))$ is the index class of $X$.
If $X$ is a oriented $n$-dimensional manifold, then we have another familiar formula
\begin{equation}\label{index AS}
    \textrm{ind}(D)=(-1)^{n(n+1)/2}(\psi^{-1}\circ\textrm{ch}(\sigma(D))\cup\mathfrak{T}(X))[X],
\end{equation}
where we note that if $X$ is an oriented manifold, then we have $\psi(u)[TX]=(-1)^{n( n-1)}u[X]$ for each $u\in H^n(X)$, $\psi: H^*(X)\rightarrow H^*(TX)$ is the Thom isomorphism.
\par
The AS theorem will be more computable if we introduce the concept of elliptic complexes. Consider a sequence of Fredholm operators,
$$0\longrightarrow\Gamma E^0\stackrel{d_0}\longrightarrow \Gamma E^1\stackrel{d_1}\longrightarrow\cdots\stackrel{d_{n-1}}\longrightarrow\Gamma E^n\longrightarrow 0 $$
where $\Gamma E^i$ is a sequence of vector bundle on a compact base manifold. The sequence $(E^i, d_i)$ is called elliptic complex if $d_i$ satisfying $d_i\circ d_{i-1}=0$ for any $i$. Let $(E,d)$ be an  elliptic complex over an $n$-dimensional
compact manifold without boundary. The index of this complex is given by
\begin{equation}\label{AS1}
    \textrm{ind}{D}=(-1)^{n(n+1)/2}\int_{M}\textrm{ch}((-1)^r\oplus E_r))\frac{\textrm{td}(TX\otimes_{\mathbb{R}}\mathbb{C})}{e(TX)}
\end{equation}
where $e(TX)$ is the Euler class of $X$\cite{nakahara2003geometry}.
\par
In physics, the Dirac Hamiltonian $H$ with chiral symmetry possesses an index, it relates the integral of the anomaly polynomial through the formula

\begin{equation}\label{anomaly and AS}
\textrm{ind}(H)=\int_{M_D}\Omega_{D}(\mathcal {R},\mathcal {F}),
\end{equation}
where $\Omega_D(\mathcal {R},\mathcal {F})$ is the anomaly polynomial\cite{ryu2012electromagnetic} which can be given by
\begin{equation}\label{anomyly polynomial}
    \Omega_{D}(\mathcal {R},\mathcal {F})=(\textrm{ch}(\mathcal {F}) \hat{A}(\mathcal {R}))|_D
\end{equation}
where $\hat{A}(\mathcal {R})$ is the so-called Dirac genus\cite{nakahara2003geometry}, and $\mathcal {R}$ is the usual Riemann curvature two-forms. The anomaly is an integer is amazing and deep, it reflects the global topological properties of space-time manifold the physic system acts on.
If we denote $\nu^+ (\nu^-)$ as the number of zero-energy modes of chirality $+ (-)$ for Dirac Hamiltonian, the AS theorem also can be expressed
\begin{equation}\label{nu AS}
    \nu^+-\nu^-=\int_{M_D}\Omega_{D}(\mathcal {R},\mathcal {F})
\end{equation}
This is a beautiful and profound relation between mathematics and physics, the elliptic operator Dirac Hamiltonian can reflect the topology properties of the fiber bundles that it act on. So some robust quantum anomalies in physics is just the property of the elliptic operator used in physics, such as Dirac operator, Laplace operator and so on.
\section{Topology in Physics}
In this section, we apply the topological invariants discussed above to study the robust physical phenomenons which protected by topology property of Hamiltonian or field theory. The robust quantization of QIHE, QSHE in 2-dimensional and 3-dimensional topological insulators are extensively studied both in experiments and theory in the last years. During the process of investigate, there are mainly two routes to demonstrate the topology property in theory, they are TBT and TFT. TBT try to study the topology property of Bloch Hamiltons through characteristic class, which mainly involves single-particle non-interacting fermion systems.  We give the integral formulas about any topological insulators or superconductors, and we popularize  one-new topological invariant called  Fu-Kane invariant \cite{fu2006time}.  Compared with TBT, TFT can solve some interacting systems, can probe the topology response function directly from Chern-Simons effect action by coupling an external field like electromagnetic fields or gravitation field. We give a general relationship between interacting topological invariants which expressed with Green's function and non-interacting ones using Chern-number of valence band, and establish the relationships of various topological invariants extensively used in this field. In addition, we discuss the unified phase-space Chern-Smions theories, and obtain the generating function of all Chern-Simons effective action in any space-time dimensions.  Following that, we review the important bulk-boundary correspondence expressing with something like AS index theorem. What's more, we systematically study the 2-complex and 8-real classifying spaces using loop space discussion, and establish the relations of them for complex and real cases, respectively. Finally, we give all gapped free-fermions systems classification in any symmetry classes and for any base manifold, i.e., point space, sphere, and torus base manifold.
\subsection{Topological Band Theory}
As discussed in the subsection of fibre bundle, the usual Bloch-Hamiltonian can be seen as the fibre bundles over the base manifold $T^d$ for lattice model, where $T^d$ is the $d$-dimensional space torus. It will be replaced by $S^d$ $d$-dimensional space spheres for continuum model. If the base manifold replaced by $\mathbb{R}^d$, the total space can be described as a total product space $\mathbb{R}^d\times H(k)$, so it stands for a trivial bundle, there is no chance have a non-trivial TSMs. Now it's time to state the problem of TBT, how to describe the non-trivial topology property for fiber bundles $P(T^d,G)$ or $P(S^d,G)$, $G$ is the structural group of fiber bundles, also called the symmetry group of Bloch-Hamiltonian. In order to apply  Chern-characteristic to classify the different fiber bundles, let define the non-Aelian Berry connection 1-form for occupied bands like following:
\begin{equation}\label{define berry conncetion}
    \mathcal {A}^{ab}(k)= \mathcal {A}^{ab}_{\mu}\textrm{d}k_{\mu}=\langle u_a^-(k) |du_b^-(k)\rangle, ~~~\mu=1,2,\cdots,d,
\end{equation}
where $a,b$ are the valence band index of insulator, $d$ is the dimension of base manifold and  $u_a^-(k)$ is the valence band wavefunction. Corresponding the Berry curvature 2-forms are defined by
\begin{equation}\label{define berry curvature}
   \mathcal {F}^{ab}(k)=\textrm{d}\mathcal {A}^{ab}+(\mathcal {A}^2)^{ab}=\frac{1}{2}F^{ab}_{\mu\nu}(k)\textrm{d}k_{\mu}\wedge\textrm{d}k_{\nu}.
\end{equation}
After introducing explicit definitions, we can apply the topological invariants to investigate topology information that appears in physics, which can be obtained by integrating these local Berry curvature or connection forms over the base manifold.
\par
In the following, we will give and prove all integral topological invariants for any TSMs which can be demonstrated with Bloch equations or BdG ones. And we will give the $\mathbb{Z}_2$ topological invariants homotopiclly discussion using $interpolation$ used by Qi\cite{qi2008topological}. In addition, we will give this construction a topological interpretation using homotopy theory.
\subsubsection{Integral topological invariants for Complex case}
\par
For class A topological insulator (  IQHE in 2-dimension), the structure group is $U(n)$ because of no symmetry constraint on the Hamiltonian,  we can use Chern number $\textrm{Ch}_n[\mathcal{F}]$(equation (\ref{chern number})) to distinguish any two different non-trivial TSMs in even-dimension. In fact, it is just the reflection of $\pi_{d-1}(U(n))=\mathbb{Z}$ in homotopy language (\ref{bott periodicity for stable u}), we know only there are non-trivial fibre bundles in even space dimension for structural group $U(n)$ according to the 2-period Bott theorem for stable range unitary group. The classifying space is $C_0$ for class A, so the classification of fiber bundle is equivalent to the set of homotopically non-equivalent maps from base manifold $[T^d,C_0]$ or $[S^d,C_0]$. We rewrite the formula in the following for convenience,

    \begin{eqnarray}\label{chern number P}
      \textrm{Ch}_n[\mathcal{F}]
       &=&  \int_{\textrm{BZ}^{d=2n}}\frac{1}{n!}\textrm{tr}(\frac{i\mathcal{F}}{2\pi})^n\in\pi_d(C_0)=\mathbb{Z}.
    \end{eqnarray}
\par
For class AIII in odd dimensions, another topological invariants called winding number $\textrm{W}_{2n+1}[g]$ (\ref{winding number}) can be used to classify the different topological states. In this class, the Hamiltonian equips with the symmetry called chiral symmetry $\{H(k),\Gamma\}=0 , \Gamma^2=1$, where $\Gamma$ is a unitary matrix that anti-commutes with the Hamiltonian. In this case, the class AIII Hamiltonians can be brought into block off-diagonal form,
$$H=\left(
  \begin{array}{cc}
     0& q(k) \\
     q^{\dag}(k)&  0\\
  \end{array}
\right),~~~~q\in U(N),$$
in the basis in which $\Gamma$ is diagonal. The topological insulators are characterized by the winding number $\textrm{W}_{2n-1}[q]$, where substitute $g$ with $q$.
The explicit formula can be expressed like following
\begin{equation}\label{winding number P}
    \textrm{W}_{2n+1}[q]=\int_{\textrm{BZ}^{2n+1}}\omega_{2n+1}[g]\in\pi_d(C_1)=\mathbb{Z},
\end{equation}
where $\omega_{2n+1}[g]$ is winding number density as defined like equation (\ref{winding number density}).
\par
The topology property of $H$ also can be obtained by studying the zero mode chirality according to  AS index theorem. In fact, $H=\hat{h}+\hat{h}^{\dag}$, $\hat{h}=q\otimes \tau_{+}$, where $\tau_{\pm}=(\tau_x\pm\tau_y)/2$, $\tau$ is the usual Pauli matrix. Thus, $\hat{h}$ is a Dirac operator in mathematic. Physically, it sends a negative chiral state to a positive one while annihilating all positive chiral states. The chirality of the zero mode are defined as the index of $H$,
\begin{equation}\label{chiral zero mode}
    \textrm{ind}(H)= \textrm{dim~ker}(\hat{h}^{\dag})-\textrm{dim~ker}(\hat{h}).
\end{equation}
Therefore, the analytic index equal to the topological index due to AS index theorem, $ \textrm{ind}(H)=\textrm{W}_{2n+1}[q]$.
\par
Based on the equation of (\ref{chern number and winding n}) $\textrm{Ch}_n[F]=\textrm{W}_{2n-1}[g]$, one can construct the relationship between class A in even-dimensional and class AIII in odd-dimensional topological insulators, the odd-dimensional topological insulators "lives" on the equator of sphere. The relation can be expressed with Stokes theorem,
\begin{equation}\label{relation A and AIII}
    (S^d, \textrm{ch}_n(\mathcal {F}))=(S^{d-1},\omega_{d-1}[g])=(S^{d-1},\omega_{d-1}[q])
\end{equation}
where $d$ is an even number, $S^{d-1}$ is the equator of the sphere $S^d$, and the gauge transformation function $g$ is replaced by $q$.When the base manifold is torus $T^d$, the $S^{d-1}$ will be substitute by the interface of two open sets covering as $\partial\textrm{BZ}^d$, $\partial$ is the boundary operator. The beautiful link can be demonstrated as following, $d$ is an even number,
\[\begin{CD}\label{relation}
\pi_d(C_0)\ni\textrm{Ch}_d[\mathcal {F}]  @>g\leftrightarrow q>>\textrm{W}_{d-1}[q]\in\pi_{d-1}(C_1)\\
@| @|\\
   (S^d, \textrm{ch}_d(\mathcal {F})) @>Stokes>>(S^{d-1},\omega_{d-1}[q])
\end{CD}\]
where $g\leftrightarrow q$ means the gauge transformation group $g$ in $d$-dimensional is replaced by the $(d-1)$-dimensional group $q$.
\subsubsection{Integral topological invariants for Real case}
 In physics, there are two anti-unitary symmetries can compose on the Bloch Hamiltonian of a insulator or Bogoliubov de Gennes (BdG) Hamiltonian of a superconductor. The Hamiltonian will be in the real case under the condition of complex operator of anti-unitary.
 \begin{itemize}
   \item (Time reversal) Any anti-unitary operator $T$ that commutes with band Hamiltonian according to $T^{-1} H(\textbf{k})T=H(-\textbf{k})$ qualifies as a time reversal (TR) symmetry.
   \item (Particle-hole ) Any anti-unitary operator $C$ that anti-commutes with band Hamiltonian according to $C^{-1}H(\textbf{k})C=-H(-\textbf{k})$ qualifies as a particle hole (PH) symmetry.
   \item (Chiral) Any unitary operator $S$ that commutes with band Hamiltonian according to $S^{-1}H(\textbf{k})S=-H(\textbf{k}) $ qualifies as a chiral symmetry, such as the composition $TC$.
 \end{itemize}

\begin{figure}[h]
 \scalebox{0.5}[0.5]{\includegraphics{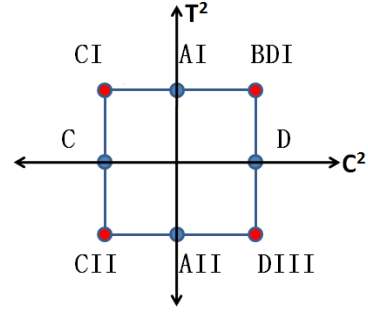}}\\
  \caption{The 8 $real$ symmetry classes that involve the anti-unitary symmetry $T$ (time reversal) and/or $C$ (particle-hole) are specified by the values of $T^2=\pm1$ and $C^2\pm1$. They can be visualized on an eight hour "clock"\cite{teo2010topological}. The chiral classes are denoted as  red, while non-chiral is indicated by blue.}\label{bott clock}
\end{figure}
The 8 $real$ symmetry classes can be arranged on a eight hour "clock" see Fig.\ref{bott clock} according to the sign of  $T^2$ and $C^2$. The four chiral classes with both PH
and TR symmetries are on the corners like (e.g. class CI has $T^2=1$ and $C^2=-1$), while the other four non-chiral symmetries classes with other PH or TR symmetry are on the axes (e.g. class AII has $T^2=-1$ but without PH symmetry). If both  PH and TR are present, we may assume they commute $[T, C]=0$.
\par
According to the periodicity table of TSMs in table \ref{table2}, we know there are four classes non-trivial topological insulators (superconductors) in one Bott period of every symmetry group for $real$ case. Now how to decide the value of the number $\mathbb{Z}$ or $\mathbb{Z}_2$ specifically ? It is a difficult problem in general, but we can establish some integral formulas to describe its non-trivial topology properties using Berry curvature or connection.  Compared with the $complex$ cases,  the integral formula will be more complicated in $real$ cases. Following we will give a general integral formula for non-chiral classes AII (AI) with TR symmetry only and classes D (C) with PH symmetry only.   The Chern number density is subject to the constraint due to the time reversal symmetry
\begin{equation}\label{AII}
    \textrm{ch}_{n}[\mathcal {F}(\textbf{k})]=(-1)^n \textrm{ch}_{n}[\mathcal {F}(-\textbf{k})],
\end{equation}
so the Chern number will be zero in the 2n-dimensional but not 4n-dimensional. However, the constraint on the Chern-number density for non-chiral classes D (C) with PH symmetry only will be replaced by
\begin{equation}\label{D}
    \textrm{ch}_{n}[\mathcal {F}(\textbf{k})]=(-1)^{n+1} \textrm{ch}_{n}[\mathcal {F}(-\textbf{k})],
\end{equation}
where the extra minus sign comes from the sign different of Chern-number density of valence band and conduction band\cite{murakami20042}, which also can be seen from equation (\ref{p+p=1}). Hence,  there are non-trivial $\mathbb{Z}$ topological states in these non-chiral classes with only PH symmetry for $(4n-2)$ dimension.
The zero Chern-number can exist for $(4n-2)$-dimension with TR symmetry or $4n$-dimension with PH symmetry, it seems we will get trivial fiber bundle.
In other words, there is no obstruction to define the Bloch wave functions continuously over the entire base space. However, TR (PH) relation between $\textbf{k}$ and $-\textbf{k}$ allows for an additional constraint so that the specified by the degree of freedom of half the Brillouin zone.  From the constraint of Chern-number, we know there is also non-zero Chern-number in the dimension $d\equiv0$ (mod 4) for TR symmetry classes and $d\equiv-2$ (mod 4) for PH symmetry classes, generally the condition is $s-d\equiv0$ (mod 4)\cite{teo2010topological,teo2011topological} for the other non-chiral classes denoted with blue color in Fig.\ref{bott clock}, where $s=0,1,2,\cdots,7$ (mod 8) denoting the class of symmetry. Following we write down the general integral formula of topological invariants according to the dimension hierachy \cite{qi2008topological,ryu2010topological} along the row in the periodic table \ref{table2}. For simplicity, we assume the base manifold is compact manifold $S^d$.
\par
\begin{description}
  \item[$\mathbb{Z}$-invariants] This applies to $s-d\equiv0$ (mod 4). For non-chiral classes like AII, the integral invariant is given by the Chern character that classifies the vector bundle of valence states

\begin{equation}\label{AII 4}
   \nu=\textrm{Ch}_{d/2}[\mathcal{F}]=\int_{S^d}\frac{1}{(\frac{d}{2})!}
\textrm{tr}(\frac{i\mathcal{F}}{2\pi})^{d/2}
\end{equation}
where $d$ is the space dimension. In fact, one will always get a even integer for the case $s-d\equiv4$ (mod 8), because the Hamiltonian will be reduced to two copies of the Hamiltonian of case   $s-d\equiv0$ (mod 4)\cite{ryu2010topological}.
\item[1st $\mathbb{Z}_2$ descendent ] This applies to case $s-d\equiv-1$ (mod 8). For AII like non-chiral symmetry classes, we can construct a Chern-Simons invariant equation (\ref{csn})
\begin{eqnarray}\label{AII 3}
          \nu_1&=&2\textrm{CS}_{d}[\mathcal{A},\mathcal{F}]\nonumber\\&=&
2\int_{S^{d}}Q_{d}(\mathcal{A},\mathcal{F})~~~~(\textrm{mod}~2)
\end{eqnarray}
where $Q_{d}(\mathcal{A},\mathcal{F})$ is the Chern-Simons form (\ref{chern-simons defenition}). As discussed in subsection Chern character, the Chern-Simons number is not gauge invariant, but it will have an integer difference after gauge transformation. Hence, the Chern-Simons invariant used above can figure out the $\mathbb{Z}_2$ index, because it will change only an even integer after gauge transformation. One can get the $\mathbb{Z}_2$ index after the calculation of mod 2.
  \item[2nd $\mathbb{Z}_2$ descendent ] This applies to case $s-d\equiv-2$ (mod 8). The integral topological invariants will be more subtle and difficulty. We will construct a similar Fu-Kane integral invariants \cite{fu2006time,teo2010topological}which  also depend  on the Berry connection $\mathcal{A}$. Because the  Chern-number will be zero (\ref{AII}), it is predicted the fibre bundle will be global trivial. But due to the involution of $\textbf{k}\leftrightarrow\textbf{-k}$ of the base manifold, there is another obstruction to globally defined the Bloch wave function. For non-chiral classes the Fu-Kane topological invariants can be expressed as
\begin{eqnarray}\label{AII 2}
   \nu_2&=& \int_{S^d_{1/2}}\frac{1}{(\frac{d}{2})!}
\textrm{tr}(\frac{i\mathcal{F}}{2\pi})^{d/2}-\oint_{\partial S^{d}_{1/2}}Q_{d-1}(\mathcal{A},\mathcal{F})\nonumber  \\
   &=&  \int_{S^d_{1/2}}\textrm{ch}_{\frac{d}{2}}(\mathcal {F})-\oint_{\partial S^{d}_{1/2}}Q_{d-1}(\mathcal{A},\mathcal{F})(\textrm{mod}~2)
\end{eqnarray}
where $S^d_{1/2}$ is a hemisphere, and $\partial S^d_{1/2}=S^{d-1}$ is its equator boundary that is close under the condition $\textbf{k}\leftrightarrow\textbf{-k}$. The Chern-Simons constructed with a gauge that satisfies the constraint\ref{T-Gauge constraint}  along the equator for class AI and AII (s = 0, 4) with TR symmetry. The gauge constraint\cite{teo2010topological} (\ref{C-Gauge constraint}) will be more complicated with only PH symmetry non-chiral classes like D, C (s = 2, 6), $G_{\textbf{k}}$ should be able to deformed  continunously into a constant matrix preserve the PH symmetry. The explicit constraints are written like following,
\begin{equation}\label{T-Gauge constraint}
    w_{mn}(\textbf{k})=\langle u^-_m(\textbf{k}) |T|u^-_n(-\textbf{k})\rangle\equiv \textrm{constant},
\end{equation}
\begin{equation}\label{C-Gauge constraint}
   G_{\textbf{k}}=\left(
     \begin{array}{cc}
       \mid &\mid \\
       u^+(\textbf{k})& u^-(\textbf{k}) \\
       \mid & \mid \\
     \end{array}
   \right)\simeq \textrm{constant},
\end{equation}
where $u^+, u^-$ are stand for conduction and valence band wave function. The topological invariant $\nu_2$ will be an $\mathbb{Z}_2$ value under these constraint conditions.
\end{description}
\par
These significant topological invariants integral formula are so beautiful and profound. Now we give an intuitive interpretation of these integral formula,  which may be useful for physicists to understand the topological meaning.  Actually, they have very clear topology properties, $\mathbb{Z}$-invariants indicate that the non-trivial global Bloch wave function can not be defined over the whole Brillioun zone. One have $\mathbb{Z}$ kinds of methods to twist the fiber bundles. In other words, it classifies the distinct different non-trivial fiber bundles in the cases $q-d\equiv0$ (mod 4). 1st $\mathbb{Z}_2$ invariants  clearly indicate there are only two distinct fiber bundles, we denote "0"  and "1" as the trivial fiber bundle and non-trivial fiber bundle, respectively. For example, one can construct a cylinder or M$\ddot{\textrm{o}}$bius strip over the base manifold $S^1$ with fiber $\mathbb{R}$, respectively. The fiber bundle cylinder have the trivial product $S^1\times \mathbb{R}$, but one cannot write a product of base manifold with fiber globally for the M$\ddot{\textrm{o}}$bius strip. So it is a non-trivial object that you can not obtain it from  continuously deformation of cylinder. In physics, it predicts there is non-trivial topological state in 3-dimension in class AII, called topological insulators. The most complicated case is the 2nd $\mathbb{Z}_2$, the integral region is the half Brillioun zone, and contains two forms integration, the first term is the "area" integral, the second term is the "line" integral. It's the high dimensional general formula like Green-formula which relates the line integral and area integral in 2-dimension, or the Gauss-formula which establish the relationship between volume integral and area integral. The first term in equation (\ref{AII 2}) is invariant under any gauge transformations, but the second term will be changed by an  any integer. However, one can prove the integer will be only even under the constraint of equation (\ref{T-Gauge constraint}) or (\ref{C-Gauge constraint}). The intrinsic topology information is contained in the boundary term or the second term in equation (\ref{AII 2}), the group of the gauge transformation will be reduced to a block off-diagonal matrix because the TR symmetry constraint, so the difference of the second term under this gauge transformation will be 2 times a winding number term. In other words, the integral formula predict a $\mathbb{Z}_2$ invariant.
\par
\begin{figure}[h]
  \scalebox{0.6} [0.5]{\includegraphics{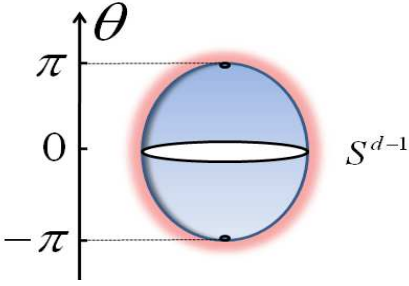}}\\
  \caption{Schematic of the suspention of $S^{d-1}$, the north hemisphere is the TR(PH) symmetry partner of the south hemisphere. The $d-1$ dimensional $H_1(\textbf{k})$ "lives " on the equator of the $d$ dimensional $H(\textbf{k},\theta)$ with $\theta\in (-\pi,\pi)$. The black points of the north and south poles are the base manifold of the constant Hamiltonians.}\label{suspention of sphere}
\end{figure}

Now, we give a derivation of these integral formula using the methods which have used in ref.\cite{qi2008topological,teo2010topological}. The $\nu_1$ topological invariant can be obtained from the one higher dimension topological insulators whose $\mathbb{Z}$ invariant formula is Chern-number. If the manifold of the $d-1$ dimensional $\nu_1$ topological state can be viewed as the equator of the $d$ dimensional $\mathbb{Z}$ topological state with the same symmetry, we can compute its topological invariant via Stokes theorem. Now we introduce one point parameter deformation $H(\textbf{k},\theta)$ that connects $H_1(\textbf{k})$ at $\theta=0$ to a constant Hamiltonian at $\theta=\pi$ . In general, $H(\textbf{k},\theta)$ for $\theta\in(0,\pi)$ does not necessarily satisfy the TR(PH) symmetry. But the symmetry can be restored by introducing a mirror image $H(\textbf{k},\theta)=T(-\textbf{k},-\theta)T^{-1} (H(\textbf{k},\theta)=-C(-\textbf{k},-\theta)C^{-1})$ for $\theta\in(-\pi,0)$. For $\theta=\pm\pi$, $\textbf{k}$ can be replaced by a single point, so the $d$ space of $(\textbf{k},\theta)$ is the suspention $S(S^{d-1})$ of the original space Fig.\ref{suspention of sphere}.
The Hamiltonian in $d$ dimensional space can be characterized by the Chern-number
\begin{equation}\label{z-z2}
    \textrm{Ch}_{d/2}[\mathcal{F}]=\int_{SS^{d-1}}\frac{1}{(\frac{d}{2})!}
\textrm{tr}(\frac{i\mathcal{F}}{2\pi})^{d/2}
\end{equation}
Due to the symmetry of TR(PH), the contributions from the two hemisphere $\theta > 0$ and $\theta < 0$ are equal. Using the equation of (\ref{chern-simons}) and Stokes theorem (\ref{stokes}) one can  therefore obtain

\begin{eqnarray}\label{v1}
   \nu_1&=&2\cdot\int_{(SS^{d-1})^+}\frac{1}{(\frac{d}{2})!}
\textrm{tr}(\frac{i\mathcal{F^+}}{2\pi})^{d/2}=2\cdot\int_{S^{d-1}}Q_{d-1}(\mathcal{A^+},\mathcal{F^+})\nonumber\\
   &=&2\cdot\textrm{CS}_{d-1}[\mathcal{A},\mathcal{F}],
\end{eqnarray}
where $\mathcal{A^+}, \mathcal{F^+}$ are the Berry gauge potential and Berry  curvature of the north  hemisphere, and the last equal is restrict on the equator. The result is agree with equation (\ref{AII 3}).  Although the $\nu_1$ is different for different interpolation $H(\textbf{k},\theta)$. However, TR(PH) symmetry requires the difference is an even integer.
\par
\begin{figure}[h]
  \scalebox{0.5} [0.5]{\includegraphics{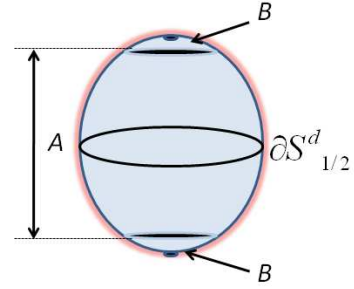}}\\
  \caption{Schematic of the base space  $S^d$. Division of $S^d$ into patches $A$ and $B$, each is closed under TR(PH) symmetry and has individual valence frame $|u_m^{A/B}\rangle$ that satisfies the gauge condition (\ref{T-Gauge constraint}) or (\ref{C-Gauge constraint}). The boundary of the half base manifold is $\partial S^d_{1/2}$, which is also the equator of sphere of $S^d$.}\label{two patches}
\end{figure}

For the derivation of the integral formula (\ref{AII 2}) is a little more difficult. Due to the Chern-number will be zero in this case, only the obstruction for the global Bloch wavefunction is the $\mathbb{Z}_2$ group action of TR(PH) symmetry on the base manifold. Hence, the gauge conditions of (\ref{T-Gauge constraint}) or (\ref{C-Gauge constraint}) will fore the valence frame to be singular at two points, depicted in the Fig.\ref{two patches}, related each other by TR(PH) symmetry.
\par
The wave functions on the two patches can be related by transition function $U(k)$ on the boundary $\partial B=S^{d-1}\cup S^{d-1}$. In homotopy language, this is the group $\pi_{d-1}(U(k))$, which can be calculated by the winding number (\ref{winding number}) on one of the spheres,

\begin{eqnarray}
   \textrm{W}_{d-1}[g]&=&\int_{S^{d-1}}\omega_{d-1}[g]\nonumber  \\
   &=& \int_{S^{d-1}}(Q_{d-1}(\mathcal{A^B},\mathcal{F^B})-Q_{d-1}(\mathcal{A^A},\mathcal{F^A})) \nonumber\\
   &=& \int_{B\bigcap S^{d}_{1/2}}\textrm{ch}_{\frac{d}{2}}(\mathcal{F^B})+ \int_{A\bigcap S^d_{1/2}}\textrm{ch}_{\frac{d}{2}}(\mathcal{F^A})\nonumber\\
   &&-\int_{\partial S^d_{1/2}}
Q_{d-1}(\mathcal{A^A},\mathcal{F^A})\label{w-cs}  \\
   &=& \int_{S^{d}_{1/2}}\textrm{ch}_{\frac{d}{2}}(\mathcal{F})-\int_{\partial S^d_{1/2}}Q_{d-1}(\mathcal{A},\mathcal{F}),
\end{eqnarray}
where we have used the equation of (\ref{chern-simons}) and (\ref{stokes}) for any $g\in U(k)$.
\par
The first curvature term of equation (\ref{AII 2})is gauge invariant. Any gauge transformation on the boundary $\partial S^{d}_{1/2}$ respecting the gauge condition in equation of (\ref{T-Gauge constraint}) or (\ref{C-Gauge constraint}) have even number and would alter the Chern-Simons integral by an even integer. The gauge condition is essential to make the formula nonvacuous.
\par
Now we can written the following map between these topological invariants
\[\begin{CD}\label{relation of desendents}
  \nu_2(\mathbb{Z}_2)@<i^*<<\nu_1(\mathbb{Z}_2) @<i^*<< \nu (\mathbb{Z})\\
\end{CD}\]
where $i^*$ restricts a model on a sphere $S^d$ onto its equator $S^{d-1}$, which is also the loop space of $S^d$.
\par
Four among the eight symmetry classes of table \ref{table2} are invariant under chiral symmetry, which is a combination of particle-hole and time-reversal symmetries. These four symmetry classes are called BDI, CI, CII and DIII. We can characterize the topological properties of chiral symmetric $\mathbb{Z}$ topological insulators by a winding number (\ref{winding number}), which is defined in terms of the block off-diagonal projector $q$ like the class AIII. As the same reasons for non-chiral classes, we can derive a $\mathbb{Z}_2$ classification by imposing the constraint of additional discrete symmetries (such as $T$ and $C$) for lower-dimensional descendants.
Fist of all, we note that the winding number density (\ref{winding number density}) is purely real, $\omega^*_{2n+1}[q]=\omega_{2n+1}[q]$, which can be checked by direct calculation. Without any additional discrete symmetries, i.e. for symmetry class AIII, the off-diagonal projector $q$ and hence the winding number density are not subject to additional constraint. However for the symmetry classes BDI, DIII, CII, and CI, the presence of time reversal and particle-hole symmetries relates the projector $q$ at wavevector $\textbf{k}$ to the one at wavevector $-\textbf{k}$.
\par
The winding number density for symmetry class DIII and CI is subject to the constraint\cite{ryu2010topological},
\begin{equation}\label{DIII CI}
    \omega_{2n+1}[q(\textbf{k})]=(-1)^{n+1}\omega^*_{2n+1}[q(-\textbf{k})].
\end{equation}
Consequently, the winding number $\textrm{W}_{4n+1}[q]$ is vanishing in $d=4n+1$ dimensions, i.e. there exists no non-trivial topological state characterized by an integer winding number in (4n+1)-dimensional systems belong to symmetry class DIII or CI (table \ref{table2}). Conversely, in $d=(4n+3)$ dimensions both in class DIII and CI there are topologically nontrivial states. While for classes BDI and CII, the winding number density is constrained by \cite{ryu2010topological}
\begin{equation}\label{BDI CII}
     \omega_{2n+1}[q(\textbf{k})]=(-1)^{n}\omega^*_{2n+1}[q(-\textbf{k})].
\end{equation}
As a result, the winding number $\textrm{W}_{4n-1}[q]$ is vanishing in $d=4n-1$ dimensions, i.e. there are no nontrivial $(4n-1)$-dimensional $\mathbb{Z}$ topological insulators belonging to symmetry class BDI or CII (table \ref{table2}). On the other hand, in (4n+1) dimensions there exist $\mathbb{Z}$ topological states in both classes BDI and CII.  The non-trivial $\mathbb{Z}$ topological states in all chiral classes exist in the condition of $s-d\equiv0$ (mod 4), which is identity with the conditions for integer $\mathbb{Z}$  topological states of non-chiral classes. Like the non-chiral classes, we will give the integral topological invariants formula for chiral classes with base manifold $S^d$.
\begin{description}
  \item[$\mathbb{Z}$-invariants] This applies to $s-d\equiv0$ (mod 4). For chiral classes, the integral invariant is given by the winding number that characterized the stable homotopy group $\pi_d(U(n))$
\begin{eqnarray}\label{DIII CI3}
   && \tilde{\nu}=\textrm{W}_d[q]\nonumber\\
&=&\int_{S^d}\omega_d[q] \nonumber\\
   &=& \frac{(-1)^{\frac{d-1}{2}}(\frac{d-1}{2})!}{d!}
(\frac{i}{2\pi})^{\frac{d+1}{2}}\int_{S^d}\textrm{tr}(q^{-1}\textrm{d}q)^d
\end{eqnarray}
where $d$ is the space dimension, $q$ is the projector operator\cite{ryu2010topological}. In fact, one will always get an even integer for the case $s-d\equiv4$ (mod 8), because the Hamiltonian will be reduced to two copies of the Hamiltonian of case   $s-d\equiv0$ (mod 4)\cite{ryu2010topological}.
\item[1st $\mathbb{Z}_2$ descendent ] This applies to case $s-d\equiv-1$ (mod 8). For  $s$ chiral symmetry class, it is identical to that of the 2nd descendent in the non-chiral class $s+1$, $\pi_0{(R_{s-d})}=\pi_0{(R_{(s+1)-(d+1)})}$. So the chiral class can be characterized by the 2nd $\mathbb{Z}_2$ in non-chiral classes integral topological invariant called Fu-Kane invariant by forgetting either TR or PH symmetry.
\begin{eqnarray}\label{DIII CI2}
   \tilde{\nu}_1&=& \int_{S^d_{1/2}}\frac{1}{(\frac{d}{2})!}
\textrm{tr}(\frac{i\mathcal{F}}{2\pi})^{d/2}-\oint_{\partial S^{d}_{1/2}}Q_{d-1}(\mathcal{A},\mathcal{F})\nonumber  \\
   &=&  \int_{S^d_{1/2}}\textrm{ch}_{\frac{d}{2}}(\mathcal {F})-\oint_{\partial S^{d}_{1/2}}Q_{d-1}(\mathcal{A},\mathcal{F})(\textrm{mod}~2)
\end{eqnarray}
where $S^d_{1/2}$ is a hemisphere, and $\partial S^d_{1/2}=S^{d-1}$ is its equator boundary that is close under the condition $\textbf{k}\leftrightarrow\textbf{-k}$.

As the same reason for the non-chiral classes, Fu-Kane integral topological invariants will figure out the $\mathbb{Z}_2$ index under the constraint of gauge condition (\ref{C-Gauge constraint}) or (\ref{T-Gauge constraint}).
  \item[2nd $\mathbb{Z}_2$ descendent ] This applies to case $s-d\equiv-2$ (mod 8).  For chiral classes, we can construct the Chern-Simons invariant with Berry connection $\mathcal{A}^c$ of valence frame of chiral classes,

\begin{eqnarray}\label{DIII CI1}
    \tilde{\nu_2}&=& 2\textrm{CS}_{d}[\mathcal{A}^c,\mathcal{F}^c]\nonumber\\
   &=& 2\int_{S^{d}}Q_{d}(\mathcal{A}^c,\mathcal{F}^c)\nonumber  \\
   &=&2\int_{S^{d}}\frac{1}{2}\omega_{d}[q] \label{chiral CS W} \\
   &=& \textrm{W}_d[q]~~~(\textrm{mod}~ 2)
\end{eqnarray}
where $Q_{d}(\mathcal{A}^c,\mathcal{F}^c)$ is the Chern-Simons form, and the gauge constraint taken on over the base space $S^d$. The factor of $\frac{1}{2}$ comes out because of the Berry gauge of valence frame is identity to $\mathcal{A}^c=\frac{1}{2}q^{\dagger}\textrm{d}q$ for chiral classes\cite{ryu2010topological,teo2010topological}.  The valence frame of chiral classes can be chosen to be
\begin{equation}\label{valence frame of chiral class}
    u(\textbf{k})=\frac{1}{\sqrt{2}}\left(
                                      \begin{array}{c}
                                        q(\textbf{k}) \\
                                         -\mathbb{I}\\
                                      \end{array}
                                    \right).
\end{equation}
This corresponds the Berry connection $\mathcal{A}^c=u^{\dagger}\textrm{d}u=\frac{1}{2}q^{\dagger}\textrm{d}q$, so we have
\begin{eqnarray}
  \textrm{d}\mathbb{I}&=& \textrm{d}(q^{-1}q) =(\textrm{d}q) q^{-1}+q\textrm{d}q^{-1},\nonumber \\
   d\mathcal{A}^c&=&\frac{1}{2}\textrm{d}(q^{-1}\textrm{d}q)= -\frac{1}{2}(q^{-1}\textrm{d}q)(q^{-1}\textrm{d}q),\nonumber \\
  \mathcal{F}^c_t &=&t\textrm{d}\mathcal{A}^c+t^2(\mathcal{A}^c)^2=
(-\frac{t}{2}+\frac{t^2}{4})(q^{-1}\textrm{d}q)^2.
\end{eqnarray}
Substituting above chiral Berry connection $\mathcal{A}^c$ and $\mathcal{F}^c_t $ into $Q_{d}(\mathcal{A}^c,\mathcal{F}^c)$ we will get
\begin{equation}\label{Q-W}
    Q_{d}(\mathcal{A}^c,\mathcal{F}^c)=\frac{1}{2}\omega_d[q],
\end{equation}
which is half of the winding number density.

\end{description}
\par
Until now, we have given all the explicit integral topological invariants which characterize the twist of the fiber bundles for chiral and non-chiral $real$ classes. The structure group of $real$ classes are more complicated than $U(n)$, so we construct the topological invariants through the constraint of symmetry imposing on the Chern-number ,winding number and Chern-Simons invariants whose gauge group is still $U(n)$ rather than directly using the corresponding Lie group of each $real$ classes. The involution of $\textbf{k}\leftrightarrow-\textbf{k}$ can be realized by imposing the gauge condition like (\ref{C-Gauge constraint}) for PH symmetry or (\ref{T-Gauge constraint}) for TR symmetry.
\subsubsection{$\mathbb{Z}_2$ classification in Homotopy language}
\par
Next, let us analysis the dimensional hierarchy in homotopy language. We show how the $\mathbb{Z}_2$ topological insulators (superconductors) can be derived as lower dimensional descendants of parent $\mathbb{Z}$ topological insulators (superconductors). The general idea is to construct an interpolation between lower dimensional topological insulators (superconductors) Hamiltonians, then we can compare the different of any two lower-dimensional Hamiltonians by investing the properties of homotopy maps (\ref{homotopy}).
\par
Assuming the parent $\mathbb{Z}$ topological Hamiltonian in $d$-dimension can be characterized by Chern number for non-chiral classes. Then the homotopy maps or interpolations between any $(d-1)$-dimensional Hamiltonian in the same symmetry class are also defined in $d$-dimension as the same as the parent Hamiltonian. Consequently, a natural mind is that can we study the properties of homotopy maps via $d$-dimensional Chern-number? Fortunately, we can construct those homotopy maps whose symmetry class is the same as the $(d-1)$-dimensional topological states by usual homotopy map with its symmetry transformed partner.
\par
Now we will show how to define a $\mathbb{Z}_2$ invariant for AII  class, the discussion is very similar for other non-chiral classes. For any two band insulators $h_1(\textbf{k})$ and $h_2(\textbf{k})$, a homotopy map $h(\textbf{k},\theta)$ can be defined, satisfying
\begin{equation}\label{interpolation}
    h(\textbf{k},0)=h_1(\textbf{k}),~~h(\textbf{k},\pi)=h_2(\textbf{k}).
\end{equation}
If we define $[h(\textbf{k},\theta)]$ as the set of homotopically equivalence classes, then the classification of $d-1$-dimensional topological insulators is tantamount to find how many element in $[h(\textbf{k},\theta)]$. The homotopy map will be in the same symmetry class if we define for $\theta\in[\pi,2\pi]$
\begin{equation}\label{partner of interpolation}
    T^{-1}h(\textbf{k},\theta)T=h(-\textbf{k},-\theta)
\end{equation}
and $h(\textbf{k},\theta)$ is gapped for any $\theta\in[\pi,2\pi]$. Since the interpolation is periodic in $\theta$ just like $\textbf{k}$, so we can characterize the homotopy maps by Chern-number of Berry connection can be defined in the $(\textbf{k},\theta)$ $d$-dimension space. Two different homotopy maps $h(\textbf{k},\theta)$ and $h^{'}(\textbf{k},\theta)$ generally give different Chern-numbers, $\textrm{Ch}_{d/2}(h(\textbf{k},\theta))\neq\textrm{Ch}_{d/2}(h^{'}(\textbf{k},\theta))$.
However, we can show that symmetry constraint (\ref{partner of interpolation}) will leads to
\begin{equation}\label{the difference of Chern number}
    \textrm{Ch}_{d/2}(h(\textbf{k},\theta))-\textrm{Ch}_{d/2}(h^{'}(\textbf{k},\theta))=0~~(\textrm{mod} ~2),
\end{equation}
for any two interpolations. So we get the isomorphism of group $\textrm{Ch }\textrm{mod}~2: [h(\textbf{k},\theta)]\leftrightarrow\mathbb{Z}_2$, there are only two elements in $[h(\textbf{k},\theta)]$, which means that there are only non-trivial and trivial two states in $d-1$ dimension.
To prove equation (\ref{the difference of Chern number}) we introduce two new homotopy maps $g_1(\textbf{k},\theta)$ and $g_2(\textbf{k},\theta)$ that transform into each other under TR symmetry operations Fig.\ref{interpolation}
\begin{equation}\label{g1}
     g_1(\textbf{k},\theta)=\left\{\begin{array}{cr}
                                h(\textbf{k},\theta),& \theta\in[0,\pi], \\
                               h^{'}(\textbf{k},2\pi-\theta),&\theta\in[\pi,2\pi],
                            \end{array}\right.
\end{equation}
\begin{equation}\label{g2}
    g_2(\textbf{k},\theta)=\left\{\begin{array}{cr}
                                h^{'}(\textbf{k},2\pi-\theta),&\theta\in[0,\pi],\\
                               h(\textbf{k},\theta),& \theta\in[\pi,2\pi].
                            \end{array}\right.
\end{equation}
These are recombinations of the maps $h(\textbf{k},\theta)$ and $h^{'}(\textbf{k},\theta)$ with the following relation
\begin{equation}\label{chern number of g and h}
     \textrm{Ch}_{d/2}(h)-\textrm{Ch}_{d/2}(h^{'})=
 \textrm{Ch}_{d/2}(g_1)+\textrm{Ch}_{d/2}(g_2),
\end{equation}
which can be intuitively interpreted as the "area", it equals the area enclosed  by closed blue line minus the area  enclosed by closed red line like in Fig.\ref{interpolation} (a). However, we can also calculate the area by adding two areas of enclosed by the closed blue line and red line like in Fig.\ref{interpolation} (b). Consequently, the equation (\ref{chern number of g and h}) is correct.
\begin{figure}[h]
 \scalebox{0.4} [0.4]{\includegraphics{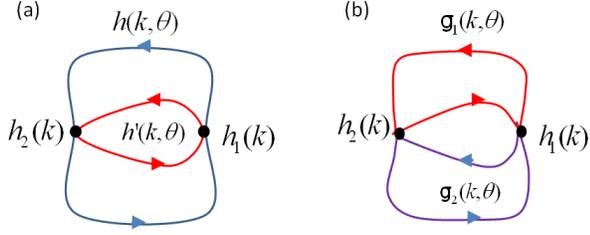}}\\
  \caption{(a) Two different homotopy maps, each colored red and blue, respectively. In (b), the original homotopy maps are rearranged into two different interpolations denoted as $g_1$ and $g_2$, which is the symmetry partner of each other.}\label{interpolation}
\end{figure}
Now according to the result of the equation (\ref{AII}), we know for TR symmetry we found $\textrm{ch}_{n}[\mathcal {F}(\textbf{k})]= \textrm{ch}_{n}[\mathcal {F}(-\textbf{k})],$ in $d=4n$ spatial dimensions, with this and equation (\ref{chern number}) we obtain
\begin{eqnarray}
  \textrm{Ch}_{n}(g_1) &=&\int \textrm{ch}_{n}(g_1(\textbf{k},\theta))  \\
   &=&\int \textrm{ch}_{n}(g_1(-\textbf{k},-\theta))   \\
   &=&\int \textrm{ch}_{n}(g_2(\textbf{k},\theta))=  \textrm{Ch}_{n}(g_2)
\end{eqnarray}
where $\textrm{d}^{4n-1}k\textrm{d}\theta$ is the volume form, and we made use of the fact that $g_1$ transforms into $g_2$ under the TR symmetry operations of class AII. In conclusion, we have shown that $\textrm{Ch}_{d/2}(h(\textbf{k},\theta))-\textrm{Ch}_{d/2}(h^{'}(\textbf{k},\theta))
=2\textrm{Ch}_{d/2}(g_1(\textbf{k},\theta))\in 2\mathbb{Z}$ for any two interpolations. Hence we can define a relative invariant for the
$4n-1$-dimensional Hamiltonian $h_1(\textbf{k})$ and $h_2(\textbf{k})$
\begin{equation}\label{definition of relative z2}
    \nu_{d-1}[h_1(\textbf{k}),h_2(\textbf{k})]=(-1)^{\textrm{Ch}_{d/2}(h(\textbf{k},\theta))},
\end{equation}
which is independent of the particular choice of the interpolation $h(\textbf{k},\theta)$ between $h_1(\textbf{k})$ and $h_2(\textbf{k})$. Once we have identified a 'vacuum' Hamiltonian $h_0(\textbf{k})\equiv h_0$, we can construct with equation (\ref{definition of relative z2}) a $\mathbb{Z}_2$ invariant: nontrivial Hamiltonians are characterized by $ \nu_{d-1}[h_1(\textbf{k}),h_0]=-1$, whereas trivial ones satisfy $\nu_{d-1}[h_1(\textbf{k}),h_0]=1$. In physics, the $\nu_{3}[h_1(\textbf{k}),h_0]=-1$ indicate there is non-trivial topological TR symmetry protected insulators in 3-dimension, which is just the usual topological insulators, precisely the strong topological insulators. As we all know, there is still another topological insulator QSHE in 2-dimension, so we can also try to obtain the 2nd-descendants through this homotopy discussion.
\par
The dimensional reduction procedure presented can be repeated once more to obtain a $\mathbb{Z}_2$ classification of the second descendants. The idea is very easy, we can construct two-dimensional homotopy map between any two $(d-2)$-dimension Hamiltonians in class AII, then study the properties of homotopy maps through topological invariant $\nu_{d-1}[h_1(\textbf{k}),h_2(\textbf{k})]$. Similarly, we consider two $(d-2)$-dimensional Hamiltonians $h_1(\textbf{k})$ and $h_2(\textbf{k})$ with TR symmetry. Now we define a homotopy map $h(\textbf{k},\theta)$, $\theta\in[0,2\pi]$

\begin{eqnarray}\label{interpolation 2}
  && h(\textbf{k},0)=h_1(\textbf{k}),~~h(\textbf{k},\pi)=h_2(\textbf{k}),\nonumber\\
 &&T^{-1}h(\textbf{k},\theta)T=h(-\textbf{k},-\theta).
\end{eqnarray}
We can interpret $h(\textbf{k},\theta)$ as a $(d-1)$-dimension Hamiltonian belonging to the same symmetry class AII. Therefore, for any two deformations $h(\textbf{k},\theta)$ and $h^{'}(\textbf{k},\theta)$ of form (\ref{interpolation 2}) a relative invariant $\nu_{d-1}[h(\textbf{k}),h^{'}(\textbf{k})]$ can be defined. It turns out that due to condition $(\ref{interpolation 2})$ the invariant $\nu_{d-1}[h(\textbf{k}),h^{'}(\textbf{k})]$ is independent of the particular choice of interpolations, i.e. $\nu_{d-1}[h(\textbf{k}),h^{'}(\textbf{k})]$=1 for any two homotopy maps $h(\textbf{k},\theta)$ and $h^{'}(\textbf{k},\theta)$ satisfying the condition (\ref{interpolation 2}). To prove this, we consider the 2 parameters homotopy map $h(\textbf{k},\theta,\varphi)$ between above 1 parameter homotopy maps $h(\textbf{k},\theta)$ and $h^{'}(\textbf{k},\theta)$ with
\begin{eqnarray}
   &&h(\textbf{k},\theta,0)=h(\textbf{k},\theta),~~h(\textbf{k},\theta,\pi)=h^{'}(\textbf{k},\theta),\nonumber \\
   &&h(\textbf{k},0,\varphi)= h_1(\textbf{k}),~~h(\textbf{k},\pi,\varphi)=h_2(\textbf{k}), \nonumber\\
   && T^{-1}h(\textbf{k},\theta,\varphi)T=h(-\textbf{k},-\theta, -\varphi).
\end{eqnarray}
This represents a $d$-dimensional insulator Hamiltonian with the same symmetry, for which  the Chern-number can be defined $\textrm{Ch}_{d/2}(h(\textbf{k},\theta, \varphi))$. We note that $h(\textbf{k},\theta,\varphi)$ is not only a homotopy map between $h(\textbf{k},\theta)$ and $h^{'}(\textbf{k},\theta)$, but can also be viewed as a continuous homotopy map between $h(\textbf{k},0,\varphi)= h_1(\textbf{k})$ and $h(\textbf{k},\pi,\varphi)=h_2(\textbf{k})$ for any $\varphi\in[0,2\pi]$. Therefore we find
$\nu_{d-1}[h(\textbf{k},\theta),h^{'}(\textbf{k},\theta)]
=\nu_{d-1}[h(\textbf{k},\theta,0),h(\textbf{k},\theta,\pi)]
=\nu_{d-1}[h(\textbf{k},0,\varphi),h(\textbf{k},\pi,\varphi)]$.
Since $h(\textbf{k},0,\varphi)=h_1(\textbf{k})$ and $h(\textbf{k},\pi,\varphi)=h_2(\textbf{k})$ are independent of $\varphi$, we find that $\nu_{d-1}[h(\textbf{k},0,\varphi),h(\textbf{k},\pi,\varphi)]=(-1)^{\textrm{Ch}_{d/2}(h)}=1$
due to Chern-number $\textrm{Ch}_{d/2}(h)$ is zero.
Hence, we have shown that $\nu_{d-1}[h(\textbf{k},\theta),h^{'}(\textbf{k},\theta)]$ only depends on $h_1(\textbf{k})$ and $h_2(\textbf{k})$. Therefore, $\nu_{d-2}[h_1(\textbf{k}),h_2(\textbf{k})]\equiv\nu_{d-1}[h(\textbf{k},\theta),h_0]$
 together with a reference vacuum $h_0$ constitutes a well-defined $\mathbb{Z}_2$ invariant in $(d-1)$-dimensions. By defining a constant Hamiltonian $h_0$ as a reference, all $(d-2)$-dimensional insulators are classified by the value of $\nu_{d-2}[h_0,h(\textbf{k})]$. An insulator with $\nu_{d-2}[h_0,h(\textbf{k})]=-1$ cannot be adiabatically deformed to the trivial Hamiltonian $h_0$ without breaking time-reversal symmetry.
\par
The other non-chiral symmetry classes topological insulators can be similarly discussed using this procedure talked above. The homotopy discussion of dimensional hierarchy in chiral classes also can be similarly discussed with the parent $\mathbb{Z}$ topological insulators characterized by winding number. This has been performed previously in\cite{ryu2010topological}, the readers who are interested can refer\cite{ryu2010topological} for details.

\subsubsection{General Homotopy discussion}
\par
Now we will give a mathematic way to interpret this homotopy analysis. Here we only show a simple argument and the details will be published elsewhere in the future. Assuming the manifold of  insulator Hamiltonian in symmetry class $s$  and $d$ spatial dimension is $h^d_{s}(\textbf{k})$, the path space $\Omega(h^d_s,p,q)$ is the set of paths  $\gamma(t)$ on the manifold (see homotopy subsection) that from a point $p=\gamma(0)$ to point $q=\gamma(1)$.  We also can construct a set denoting $\Omega_h(h^{d}_{s}, h_1, h_2)$ of homotopy maps $h^{d}_{s}(\textbf{k},\theta)$ from $(d-1)$-dimensional Hamiltonian $h^{d}_{s}(\textbf{k},0)=h_1$ to $h^{d}_{s}(\textbf{k},\pi)=h_2$. Now  one of the paths can identify  with one homotopy maps of $h^{d}_{s}(\textbf{k},\theta)$ if  $p=h_1$ and $q=h_2$, so we can construct a map
\begin{equation}\label{path space and homotopy set}
    F: \Omega(h^d_s,p,q)\rightarrow \Omega_h(h^{d}_{s}, h_1, h_2),
\end{equation}
from the paths space to homotopy maps of any two $(d-1)$-dimensional Hamiltonians, it is easy to know this is a isomorphism. Combining with the discrete symmetry partner $h^{d}_{s}(-\textbf{k},-\theta)$, this will correspond to a loop in manifold $h^d_s$, denoted as $\Omega_h(h^{d}_{s}, h_1)$ or $\Omega(h^d_s,p)$ with base point $p=h_1$. So we can study the homotopoy properties of loop space $\Omega(h^{d}_{s}, p)$ to classify the $(d-1)$-dimensional Hamiltonians,
\begin{equation}\label{homotopy of loop of interpolation}
    \pi_i(\Omega(h^{d}_{s}, p))=\pi_i(\Omega h^d_s)=\pi_{i+1}(h^d_s)
\end{equation}
where $\Omega$ is the loop space operation and we have used the equation (\ref{loop space}). If we use the label of classify space $R_{s-d}$ denote the $d$-dimensional insulator with symmetry $s$, then we obtained
\begin{equation}\label{loop and homotopy}
  \pi_0(\Omega_h(R_{s-d}, h_1))=\pi_{1}(R_{s-d})=\pi_0(R_{s-(d-1)})
\end{equation}
where we have used $\pi_{i-1}(R_{s-d})=\pi_{i+1}(R_{s-(d-1)})$ for any integer $i$. So the $(d-1)$-dimensional insulator can be classified with $\pi_0(\Omega_h(R_{s-d}, h_1))$ or $\pi_0(R_{s-(d-1)})$, which is consist with the subsection of homotopy.
\par
In a similar manner, we can construct $\Omega_h(R_{s-(d-1)}, h_1)$ of the homotopy maps which form the loop space of  $(d-1)$-dimensional Hamiltonian manifold, with $(d-2)$-dimensional $h_1$ as the base point.
As the same logic, we can study the properties of these homotopy maps by studying the zero homotopy group of $\Omega_h(R_{s-(d-1)}, h_1)$,
\begin{equation}\label{the 2th descendants}
    \pi_0(\Omega_h(R_{s-(d-1)}, h_1))=\pi_0(R_{s-(d-2)}),
\end{equation}
which is also consist with the subsection of homotopy classification. Because the loop space  and double loop space of $R_{s-d}$  is non-trivial for $(s-q)\equiv 0$ (mod 4) (see table \ref{table1}), there are the so called 1st and 2nd descendants talked above, which also explain  why the homotopy discussion can only exist in $(d-1)$ and $(d-2)$-dimension. While for $\mathbb{Z}$ invariant satisfying $(s-q)\equiv 4$ (mod 8), there are no non-trivial loop space manifold following, so there are no $\mathbb{Z}_2$ descendants.
\par
For example, the parents of TR symmetry class $s=4$ is $4n$-dimensional insulators, the 4D quantum hall effect\cite{zhang2001four} in this class with $d=4$. The topological invariant is $\pi_0(R_{4-4})=\mathbb{Z}$, the 1st descendants can be described with a $\mathbb{Z}_2$ index, which can be obtained by $\pi_0(R_{4-3})=\mathbb{Z}_2$, the 2nd descendants is characterized by $\pi_0(R_{4-2})=\mathbb{Z}_2$.
\par
In conclusion, the homotopy maps discussed in the last subsubsection are nothing but establishing an explicit way to study the loop spaces of the parent space. It can be easily understood in loop space language in this subsubsection. In physics, the loop space can be interpreted as an adiabatic evolution of $\theta$ from $0$ to $2\pi$ , which defines a cycle of adiabatic pumping in $h(\textbf{k},\theta)$.
\par
The TBT is a beautiful theory since it studies the topology properties of Bloch Hamiltonian or BdG superconductors, but we cannot directly establish the relationships between the topological responses of physical system and these  topological invariants. Next, we will discuss TFT which describing the topological responses directly through coupling external field such as electromagnetic field or gravitational field. Through this theory, we can obtain the quantization  Hall conductance of IQHE, topological magnetoelectric response and so on.
\subsection{Topological Field Theory}
In this subsection, we mainly discuss the topological properties of  Chern-Simons topological field theory which  has been extensively used in topological insulators\cite{qi2008topological}. It is a powerful effective theory, both the integer and the fractional quantum hall effect can be described by the topological Chern-Simons field theory in $2+1$ dimensions space-time space\cite{zhang1989effective}. This effective topological field theory captures all physically measurable topological effects, including the quantization of the Hall conductance, the fractional charge, and the statistics of quasiparticles\cite{zhang1992chern}. In recent years, the time reversal invariant (TRI) insulators have found also can be described by Chern-Simons field theory\cite{bernevig2002effective} and the dimensional reduction of it\cite{qi2008topological}. For example, the field theory of  QSHE insulators and TRI topological insulators in (3+1) dimension can be obtained by dimensional reduction from the parent Chern-Simons topological field theory in (4+1) dimensional space-time.
 \par

In the following, we will show the Chern-Simons effective theory can be appearance naturally in physics. And we will give a compact formula of Chern-Simons field theory with the convenient differential forms, which is equivalence to the familiar Chern-Simons field theory that is expressed with various components and anti-symmetric tensor.  From this compact form, we obtain the general connection between non-interacting Chern-Simons field theory with the factor of Chern-number $(\textrm{Ch}_n)$ of valence fiber bundle and interacting ones with the factor of winding number of Greens' functions $( N_{2n+1}[G])$.  In addition, we find the general connection of topological invariants that describe different Dirac models through the projector operator. What's more,  we find the  beautiful mathematic corresponding topological invariants for various Dirac models. The topological invariant in lattice Dirac models of topological insulators are extended to higher dimension, and we find it is nothing but the Gauss map number, the special case for $d$=2, and $d=4$ have been discussed\cite{qi2008topological}. In addition, we will discuss the general phase-space Chern-Simons theory defined in ref.(\cite{qi2008topological}), which gives a unified theory of all topological insulators and contains all experimentally observable topological effects. Finally, we briefly review some important results about bulk-boundary correspondence, and introduce the beautiful index theorem\cite{volovik2003universe} that relates the chirality gapless excitation states of the edge and the topological invariants of bulk.
\subsubsection{The Chern-Simons field theory}
\begin{figure}[h]
 \scalebox{0.5} [0.4]{\includegraphics{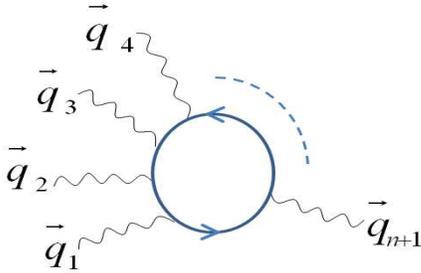}}\\
  \caption{ The one loop Feynman diagram in $2n+1$ dimensions contributing to the induced Chern-Simons action for Abelian gauge fields. The loop is a fermion propagator and the wavy lines are external legs corresponding to the gauge field.}\label{Feynman loop}
\end{figure}

Field theory of electrons interacting with external Abelian gauge field $A_{\mu}$ is given by the Dirac Lagrangian density\cite{dayi2011effective}
\begin{equation}\label{Lagrangian density}
    \mathscr{L}(\psi,\overline{\psi},A)=\overline{\psi}[\gamma^{\mu}(p_{\mu}+A_{\mu})-m]\psi
\end{equation}
where $\mu=0,1,\ldots, d$. By integrating out the fermionic degrees of freedom in the related path integral and using the one loop approximation in the weak field or long wavelength limit for the gauge fields, one will get the Abelian Chern-Simons action in $D=d+1=(2n+1)$  continuous Euclidean dimensions like following \cite{golterman1993chern,qi2008topological,dayi2011effective}
\begin{eqnarray}\label{Chern-Simons field}
  S_{2n+1}&=&c_n\cdot\epsilon^{\mu_1\mu_2\cdots\mu_{2n+1}}\int\textrm{d}^{2n+1}x \nonumber\\
   &&\times A_{\mu_1}\partial_{\mu_2}A_{\mu_3}\cdots\partial_{\mu_{2n}}A_{\mu_{2n+1}}
\end{eqnarray}
with the coefficient $c_n$ which can be computed by calculating the relevant portion of the Fig.\ref{Feynman loop} and note that Ward identity,
\begin{eqnarray}\label{cn coefficient}
  c_n &=& \frac{(-i)^{n+1}\epsilon^{\mu_1\mu_2\cdots\mu_{2n+1}}}{(n+1)(2n+1)!}
\int\frac{\textrm{d}^{2n+1}}{(2\pi)^{2n+1}}\times\textrm{Tr}\nonumber\\
   &&[(G\partial_{\mu_1}G^{-1})(G\partial_{\mu_2}G^{-1})\cdots (G\partial_{\mu_{2n+1}}G^{-1})],
\end{eqnarray}
where $\epsilon^{\mu_1\mu_2\cdots\mu_{2n+1}}$ is totally antisymmetry tensor, $A$ is the external Abelian gauge field, $x$ is the space time coordinates, $k$ is the  momentum-frequency space and $G\equiv G(\textbf{k},\omega)$ is the imaginary-time  single-particle Green's function of a fully interacting insulator.
\par
This Chern-Simons field formula is not so rigorous that one cannot see the topological properties at first time because of many components  and indices used in the expression. Now we will give the beautiful formula of this expression with differential forms, it's essential simplicity because one can replace  the integral volume ,  the sum  of antisymmetry tensor and  corresponding components by the simple wedge product. For example, the components expression (\ref{Chern-Simons field})can be replaced by
\begin{equation}\label{chern-simons field 1}
     S_{2n+1}=c_n\cdot\int A\wedge\overbrace{\textrm{d}A\cdots\wedge\textrm{d}A}^{n}
=c_n\int A(\textrm{d}A)^n,
\end{equation}
where $\wedge$ is wedge product, $\textrm{d}$ is the exterior differential operator. As the same reason, the coefficient $c_n$ also can be written an compact expression as
\begin{equation}\label{cn coefficient1}
    c_n=\frac{(-i)^{n+1}}{(n+1)(2n+1)!(2\pi)^{2n+1}}\int\textrm{Tr}(G\textrm{d}G^{-1})^{2n+1}.
\end{equation}
However, this transformation of expressions is not a useless  mathematic tricks, we will later show that the compact forms are very convenient for deducing more complexity expressions. Next we will use winding number (\ref{winding number}) of Green's function and Chern-Simons topological invariant (\ref{abelian}) to reexpress the equation (\ref{Chern-Simons field}),
\begin{eqnarray}
 &&S_{2n+1}\nonumber\\
   &=& \frac{(-i)^{n+1}}{(n+1)(2n+1)!(2\pi)^{2n+1}}\int\textrm{Tr}(G\textrm{d}G^{-1})^{2n+1}\int A(\textrm{d}A)^n\nonumber  \\
   &=&-2\pi \frac{(-i)^{n+1}n!}{(2n+1)!}\int\frac{1}{(2\pi)^{n+1}}\textrm{Tr}(G\textrm{d}G^{-1})^{2n+1}\nonumber\\ &&\int \frac{1}{(n+1)!(2\pi)^{n+1}}A(\textrm{d}A)^n \nonumber  \\
   &=&-2\pi\textrm{W}_{2n+1}[G^{-1}]\cdot (-i)^{n+1} \textrm{CS}_{2n+1}[A] \nonumber  \\
   &=&2\pi\cdot\textrm{N}_{2n+1}[G]\cdot\textrm{CS}_{2n+1}[A']
\end{eqnarray}
where we have used the winding number definition (\ref{winding number}), redefine an invariant $\textrm{N}_{2n+1}[G]=-\textrm{W}_{2n+1}[G^{-1}]$ which actually equal to the winding number and define $A^{'}=-iA$.  $N_{2n+1}[G]$ is a significant topological invariants in CMP\cite{volovik2003universe,wang2010topological,qi2011topological,zubkov2012momentum},
which has been used to investigate the robust properties of ground states. In order to simplify the label of external gauge field $A^{'}$, we still denote $A$ in the final form for convenience. Consequently, the Chern-Simons field theory of insulators with interacting can be reexpressed in a more compact form,
\begin{equation}\label{compact Chern-Simons field theory}
    S_{2n+1}=2\pi\cdot\textrm{N}_{2n+1}[G]\cdot\textrm{CS}_{2n+1}[A].
\end{equation}
One can quickly know the topological properties of this effective field theory from above compact form (\ref{compact Chern-Simons field theory}), this compact form means
\begin{equation}\label{Chen-Simons field theory property}
    S_{2n+1}=2\pi  ~~(\textrm{mod} ~1),
\end{equation}
$S_{2n+1}$ is nothing but an integer times $2\pi $, $S_{2n+1}=2n\pi$, $n$ is any integer number. We can get the response of fermionic system trough the motion equation\cite{qi2008topological}

\begin{eqnarray}
   &&J_{2n+1}\nonumber\\
 &=&\frac{\delta S_{2n+1}}{\delta A}  \nonumber\\
   &=&\frac{2\pi }{\delta A}\frac{\textrm{N}_{2n+1}[G]}{(n+1)!(2\pi)^{n+1}}\int \delta A \wedge(\textrm{d}A)^n +nA\wedge\delta (\textrm{d}A)(\textrm{d}A)^{n-1} \nonumber  \\
   &=&\frac{2\pi }{\delta A}\frac{\textrm{N}_{2n+1}[G]}{(n+1)!(2\pi)^{n+1}}\int \delta A \wedge(\textrm{d}A)^n -n\textrm{d}A\wedge\delta A(\textrm{d}A)^{n-1}\nonumber  \\
   &=&  \frac{\textrm{N}_{2n+1}[G]}{n!(2\pi)^{n}}\int(\textrm{d}A)^n,
\end{eqnarray}
where $J$ is the general "current", we have used the communication of variational operator $\delta$ and exterior differential operator $\textrm{d}$, $\textrm{d}\delta=\delta\textrm{d}$. Finally, we can define the general current density one-form $\mathcal {J}$ as
\begin{equation}\label{general current densitiy}
    \mathcal {J}_{2n+1}= \frac{\textrm{N}_{2n+1}[G]}{n!(2\pi)^{n}}\ast (\textrm{d}A)^n
\end{equation}
where $\ast$ is the Hogde star operator which makes a $p$-form become a $(n-p)$-form, $n$ is the total dimension of differential forms\cite{nakahara2003geometry}. So from the definition of $j$  and the property of Hogde star operator, we know it is a one-form. This is a beautiful general current formula, the  physical response of 2D-IQHE  and 4D-IQHE  can be reexpressed use this one-form formula  as $\mathcal {J}_3$ and $\mathcal {J}_5$.  For convenience, here we write these currents explicitly,
\begin{equation}\label{j3}
    \mathcal {J}_3^{\mu}=\frac{N_3}{2\pi}\epsilon^{\mu\nu\tau}\partial_{\nu}A_{\tau}
\end{equation}
\begin{equation}\label{j5}
     \mathcal {J}_3^{\mu}=\frac{N_5}{8\pi^2}
\epsilon^{\mu\nu\rho\sigma\tau}\partial_{\nu}A_{\rho}\partial_{\sigma}A_{\tau}
\end{equation}
these components formula both are consistent with the previous work\cite{qi2008topological}. Later we will show the topological invariants $\textrm{N}_{2n+1}[G]$ will be equal to the Chern-number of non-interacting system generally, $\textrm{N}_{2n+1}[G_0]
=\textrm{Ch}_n$, where $G_{0}$ is the Green's function of non-interacting system.
In other words, we can write the beautiful  general current formula for non-interacting insulators system as
\begin{equation}\label{non-interacint j}
    j_{2n+1}=\frac{\textrm{Ch}_n(\mathcal {F})}{n!(2\pi)^{n}}\ast (\textrm{d}A)^n
\end{equation}
where $\mathcal {F}$ is Berry curvature defined as (\ref{define berry curvature}) in momentum space.
\par
This low-energy effective Chern-Simons field theory can provide nearly all physical response\cite{zhang1992chern,qi2008topological}, we will obtain the important $\mathbb{Z}_2$ topological insulators via 'dimensional reduction' by compactifying one or more dimensions. In the following section, we will prove some important topological invariants used in this field, and obtain some significant relationship between them. In addition, we will give some special model topological invariants and explaining the physical meaning.
\subsubsection{General topological invariants and their relation}
 In this subsection, we will show how to derive the Chern-number of non-interacting systems from the topological invariants $\textrm{N}_{2n+1}[G]$, and next give a general formula for a special Dirac lattice model.
 The important step is to use the projector operators of valence band $P_{G}$  and conduction band $P_E$, the same trick has been used in the ref \cite{murakami20042} and \cite{qi2008topological}. The Hamiltonian and Green's function both can be expressed use these projectors, the process of which is equal to band "flat" but preserve the topology properties. The final form is actually homotopically equivalent to the original model only if the band gap doesn't close, one can explicitly construct this homotopy relation as in ref \cite{qi2008topological}. Finally, one can obtain the non-interacting Chern-number by performing again the integration over the frequency.
\par
With the projector operators, one can write the Green's function in a simple form
\begin{eqnarray}
  G_0(\textbf{k},\omega) &=&[i\omega-\epsilon_GP_{G}(\textbf{k})-\epsilon_E P_{E}(\textbf{k})]^{-1}\nonumber \\
   &=& \frac{P_{G}(\textbf{k})}{i\omega-\epsilon_G}+\frac{P_{E}(\textbf{k})}{i\omega-\epsilon_E},
\end{eqnarray}
where $\epsilon_G,\epsilon_E$ are the energy of the ground state and excited state, respectively. Next, we introduce some important properties required to prove the general relation. As the same as usual projector, we have $P_{G}+P_{E}\equiv I, P_G^2=P_G, P_E^2=P_E, P_EP_G= 0$. In addition, from this identity and the property of exterior differential operator $\textrm{d}$ we obtain
\begin{gather}\label{projector}
    \textrm{d}P_G+\textrm{d}P_E=0\\
    \textrm{d}P_EP_G+P_E\textrm{d}P_G=0 \label{2projector}\\
    P_E\textrm{d}P_GP_E=P_G\textrm{d}P_GP_G=0,\label{2projector relation}
\end{gather}
where (\ref{2projector relation}) can be obtained by times $P_G$ from the left side of (\ref{2projector}) or $P_E$ from the right side of (\ref{2projector}).
 And the inverse of Green's function reads as $G(\textbf{k},\omega)^{-1}=i\omega-\epsilon_GP_{G}(\textbf{k})-\epsilon_E P_{E}(\textbf{k})$, where $\epsilon_E>0 $ is the conduction energy of "flat" model and $\epsilon_G<0 $ is  the corresponding ground state energy. After introducing these required knowledge, one more thing we need is the curvature 2-form  which expressed with projectors.  As a matter of fact, one can  obtain the following identities  by choosing an explicit basis\cite{qi2008topological},
\begin{equation}\label{p-curvature}
   \mathcal{F}=P\textrm{d}P\wedge \textrm{d}PP.
\end{equation}
 It is time to put  all of above into the form of $\textrm{N}_{2n+1}[G]$, here we give the explicit calculation as following,
\begin{eqnarray}
   &&N_{2n+1}[G_0]=-\int_{BZ\times\mathbb{R}^{\omega}}\omega_{2n+1}[G_0^{-1}] \nonumber \\
   &=& \frac{(-1)^{n+1}n!}{(2n+1)!}\left(\frac{i}{2\pi}\right)^{n+1}
    \int_{BZ\times\mathbb{R}^{\omega}}\textrm{tr}[(G_0\textrm{d}G_0^{-1})^{2n+1}]\nonumber\\
   &=&\frac{(-1)^{n+1}n!}{(2n+1)!}\left(\frac{i}{2\pi}\right)^{n+1}\int\textrm{d}(i\omega)
   \frac{(2n+1)(\epsilon_{G}-\epsilon_{E})^{2n}}{(i\omega-\epsilon_{G})^{n+1}
   (i\omega-\epsilon_{E})^{n}}\nonumber\\&&\cdot\int_{BZ}\textrm{tr}[P_G(\textrm{d}P_G\wedge
   \textrm{d}P_G)^nP_G]+(G\leftrightarrow E)\nonumber\\
   &=&\frac{n!}{(2n)!}\left(\frac{-i}{2\pi}\right)^{n+1}(2\pi i)\frac{(-1)^n(2n)!}{2(n!)^2}\nonumber\\&&\cdot\int_{BZ}\textrm{tr}[P_G(\textrm{d}P_G\wedge
   \textrm{d}P_G)^nP_G]-(G\leftrightarrow E) \label{residue} \\
   &=&\frac{1}{2n!}\left(\frac{i}{2\pi}\right)^n\int_{BZ}\textrm{tr}[(P_G-P_E)(\textrm{d}P_G\wedge
   \textrm{d}P_G)^n]\nonumber\\
   &=& \frac{1}{2n!}\left(\frac{i}{2\pi}\right)^n\int_{BZ}2\textrm{tr}[P_G(\textrm{d}P_G\wedge
   \textrm{d}P_G)^n]\label{p+p=1}\\
   &=& \int_{BZ}\frac{1}{n!}\textrm{tr}[\frac{i P_G(\textrm{d}P_G\wedge
   \textrm{d}P_G)P_G}{2\pi}]^n \\
   &=& \int_{BZ}\frac{1}{n!}\textrm{tr} (\frac{i\mathcal{F}}{2\pi})^n \\
   &=& \textrm{Ch}_n[\mathcal{F}],
\end{eqnarray}
where in the equation (\ref{residue}), the residue theorem has been used, equation (\ref{p+p=1}) is correct due to the totally derivative of $\textrm{tr}(\textrm{d}P_G\wedge\textrm{d}P_G)^n$ when substituting $P_{G}+P_{E}\equiv \mathbb{I}$, and  finally we have used equation (\ref{p-curvature}). In summary, carrying out the integral over $\omega$ firstly, and then through some tedious algebra calculation, we obtain
\begin{equation}\label{N=Ch}
    N_{2n+1}[G_0]= \textrm{Ch}_n[\mathcal{F}]
\end{equation}
This process establish the relationship for the dimensional hierarchy of Chern number. In other words, we have the following map,
\begin{equation}\label{chern-chern}
   N_{2n+1}[G] =\textrm{Ch}_{n+1}[\mathcal{F'}]\stackrel{P_G}{\longrightarrow} \textrm{Ch}_n[\mathcal{F}]=N_{2n+1}[G_0],
\end{equation}
where we have used equation (\ref{chern number and winding n}), $\mathcal{F'}$  and $\mathcal{F}$ are Berry curvature forms of fire bundle $ P(T^d,P_E \otimes P_G)$ and $P(T^d,P_G)$, here $T^d$ is $d$-dimensional BZ, $P_E$ and $P_G$ are viewed as Lie groups.
\par
Secondly, we propose the topological invariant $N_{2n+1}[G_0]$ is identity to the Gauss map number, also called  the generalized winding number $W_G$ \cite{duan2000topological,qi2008topological} for the compact Green's function,
\begin{equation}\label{Gauss map number}
    G_0^{'}=\frac{i\omega+d_a\Gamma^a}{(i\omega)^2-d_a d^a},
\end{equation}
where repeated indices mean summation, and $\Gamma^a$ is Clifford algebra, satisfying,
\begin{equation}\label{Clifford Algebra}
    \{\Gamma^a ,\Gamma^b\}=2\delta_{ab}\mathbb{I}
\end{equation}
with $\mathbb{I}$ as the identity matrix, and $\{,\}$ is the anti-communication operator.
This Green's function is a generalized Dirac model, can be used to demonstrate Class A topological insulator or time-reversal invariant topological insulator. Assuming $d_ad^a = 1$, this condition is reasonable due to the topology property of Green's function is hold. Under this normalization condition the projection operator can be written
simply as
\begin{equation}\label{1-projector}
    P_{\pm}=\frac{1}{2}[1\pm d_a\Gamma^a].
\end{equation}
Substituting $P_+$ in equation (\ref{p+p=1}) by (\ref{1-projector}), we can obtain
\begin{eqnarray}
   && N_{2n+1}[G_0^{'}]=\frac{1}{n!}\left(\frac{i}{2\pi}\right)^n\int_{BZ}\textrm{tr} P_+\overbrace{\textrm{d}P_+\wedge\textrm{d}P_+\cdots\wedge\textrm{d}P_+}^{2n}\nonumber\\
   &=&\frac{1}{n!}\left(\frac{i}{2\pi}\right)^n\frac{1}{2^{2n+1}}\int_{BZ}d^{a_1}\textrm{d}d^{a_2}\wedge\cdots\wedge
   \textrm{d}d^{a_{2n+1}}\nonumber\\
   && \cdot\textrm{tr}[\Gamma_{a_1}\Gamma_{a_2}\cdots\Gamma_{a_{2n+1}}] \nonumber \\
   &=&\frac{1}{2^{2n+1}n!}\left(\frac{i}{2\pi}\right)^n\int_{BZ}d^{a_1}\textrm{d}d^{a_2}\wedge\cdots\wedge
   \textrm{d}d^{a_{2n+1}}\nonumber\\
   &&\cdot(-2i)^n\in_{a_1a_2\cdots a_{2n+1}}\\
   &=&\frac{\in_{a_1a_1\cdots a_{2n+1}}}{2^{2n+1}n!\pi^n}\int_{BZ}d^{a_1}\textrm{d}d^{a_2}\wedge\cdots\wedge
   \textrm{d}d^{a_{2n+1}}\nonumber  \\
   &=&\frac{\in_{a_1a_2\cdots a_{2n+1}}}{A(S^{2n})(2n)!}\int_{BZ}d^{a_1}\textrm{d}d^{a_2}\wedge\cdots\wedge
   \textrm{d}d^{a_{2n+1}}\nonumber  \\
   &=&W_{G}(2n),
\end{eqnarray}
where $\in_{a_0a_1\cdots a_{2n+1}}$ is totally antisymmetry tensor, $A(S^{2n})$ is the area of the 2n dimensional sphere, and the non-zeros $\Gamma$ matrixes is just 2n+1 different Gamma matrix multiplied because of the antisymmetry of wedge product. In addition, one can use an implicit representation of Gamma matrix\cite{ryu2010topological}to calculate the trace, in fact, $\Gamma_{a_1}\Gamma_{a_2}\cdots\Gamma_{a_{2n+1}}$ is a constant matrix according to Schur's theorem\cite{ma2007group}. This formula is just the general winding number of mapping from the BZ $T^{2n}$ to the sphere $S^{2n}$, which is used in ref\cite{qi2008topological,PhysRevB.74.085308}.
\par
 To make the physical picture clearer, the simplest case of a two-band model can be studied as an example\cite{PhysRevB.74.085308,qi2008topological}. The Hamiltonian of a two-band model can be generally written as
\begin{equation}\label{two-band model}
    h(\textbf{k})=d_{a}\sigma^a+\epsilon(\textbf{k})\mathbb{I},
\end{equation}
where $\mathbb{I}$ is the $2\times2$ identity matrix and $\sigma^a$ are the three Pauli matrices, this model is called quantum anomalous Hall effect (QAHE)\cite{PhysRevB.74.085308}. We can immediately write the topological invariants using $W_G(2)$ as
\begin{eqnarray}
   W_G(2)&=& \frac{\in_{123}}{A(S^2)2!}\int_{BZ}d^{1}\wedge\textrm{d}d^{2}\wedge \textrm{d} d^{3} \nonumber\\
   &=&\frac{1}{4\pi}\int_{BZ}\in_{123}d^{1} \frac{\partial d^{2}}{\partial k_x} \frac{\partial d^{3}}{\partial k_y} \textrm{d}k_x \textrm{d}k_y\nonumber\\
   &=&\frac{1}{4\pi}\int_{BZ}\textrm{d}k_x \textrm{d}k_y \vec{d}\cdot\frac{\partial\vec{d}}{\partial k_x}\times\frac{\partial\vec{d}}{\partial k_y}
\end{eqnarray}
where we  have defined $\vec{d}=(d_1, d_2, d_3)$, which is consistent with ref (\cite{qi2008topological}). For general (2n+1)-dimensional Dirac lattice models, the topological invariants that characterize the topological properties all can be obtained using $W_G(2n)$.
\par
Finally, we can add a mass term to investigate the variation of the topological invariants $W_G(2n)$, this is equivalent to study the topology properties across  the interface of domain wall in physics\cite{qi2008topological}. Therefore, the vector $\vec{\phi}$ will be dependent on (2n+1)-dimensional coordinates $v=(\textbf{k},m)$, where $\textbf{k}$ is momentum space $m$ is mass coordinates.
We will obtain another important general formula with Jacobian  determinant and general $\delta$ function\cite{duan2000topological},
\begin{eqnarray}
 W_{G}(2n) &=& \frac{\in_{a_1a_2\cdots a_{2n+1}}}{A(S^{2n})(2n)!}\int_{BZ}d^{a_1}\textrm{d}d^{a_2}\wedge\cdots\wedge
   \textrm{d}d^{a_{2n+1}}\nonumber  \\
   &=&\frac{\in_{a_1a_2\cdots a_{2n+1}}}{A(S^{2n})(2n)!}\int_{BZ\times m}\textrm{d}d^{a_1}\wedge\textrm{d}d^{a_2}\wedge\cdots\wedge
   \textrm{d}d^{a_{2n+1}}\nonumber  \\
   &=&\int_{BZ\times m}\delta(\vec{\phi}(v))D\left(\frac{\phi}{v}\right)\textrm{d}^{2n+1}v \nonumber \\
   &=& \sum_{i=1}^{l}\int_{BZ\times m} \beta_i\eta_i\delta(v-z_i)\textrm{d}^{2n+1}v \nonumber\\
   &=& \sum_{i=1}^{l}\beta_i\eta_i \label{duan}
\end{eqnarray}
where we have defined  a smooth vectors $\phi^a(v)$ satisfying,
\begin{equation}\label{phi}
    d^a(v)=\frac{\phi^a(v)}{||\phi||},~~~||\phi||=\phi_a\phi^a,
\end{equation}
and $D\left(\phi/v\right)$ is the usual Jacobian determinant of $\vec{\phi}$ with respect to $v$
\begin{equation}\label{Jacobian}
    \in^{a_1a_2\cdots a_{2n+1}}D\left(\frac{\phi}{v}\right)\textrm{d}^{2n+1}v=\textrm{d}d^{a_1}\wedge\textrm{d}d^{a_2}\wedge\cdots\wedge
   \textrm{d}d^{a_{2n+1}}.
\end{equation}
And we use the ordinary theory of $\delta$-function\cite{schwarz2002topology} that
\begin{equation}\label{dirac function}
    \delta(\vec{\phi})=\sum_{i=1}^{l}\frac{\beta_i\eta_i\delta(v-z_i)}
{D(\phi/v)|_{z_i}}, ~~~~\eta_i=\textrm{sgn}D(\phi/v)|_{z_i}=\pm1,
\end{equation}
we obtain
\begin{equation}\label{delta and dirac function}
    \delta(\vec{\phi}(v))D\left(\frac{\phi}{v}\right)=\sum_{i=1}^{l}\beta_i\eta_i\delta(v-z_i)
\end{equation}
where $\beta_i$ is the Hopf indices and $\eta_i$ are the Brouwer degrees. Finally, using the method of $\phi$-mapping topological current theory proposed by Duan\cite{duan1993topological,duan2000topological,duan2003many}, one will get the final result (\ref{duan}), which is essential the result of Poincar$\acute{\textrm{e}}$-Hopf theorem in mathematic.
\par
The equation of $(\ref{duan})$ indicate the topological invariants $W_G(2n)$ can be only changed at the critical values, $\phi=0$.  In physics, at the critical points the Dirac system becomes gapless, $d_ad^a=0$. Now let me study a particular two band model introduced in ref \cite{PhysRevB.74.085308}, which is given by
\begin{equation}\label{h(k)}
    h(\textbf{k})=\sin k_x \sigma_x +\sin k_y \sigma_y +(m+\cos k_x +\cos k_y)\sigma_z.
\end{equation}
This Hamiltonian corresponds to form (\ref{two-band model}) with $\epsilon(\textbf{k})\equiv 0$ and $\vec{d}=(\sin k_x,\sin k_y, m+\cos k_x +\cos k_y)$.  Next, the critical points  $v=z_i$ can be listed below:

\begin{equation}\label{critical points of mass}
 v=\left\{\begin{array}{l}
             z_1=(0, 0, -2)\\
             z_2=(0, \pi, 0)\\
             z_3=(\pi, 0, 0)\\
             z_4=(\pi,\pi, 2).
          \end{array}
\right.
\end{equation}
The Brouwer degree so can be easily calculated  as following
\begin{equation}\label{Brouwer degree}
  \textrm{ sgn}D(\phi/v)|_{z_1}=\textrm{sgn}D(d/v)|_{z_1}=-1,
\end{equation}
similarly we obtain $ \textrm{sgn}D(\phi/v)|_{z_4}=1$, while for the case $m<-2$, there is non critical points so $W_G(2)=0$.
It is easy to determine $W_G(2)$ in the limit $m\rightarrow+\infty$ since the unit vector $\vec{\phi}\rightarrow(0,0,1)$ in that limit, so there is no critical values, $W_G(2)=0$. Thus we only need to study the change at each quantum critical points, namely, at critical values of $m$ where the system becomes gapless. In summary, we obtain the $W_G(2n)$ or first Chern-number is assuming Hopf indices is equal 1,
\begin{equation}
 \textrm{Ch}_1=W_G(2)=\left\{\begin{array}{cl}
             \eta_1=-1& \textrm{for} -2< m< 0\\
             \eta_4= 1& \textrm{for}~   0< m< 2 \\
            0&  \textrm{otherwise}.
          \end{array}
\right.
\end{equation}
The two critical points $z_1$ and $z_2$ is more complicated because the Jacobian determinant is zero at these values, these were called bifurcation points\cite{duan1993topological}. The branch process will arise at the bifurcation points, here the topological invariants change from $-1$ for $m<0$ to $1$ for $m>0$ at the bifurcation point $m=0$.
\par
In conclusion,  we not only give the general prove of various topological invariants which have extensively used in CMP, but also establish some important relationship between these topological invariants. In order to give an intuitive  physical interpretation, we  explicitly give  the topological invariants of the two-band QAHE model. We give the corresponding  mathematic objects for various topological invariants used in physics. The explicit map can be constructed as following,

\[\begin{CD}\label{relation}
N_{2n+1}[G]  @>P_G>>N_{2n+1}[G_0] @>P_+>>N_{2n+1}[G_0^{'}]\\
@|           @|                 @|\\
\textrm{Ch}_{n+1}[\mathcal{F'}] @>>>\textrm{Ch}_n[\mathcal{F}]      @>>>W_{G}(2n).
\end{CD}\].

\subsubsection{Topological order parameter using zero-frequency Green function}
 The topological invariant $N_{2n+1}[G]$ has the disadvantage that it involves a frequency integral. In most numerical algorithms, it's very difficult to obtain the dynamic Green's function at all frequencies. Recently, Wang\cite{PhysRevX.2.031008,wang2012topological} propose  simplified topological invariants for interacting topological insulators. They are expressed in terms of Green's function at zero frequency instead of the entire frequency domain. The key idea is to introduce a smooth deformation of $G(i\omega,k)$ parameterized by $\lambda\in[0,1]$ as follows
\begin{equation}\label{interpolation of G}
    G(i\omega,k,\lambda)=(1-\lambda)G(i\omega,k)+\lambda[i\omega+G^{-1}(0,k)]^{-1}
\end{equation}
Due to the topological property of invariant of $N(G)$, we will get that $dN(G,\lambda)/d\lambda=0$ if the deformation (\ref{interpolation of G}) does not contain singularity. Fortunately, Wang prove that deformation is smooth, it means the interacting Greens function is homotopy equivalent to zero-frequency Green function, so we have
\begin{equation}\label{NG(0)}
    N[G(i\omega,k,\lambda=0)]=N[G(i\omega,k,\lambda=1)].
\end{equation}
Therefore, we just need to calculate $N[G(i\omega,k,\lambda=1)]$, which is equivalent to calculation for an effective noninteracting system with $h_{eff}(k)=-G^{-1}(0,k)$.
\par
Following the same logic of last subsection, we will obtain the following result
\begin{equation}\label{N(G(0))=Chern}
    N[G]=\textrm{Ch}_n[\mathscr{F}],
\end{equation}
where $\mathscr{F}$ is the Berry curvature of "R-space"\cite{PhysRevX.2.031008}.
It's straightforward to generalize the non-interacting topological invariants to interacting system using topological effective Hamiltonian $h_{eff}(k)=-G^{-1}(0,k)$.

\subsubsection{Phase space Chern-Simons theories}
 The phase space Chern-Simons theories have been studied in the previous works \cite{qi2008topological}. From the mathematic viewpoint, the base manifold or base space of non-interacting Chern-Simons field theory is phase-space.  In addition,  the rules of dimensional reduction used in ref \cite{qi2008topological} will be clear  with phase space Chern-Simons theory.
\par
Now we can briefly introduce the phase-space Chern-Simons form. Firstly, we define the phase-space coordinate as $\textbf{q}=(t_0,\textbf{x},\textbf{k})$ as , where $t_0$ stands for time, $\textbf{x}$ is the real space  and $\textbf{k}$  is momentum space. And the  external Abelian gauge potential one-form in phase space can be written as  $A= A^i\textrm{d}q_i$, and non-Abelian Berry phase potential one-form is $a=a^i\textrm{d}q_i$, where $i$  is the indices of phase space coordinates. After introducing these concepts, we can write the total one-form as
\begin{equation}\label{phase space total one-form}
    \mathcal{A^{\lambda}}=\lambda A+a,
\end{equation}
where $\lambda$ is a real coefficient. Now we can substitute  the gauge one-form $A^{\lambda}$ in the definition of Chern-Simons invariants $(\ref{csn})$, one can obtain

\begin{eqnarray}\label{phase space chern-simons}
   &&\textrm{CS}_{2n+1}(\lambda)\nonumber\\ &=& \int_{V_q}Q_{2n+1}(\mathcal{A^{\lambda}},\mathcal{F^{\lambda}}) \nonumber\\
   &=&\frac{1}{n!}\left(\frac{i}{2\pi}\right)^{n+1}\int_{V_q}\int_0^1\textrm{d}
        t\ \textrm{tr}[\mathcal{A^{\lambda}}(\mathcal{F^{\lambda}}_t)^{n
}],
\end{eqnarray}
where  $V_q$ is the volume of phase space, $\mathcal{F^{\lambda}}_t$ is defined as equation (\ref{Ft}),
\begin{equation}\label{ps-Ft}
    \mathcal{F^{\lambda}}_t=t\textrm{d}\mathcal{A^{\lambda}}+t^2 (\mathcal{A^{\lambda}})^2.
\end{equation}
If we substitute (\ref{phase space total one-form}) into (\ref{ps-Ft}), then $\mathcal{F^{\lambda}}_t$ can be reexpressed as
\begin{equation}\label{ps-Ft1}
   \mathcal{F^{\lambda}}_t=\lambda t \textrm{d}A +t \textrm{d}a+t^2a^2=\lambda t \textrm{d}A +\mathcal{F}_t.
\end{equation}
The factor $i^{n+1}$ in equation (\ref{phase space chern-simons}) can be easily cancel if we redefine the $\mathcal{A^{\lambda}}$, so we just omit it in the following for convenience, and we also omit the volume of phase-space $V_q$. So the (\ref{phase space chern-simons}) can be expressed as
\begin{eqnarray}
&&\textrm{CS}_{2n+1}(\lambda)\nonumber\\
 &=&\frac{1}{n!}\left(\frac{1}{2\pi}\right)^{n+1}\int\int_0^1\textrm{d}t~\textrm{tr}[\mathcal{A^{\lambda}}(\mathcal{F^{\lambda}}_t)^{n
}] \nonumber \\
   &=&\frac{1}{n!}\left(\frac{1}{2\pi}\right)^{n+1}\int
\int_0^1\textrm{d}t~\textrm{tr}
[(\lambda A+a)(\lambda t \textrm{d}A +\mathcal{F}_t)^{n
}]  \nonumber\\
    &=& \sum_{s=0}^{[\frac{n}{2}]}C_n^s \lambda^{s+1}\frac{1}{n!}\left(\frac{1}{2\pi}\right)^{n+1}\int
\int_0^1\textrm{d}t~
\textrm{tr}[A(t\textrm{d}A)^s(\mathcal{F}_t)^{n-s}]\nonumber\\ &&+\frac{1}{n!}\left(\frac{1}{2\pi}\right)^{n+1}\int
\int_0^1\textrm{d}t~\textrm{tr}[a(\mathcal{F}_t)^{n}]\nonumber\\
   &=& \sum_{r=0}^{[\frac{n}{2}]+1}\lambda^r \textrm{CS}^r_{2n+1},
\end{eqnarray}
where $\textrm{CS}^r_{2n+1}$ stands for the mixed Chern-Simons invariant in $(2n+1)$-phase-space dimensions with $r$ powers of the external $A$ field, $C^s_{n}$ is binomial coefficient and $[n/2]$ denotes the maximal integer that doesn't exceed $n/2$ . Specifically, $\textrm{CS}^0_{2n+1}$ is a pure non-Abelian Chern-Simons term of Berry gauge one-form $a$.
\par
We can now write the "generating functional" Chern-Simons field theory $S_{2n+1}(\lambda)$ as
\begin{equation}\label{generating function of cs}
    S_{2n+1}(\lambda)=2\pi\cdot\sum_{r=0}^{[\frac{n}{2}]+1}\lambda^r \textrm{CS}^r_{2n+1}=\sum_{r=0}^{[\frac{n}{2}]+1}\lambda^r S_{2n+1}^{r},
\end{equation}
where $S^r_{2n+1}$ stands for the mixed Chern-Simons action in $(2n+1)$-phase space. Therefore, it is easy obtain $S^r_{2n+1}$ from the generating function
\begin{equation}\label{phase-space genrerating}
    S^r_{2n+1}=\frac{1}{r!}\frac{\partial^r}{\partial\lambda^r}S_{2n+1}(\lambda)|_{\lambda=0}.
\end{equation}
As the reference \cite{qi2008topological} said, (\ref{generating function of cs}) is the unified theory of all topological insulators, the effective action of family tree of topological insulators can be derived from this phase-space Chern-Simons generating function.
\par
In summary, we briefly study the phase-space Chern-Simons theory generally. And all topological insulators can be well described in a unified form in this space, which contains the parent theories and their descendants.
\subsubsection{Parity anomaly, edge states and index theorem}
The presence of terms of topological origin in the Chern-Simons theory talked above are closely related to the presence of anomaly called parity anomaly in $D=2n+1$ space-time dimensions\cite{fujikawa2004path,nakahara2003geometry}. For convenience, we relabel the Chern-Simons forms as $\Omega_{2n+1}^{(0)}$ and Chern character form as $\Omega_{2n+2}$ in this subsection, respectively. In anomaly theory, there is a significant relation called $descent~ equations$\cite{nakahara2003geometry,ryu2012electromagnetic}, the fist two are
\begin{equation}\label{descent equation}
    \Omega_{2n+2}=\textrm{d}\Omega_{2n+1}^{(0)},
~~\delta_{v}\Omega_{2n+1}^{(0)}=\textrm{d}\Omega_{2n+1}^{(1)},
\end{equation}
where $\delta_v$ is the gauge transformation in question and $\Omega_{2n+1}^{(1)}$ is linear in $v$ by definition, the first equation of (\ref{descent equation}) is nothing but (\ref{chern-simons}). Therefore, when there is a boundary in the system, the integral of the Chern-Simons term is not invariant on its own, rather, upon making use of the (\ref{descent equation}), one obtains
\begin{equation}\label{chern-simons anomaly}
    \delta_{v}\int_{M_{2n+1}}\Omega_{2n+1}^{(0)}
=\int_{M_{2n+1}}\textrm{d}\Omega_{2n+1}^{(1)}=\int_{\partial M_{2n+1}}\Omega_{2n+1}^{(1)}.
\end{equation}
This is something we are familiar with from the physics of the quantum Hall effect: the presence of the boundary term appearing on the right side of Eq.(\ref{chern-simons anomaly}) signals the presence of edge mode. So generally, the edge state can be contributed to the parity anomaly in the $D=2n+1$ space-time dimensions.
\par
The feature of the topological insulator systems is that all of them are gapped in the bulk but also supposed to have gapless excitation on the boundary. Now how can we establish the bulk-boundary correspondence by the topological invariants from their respective Green's function? The question has been answered partly by Volovik\cite{volovik2003universe} and has been extended in general dimension by Gurarie.el\cite{PhysRevB.84.125132,PhysRevB.83.085426}. Here, we only outline the result, the details can be found in Ref.\cite{PhysRevB.84.125132}. Firstly, let define the topological invariants describing the edge of a topological insulator in $D=d+1$-dimensions space-time space. If the topological insulator has an edge, translation invariance in the direction perpendicular to the edge is lost, and the good quantum number is the $(d-1)$-dimensional momentum $\textbf{k}^{d-1}$ parallel to the edge. We expect the edge to have gapless excitations at some mometum (so the edge is not an insulator), the Hamiltonian at fixed $k_{d-1}=\Lambda$ (the last component of momentum space) is gapped if $\Lambda$ is sufficiently large (as a function of remaining $d-2$ momenta), and so describes a $(d-2)$-dimensional insulator. Secondly, the bulk topological insulator with translation symmetry can be distinct by $N_D[G]$ discussed in this section, where $G$ is the single-body Green's function. Finally, we will obtain the general topological invariants relation of bulk and boundary according to Ref.\cite{PhysRevB.84.125132} as follow:
\begin{equation}\label{bulk-boundary}
    N_D=N_{D-2}(\Lambda)-N_{D-2}(-\Lambda),
\end{equation}
where $N_{D-2}(\Lambda)$ is the topological invariant which describes the $(d-2)$-dimensional topological insulator at fixed $\Lambda$.
\par
For example, the chirality of the gapless excitation states along single edge in the two-dimensional system (IQHE) has a topological interpretation
\begin{equation}\label{2d bulk-boundary}
    n_+-n_{-}=N_1(\infty)-N_1(-\infty)=N_3[G],
\end{equation}
where $n_+$ and $n_{-}$ are right movers and left movers gapless states along the edge. In general, Eq.(\ref{bulk-boundary}) is the result of AS index theorem, it give the topology  relation of chirality of the edge mode and the spectral flow which can be described with a winding number $N_D$ in a semi-infinity system.
Eq.(\ref{bulk-boundary}) is also called index theorem by Volovik\cite{volovik2003universe}, it is really close to AS index theorem, and which bridges the bulk-boundary correspondence of the topological insulator.

\subsection{Periodic Table of TSMs}
The classification of the ground state bundles is a significant issues for physics, which has been obtained through K-theory\cite{kitaev2009periodic,stone2010symmetries,wen2012symmetry} or a nonlinear $\sigma$ model of disordered fermions\cite{PhysRevB.78.195125}. Recently, Ryu.\cite{ryu2012electromagnetic} classifies topological states via gauge anomalies or axial anomalies and propose this method can be described the interacting fermions system.  In summary, they establish the period-2 for complex case of fermions system and period-8 for real case pattern of correlations from different sides, but all of them underly the mechanism of periodic table  of TSMs is an interesting mathematic  result called Bott periodicity theorem. The manifold of ground-state of various symmetry is nothing but the ten symmetry spaces discovered by $\acute{\textrm{E}}$lie Cartan in 1926. In physics, Altland and Zirnbauer \cite{PhysRevB.55.1142} recognized that there is a one-to-one correspondence between single-particle Hamiltonian and the set of symmetric spaces.
\par
In this subsection, we will classify all non-trivial TSMs in various space dimensions by topology K-theory. And we establish the periodic table \ref{table2} of TSMs from a  more physical way closely by ref\cite{wen2012symmetry} and ref\cite{kitaev2009periodic}, but we interpret this classification more "topology". In the following, we firstly introduce the classification for $d=0$ case for simplicity, and establish the classifying spaces more physical. The topology relation of classifying spaces are investigated by  minimal geodesic and loop space discussion. We will show the ten symmetric classifying spaces can  naturally come up in CMP under some physical symmetries such as time-reversal, $U(1)$ symmetry and conjugation symmetry. Next we consider the more realistic system of $d\neq 0$ dimensional free fermion systems. We obtain the classification of different models such as continuum model with base manifold $S^d$ or band insulator model with base manifold $T^d$ using topology K-theory. After that, we give the physical interpretation of the classification  by studying the properties of mass matrices $m$ of a $d$-dimensional Dirac operator.  Finally, we establish the periodic table with KR-theory in all space dimensions.
\subsubsection{Classification for $d=0$ (complex case) }
It is simple for the classification of $d=0$ case, the base manifold is just one point. So the distinct non-trivial topological states only require to investigate the topological properties of fibers i.e., the space of ground state of gapped Hamiltonians. Intuitively, the difference for a given symmetry class is the occupied orbitals, or called the dimension of the 'valence' fibers. So we should consider a system with $n$ orbitals in the $n\rightarrow\infty$ limit, it is in the stable range. In this case, the zero-dimensional gapped phases are labeled by an integer $m\in\mathbb{Z}$, where $m$ still corresponds to the number of occupied orbitals, or the dimension of fibers.

\par
Now let us firstly analyse the complex case of non-interacting fermions systems. The single-body Hamiltonian of the n-orbital system is given by the $n\times n$ Hermitian matrix $H$. If the orbitals below a certain energy are filled, we can deform the energies of those orbitals to -1 and deform the energies of other orbitals to +1 without closing the energy gap. The Hamiltonian of deformation after and before is homotopically equivalence, one can construct a homotopy map between these the "flat" model and the original model. Therefore, without generality, we can assume $H$ to satisfy
\begin{equation}\label{h2=1}
    H^2=1.
\end{equation}
 It's easy to know, such a Hermitian matrix has the form
\begin{equation}\label{Matrix of H}
    H=U_{n\times n}\left(
                     \begin{array}{cc}
                       \mathbb{I}_{l\times l}& 0 \\
                       0& -\mathbb{I}_{m\times m}  \\
                     \end{array}
                   \right)U_{n\times n}^{\dagger},
\end{equation}
where $n=l+m$ and $U_{n\times n}$ is an $n\times n$ unitary matrix. But $U_{n\times n}$ is not a one-to-one  labeling of the Hermitian matrix satisfying $H^2=1$. To obtain a one-to-one labeling, we have to modulo the redundancy group that makes the
$(\begin{array}{cc}
                       \mathbb{I}_{l\times l}& 0 \\
                       0& -\mathbb{I}_{m\times m}  \\
                     \end{array})$ invariant from the unitary group $U(l+m)$. Hence, the classifying space for complex case $C_0$ of the Hermitian matrix satisfying $H^2=1$ is given by $\cup_{m} U(l+m)/U(l)\times U(m)$, which, in the limit of $n\rightarrow\infty$ limit, has the form
\begin{equation}\label{c0}
    C_0\equiv\frac{U(l+m)}{U(l)\times U(m)}\times \mathbb{Z}.
\end{equation}
As discussed in the subsection \ref{Homotopy}, the classifying space $C_0$ is nothing but the loop space of $U(l+m)$, see the equation (\ref{homotopy of path space}). In fact, we can approximate the loop space of $U(l+m)$ by minimal geodesics from $\mathbb{I}$ to $-\mathbb{I}$: $L=e^{i\lambda H}$, where $H$ is the Hermitian matrix satisfying $H^2=1$ \cite{milnor1963morse}(note the length of the geodesic is $\sqrt{\textrm{tr}(HH^{*})}$. So that $L(0)=\mathbb{I}$ and $L(\pi)=-\mathbb{I}$.  So the topology of $C_0$ is just the loop space of $U(l+m)$. Clearly, $\pi_0(C_0)=\mathbb{Z}$, which recovers the result obtained above using a simple argument: the zero-dimensional gapped phases of free conserved fermions are labeled by integer $\mathbb{Z}$. According to the equation of (\ref{loop space}), $\pi_0(C_0)=\pi_1(C_1)$, where $C_1$ is the $U(n)$ group with $n\rightarrow\infty$.
\par
How to interpret the classifying space of $C_1$ in physics? This construction can be realized by imposing symmetry to  Hermitian matrix $H$, such as chiral symmetry, for convenience later, we denote as $\gamma$ in this subsection, satisfying $\gamma H+H\gamma=0$ and $\gamma^2=1$.  To compute the classifying space, we parametrize $C_0$ by symmetric matrix $K$, that satisfying $K^2=1$. We can prove the Hamiltonian satisfying $H^2=1$ and $\gamma H+H\gamma=0$ defines a geodesic $L(\lambda)=Ke^{i\lambda \gamma}$ from $L(0)=K$ to $L(\pi)=-K$.
In fact, the loops remain in the space of $C_0$  because of $(Ke^{i\lambda \gamma})^2=Ke^{i\lambda \gamma}Ke^{i\lambda \gamma}=e^{i\lambda K\gamma K}e^{i\lambda \gamma}=e^{-i\lambda \gamma}e^{i\lambda \gamma}=1$, and $\gamma\in C_0$. So, the classifying space $C_1$ is nothing but the loop space of $C_0$. Actually, we can construct the form of Hamiltonian explicitly as ref\cite{wen2012symmetry}
\begin{equation}\label{c1}
    H=e^{i\sigma^x \otimes A_{n\times n}}e^{i\sigma^0 \otimes B_{n\times n}}(\sigma^{z}\otimes \mathbb{I}_{n\times n})e^{-i\sigma^0 \otimes B_{n\times n}}e^{-i\sigma^x \otimes A_{n\times n}},
\end{equation}
where unitary group $e^{i\sigma^x \otimes A_{n\times n}}e^{i\sigma^0 \otimes B_{n\times n}}\in U(n)\times U(n) $ ($A_{n\times n} $ and $B_{n\times n} $ are Hermitian matrices) makes the $\gamma$ invariant. From the form of (\ref{c1}), we know its positive and negative eigenvalues are paired, which is the reflection of chiral symmetry. We see that space $C_1$ is $U(n)\times U(n)/U(n)=U(n)$. From the unitary Bott periodic theorem in subsection \ref{Homotopy}, we know there are only two kinds of classifying spaces for complex cases, $C_0$ and $C_1$.
\par
Therefore, we can classify all complex ground-state gapped fermions systems for $d=0$ according to
\begin{equation}\label{complex classification}
    \pi_0(C_q)=\left\{
                 \begin{array}{cc}
                   \mathbb{Z}, & q=0 ~~ \textrm{mod}~ 2, \\
                    0,& q=0 ~~ \textrm{mod}~ 2. \\
                 \end{array}
               \right.
\end{equation}
The topology K-theory will be like following
$\textrm{KU}(pt)=\pi_0(C_0)$ and $\textrm{KU}^{-1}(pt)=\textrm{KU}(S)=\pi_1(C_0)=\pi_0(\Omega C_0)=\pi_0(C_1)$, where $S$ is the suspention\cite{hatcher2003vector} operator, $\Omega$ is the loop space operator, and $pt$ stands for one point, i.e., the zero dimensional case. So the classification of ground-state fiber bundle on one-point base manifold can be reexpressed as
\begin{equation}\label{complex classification k-theory}
    \textrm{KU}^{-q}(pt)=\pi_0(C_q)=\left\{
                 \begin{array}{cc}
                   \mathbb{Z}, & q=0 ~~ \textrm{mod}~ 2, \\
                    0,& q=0 ~~ \textrm{mod}~ 2, \\
                 \end{array}
               \right.
\end{equation}
where we have used the Bott periodic theorem of $\textrm{KU}^{-q}=\textrm{KU}^{-q-2}$ (\ref{bott of ku}).
\subsubsection{Classification for $d=0$ (real case) }
\par
When the fermion number is not conserved and/or when there is a time-reversal symmetry, the gapped phases of noninteracting fermions are classified differently. However, using the idea and approaches similar to the above discussion, we can also obtain a classification. Because we will not be assuming charge conservation, it is convenient to express the complex fermion operator $\hat{c}_j$ in terms of real fermionic operators (Majorana fermions $\hat{\eta}_i^{\dag}=\hat{\eta}_i$), $\hat{c}_j=[\hat{\eta}_{2j}+i\hat{\eta}_{2j+1}]/2$. The Hamiltonian $\hat{H}$ can be written as
\begin{equation}\label{real case H}
    \hat{H}=\frac{i}{4}\sum_{ij}A_{ij}\hat{\eta}_{i}\hat{\eta}_{j},
\end{equation}
where $A$ is a real antisymmetric matrix, $A_{ij}=-A_{ji}$.
Any real antisymmetric matrix can be written in the form
\begin{equation}\label{Matrix of A}
    A=O^{T}\left(
                   \begin{array}{ccccc}
                                   0& -\lambda_1 &  &  &  \\
                                  \lambda_1& 0 &  &  &  \\
                                   &  & 0& -\lambda_2 &  \\
                                   &  &  \lambda_2& 0 &  \\
                                   &  &  &  & \ddots
                                \end{array}
                   \right)O,
\end{equation}
where $O$ is an orthogonal matrix and the $\lambda_i$'s are positive. The eigenvalues of $A$ comes in pairs $\pm i\lambda_i$; thus, if we considering gapped systems, $\lambda_i$ is always nonzero. We can shift all $\lambda_i$ to 1 without closing the gap and change the phase of the state. Thus, we can set $A^2=-1$. This is usually called "spectrum flattening". Then, we can write
$A=O^TJ_1O$, where $J_1=\varepsilon\otimes \mathbb{I}$, $\varepsilon\equiv-i\sigma^y$.
\par
The possible choices of $A_{ij}$ correspond to the possible choice of $O\in O(2N)$, modulo $O$, which commute with the matrix $J_1$. This is nothing but the complex structural space $\Omega_1$ (seen \ref{omega1}), which is denoted as $R_0^0$  or $R_2$ in ref.\cite{wen2012symmetry}. The classifying space $\Omega_1$ can be viewed as the loop space of the $O(2N)$ from topological point. Actually, if we have a antisymmetric matrix $A$ satisfying $A^2=-1$, then we can construct a curve $L(\lambda)= e^{\lambda A}$ in $O(2N)$. And we have $e^{\lambda A^{T}}e^{\lambda A}=e^{\lambda (A^T+A)}=1$ and $L(0)=\mathbb{I}$, $L(\pi)=-\mathbb{I}$. Such a curve is, in fact, a minimal geodesic from $\mathbb{I}$ to $-\mathbb{I}$. Each such geodesic can be represented by its midpoint $L(\pi/2) = A$, so the space of such geodesics is equivalent to the space of matrices $A$ satisfying $A^2=-1$ and $A^T+A=0$. Therefore, we obtain $\Omega_1=\Omega^d(\Omega_0)=\Omega^d(O)=\Omega(O)$\cite{milnor1963morse}, where $\Omega_0=R_1$\cite{kitaev2009periodic}, $\Omega^d$ is the minimal geodesic space operator  and $\Omega$ is the loop space operator(seen subsection \ref{Homotopy}). Finally, we obtain $R_2=\Omega(R_1)$, which means the topology of classifying space of $R_2$ is nothing but the loop space of $R_1$.
\par
Now suppose that we restrict ourselves to time-reversal-invariant systems and, furthermore, to those time-reversal-invariant systems that satisfy $T^2=-1$, where $T$ is the antiunitary operator generating time reversal. Then following Ref.\cite{kitaev2009periodic,wen2012symmetry}, we write $Tc_iT^{-1}=(J_2)_{ij}c_j$. The matrix $J_2$ is antisymmetric and satisfies $J_2^2=-1$. $T$ invariance of the Hamiltonian requires $J_2A=-AJ_2$. So the choices of $A^2=-1$ is equivalent to the complex structural space that anticommuting with fixed $\rho_1=J_2$ denoted as $\Omega_2$ (seen \ref{omega2}) and $R_0^1$ or $R_3$ in Ref.\cite{wen2012symmetry}.
In topology, the space of matrices $A$ satisfying $A^2=-1$ and $A\rho_1+\rho_1A=0$ is equivaalent to the space of geodesics $L(\lambda)=\rho_1 e^{\lambda X}$ in $\Omega_1$, where we define $A=\rho_1X$. In fact, $L(\lambda)$ is a curve from point $L(0)=\rho_1$ to point $L(\pi)=-\rho_1$, and $(\rho_1e^{\lambda X})^2 = \rho_1\rho_1\rho_1^{-1}e^{\lambda X}\rho_1e^{\lambda X}=-e^{\lambda(\rho_1^{-1}X\rho_1+X)}=-\mathbb{I}$, so the geodesic $L(\lambda)$ is in space $\Omega_1$, and the midpoint is $L(\pi/2)=\rho_1X=A$. Therefore, we have $R_3=\Omega_2=\Omega^d(\Omega_1)=\Omega R_2$. That is to say, the classifying space $R_3$ has the topology $\Omega(R_2)=\Omega^2(R_1)$, which is the double loop space of $R_1$.
\par
If there are time-reversal and $U(1)$ symmetries $Q=J_1$ \cite{kitaev2009periodic,wen2012symmetry}, then we will require to find the matrix $A$ satisfying
\begin{eqnarray}
   A^2&=&-1, A\rho_i+\rho_iA=0, \rho_1^2=\rho_2^2=-1,\nonumber \\
  \rho_1&=&T, \rho_2=TQ, \rho_1\rho_2+\rho_2\rho_1=0.
\end{eqnarray}
The space of matrix $A$ above is nothing but the space $\Omega_3$ (seen \ref{omega3}), which is the geodesic space $\Omega^d(\Omega_2)$ or the loop space  $\Omega(\Omega_2)$. Assuming the complex structural $A = \rho_2 X$, then $X$ is a complex structural which anti-commutes with $\rho_2$ but commutes with $\rho_1$. Hence, the mapping $L(\lambda)=\rho_2e^{\lambda X}$ defines a geodesic from $L(0)=\rho_2$ to point $\L(\pi)=-\rho_2$, $\rho_2e^{\lambda X}$ anticommutes with $\rho_1$ and is a complex structural. So it is the minimal geodesic of $\Omega_2$, with the midpoint $L(\pi/2)=\rho_2 X=A\in\Omega_3$, so we have the homomorphism $R_4=\Omega^d(\Omega_2)\simeq\Omega(\Omega_2)\simeq\Omega_3=\Omega(R_3)$\cite{wen2012symmetry}.
From the same logic above, we know the topology of classifying space $R_4$ is the loop space of $R_3$, and $R_4=\Omega(R_3)=\Omega^2(R_2)=\Omega^3(R_1)$.
\par
If there are $U(1)$, time-reversal, and charge-conjugation symmetries, we need to calculate the topology of the matrix $A$ satisfying\cite{wen2012symmetry}
\begin{eqnarray}
  A^2&=&-1, A\rho_i+\rho_iA=0, \rho_i^2|_{i=1,2,3}=-1,\nonumber   \\
 \rho_1&=&T, \rho_2=TQ, \rho_3=TC, \rho_i\rho_j+\rho_j\rho_i|_{i\neq j}=0.
\end{eqnarray}
where $C$ is the charge-conjugation symmetries\cite{wen2012symmetry}. As the same reason as before, we will obtain $R_5=\Omega(R_4)$ through the loop space discussion by minimal geodesic approximation.
\par
The pattern is pretty obvious, we have $p$ predefined real matrices $\rho_1,\cdots, \rho_p$ satisfying $\rho_i\rho_j+\rho_j\rho_i=-2\delta_{ij}\mathbb{I}$, and look for all possible choices of another complex structural $\rho_{p+1}=A$ (where we have used the same label $p$ as Ref.\cite{kitaev2009periodic} for convenience discussed later). In general, this structure is called Clifford algebra $Cl(p+1,0)$ in mathematics. To classify free-fermion Hamiltonians, we consider representations of $Cl(p+1,0)$ with fixed action of $\rho_1,\cdots, \rho_p$, which is called the Clifford extension problem with $q$ negative generators\cite{kitaev2009periodic}. In addition, from the topology view point discussion in section \ref{Homotopy}, this process is equivalent to find the space  $\Omega_{p+1}$, which is the space of complex structural $\rho_{p+1}$ set that anti-commuting with fixed $\{\rho_i\}_1^p$. The space of those $A$ matrices is denoted as $R_0^p$ according to Ref.\cite{wen2012symmetry}, and satisfying $R_0^p=R_{p+2}=\Omega_{p+1}$\cite{wen2012symmetry}. And we have the periodic relation $R_{p+8}=R_p$ due to the Bott periodicity theorem $\Omega_{p+8}=\Omega_p$, the classifying space have the topology relation $R_{p+1}=\Omega(R_p)$ based on the discussion above. In summary, we have obtain the general topology of classifying space $R_{p+2}$ from the explicit Clifford algebra $Cl(p+1,0)$ representations and the minimal geodesic discussion.
\par
There is another problem we will meet is real  symmetric matrices $\rho_1,\cdots, \rho_q$ satisfying $\rho_i\rho_j+\rho_j\rho_i=2\delta_{ij}\mathbb{I}$. Physically, this is corresponding to the cases such as $\rho_1^2=T^2=1$ \cite{wen2012symmetry}. So, in general, we can consider a real antisymmetric matrix $A$ that satisfies (for fixed real matrices $\rho_i$, $i=1,\cdots,p+q$)
\begin{eqnarray}
  A &=&\rho_{p+q+1}, ~~\rho_j\rho_i+\rho_i\rho_j=|_{i\neq j}0, \nonumber  \\
   \rho_i^2&=&|_{i=1,\ldots,q}1, ~~\rho_i^2=|_{i=q+1,\ldots,q+p+1} -1.
\end{eqnarray}
The space of those $A$ matrices is denoted as $R_q^p$ ( An alternative notation is used in Ref.\cite{wen2012symmetry}, where the positive generators are listed first and the parameters $p$ and $q$ are swapped. ). In nature, the matrix $A\in R_q^p$ satisfies the Clifford algebra $Cl(p+1,q)$ in mathematics. How can we know the topology properties of the matrices $A$? As in the subsection \ref{Homotopy}, we can define the space $\overline{\Omega}_k(16n)$ as the real matrix $\tilde{J}$ set satisfying $\tilde{J}^2=1$ and anti-commuting with fixed $\{\tilde{J}\}_1^{k-1}$ which satisfy $\tilde{J}_i\tilde{J}_j+\tilde{J}_j\tilde{J}_i=2\delta_{ij}\mathbb{I}$. Using the same logic in subsection \ref{Homotopy}, we will obtain
\begin{equation}\label{Omega space for j^2=1}
    \overline{\Omega}_{i+2}=\Omega_i, ~~R_0^p=R_{p+2}.
\end{equation}
The space $\overline{\Omega}_i$ has the same direction as $\Omega_i$ for choices $J_i$, but it is offset by 2 as in the Ref.\cite{stone2010symmetries}. This property also can be obtained by construction a map from $R^p_0$ to $R_{p+2}$ as in Ref.\cite{wen2012symmetry}, but the approach here is more geometrical. What's more, we can construct an explicit map between $R_q^p$ and $R_{q+1}^{p+1}$ as in Ref.\cite{wen2012symmetry},
\begin{equation}\label{1,1 period of classifying space}
    R_q^p=R_{q+1}^{p+1},
\end{equation}
this (1,1) periodicity theorem may have a deep relation with the (1,1) periodicity theorem $\widetilde{\textrm{KR}}^{p,q}=\widetilde{\textrm{KR}}^{p+1,q+1}$ of $\widetilde{\textrm{KR}}$-theory in section \ref{Topoloy K-theory}\cite{atiyah1988michael}.
Combing the equation (\ref{Omega space for j^2=1}) and (\ref{1,1 period of classifying space}) and the 8-period of Bott theorem of $R_q$, we can show
\begin{equation}\label{R and Rq}
    R_q^p=R_{p-q+2~ \textrm{mod}~ 8}.
\end{equation}
So we can study the space $R^p_q$ via the space $R_{p-q+2~ \textrm{mod}~ 8}$ which have only positive generators.
\par
In K-theory, the problem is formulated in terms of difference objects $(E, F, w)$, where $E$, and $F$ are representations of $Cl(0,q+1)$, and $w$ is a linear orthogonal map that identifies them as $Cl(0,q)$ representations, see \cite{karoubi2008k}. Without loss of generality, we fix $F$ to be a sum of several copies of the regular representation (which corresponds to a trivial Hamiltonian) and $w$ the identity map. Such difference objects form the classifying space $R_q$. The Abelian group of equivalence classes of difference objects for point space or $d=0$ base manifold is $\widetilde{\textrm{KR}}^{0,q}(pt)
=\widetilde{\textrm{KO}}^{-q}(pt)=\pi(R_q)$. Therefore, the classification of gapped free-fermion Hamiltonian in $d=0$ for real case can be obtained by calculating the element of $\widetilde{\textrm{KO}}^{-q}(pt)=\pi_0(R_q)$, where $R_q$ is the classifying space or symmetric space.
\par
In summary, we have constructed the 8 real classifying spaces for different symmetry such as time-reversal, $U(1)$ symmetry or charge-conjugation. We study the topology properties of $R_q^p$ by transforming to $R_q$ spaces through equation (\ref{1,1 period of classifying space}) and (\ref{Omega space for j^2=1}). And we establish the topology relationship between different classifying space $R_q=\Omega^i(R_{q-i})$ by minimal geodesic discussion. Actually, $R_{q+1}$ is a totally submanifold of $R_q$. Finally, we use $\widetilde{\textrm{KR}}^{0,q}(pt)$ to classify all topological phases in $d=0$ for various symmetry classifying space denoted with $q$. The table can be constructed as in the table $\ref{table1}$ in subsection \ref{Homotopy}. The relationship between usual classifying spaces obtained according to the discrete symmetries\cite{ryu2010topological} imposing on the complex Hermitian matrix and our real Clifford algebra discussion can be established according to Ref.\cite{stone2010symmetries}, we omit this discussion here for simplicity.
\subsubsection{Classification for $d\neq0$ (complex and real cases)}
After introducing the classifying spaces $C_q$ in $complex~case$, we can consider $d$-dimensional free conserved fermion systems and their gapped ground states. Note for continuous system, the base manifold can be equivalent to $S^d$ by one-point compactification, where $S^d$ represents $d$-dimensional sphere in momentum space. For fiber bundles with fiber $C_q$ and base manifold $S^d$, the distinct topological phases can be obtained by
\begin{equation}\label{kus}
    \widetilde{\textrm{KU}}^{-q}(S^d)=\pi_d(C_q)=\pi_0(C_{q+d}).
\end{equation}
If the system has translational symmetry, this is corresponding to the band insulators, thus, the base manifold will be $d$-dimensional torus $T^d$. However, the new classification can be obtained from $\pi_0(C_{q+d})$. For the free-fermion systems in symmetry class $C_q$ and translation symmetry, their gapped phases are classified by by\cite{kitaev2009periodic,wen2012symmetry}
\begin{equation}\label{kuT}
    \widetilde{\textrm{KU}}^{-q}(T^d)\cong \bigoplus_{s=0}^{d-1}\left(\begin{array}{c}
                                                              d \\
                                                              s
                                                            \end{array}\right)
\pi_0(C_{q+s})=\bigoplus_{s=0}^{d-1}\left(\begin{array}{c}
                                                              d \\
                                                              s
                                                            \end{array}\right)
\pi_0(C_{q-s}),
\end{equation}
where $(\begin{array}{c}
         d \\
         s
       \end{array})
$ is the binomial coefficient, the last equation satisfies due to the 2 Bott periodicity theorem for complex classifying spaces $C_q$.
\par
The above is the classification for complex cases. For real classes, we have a similar classification for $\bar{S}^d$ base manifold according to (\ref{Sphere model}):
\begin{equation}\label{kos}
    \widetilde{\textrm{KR}}^{0,q}(\bar{S}^d)
=\widetilde{\textrm{KR}}^{d,q}(pt)=\widetilde{\textrm{KO}}^{0,d-q}(pt)\cong\pi_0(R_{q-d}).
\end{equation}
Here we have used the (1,1) isomorphism in subsection \ref{Topoloy K-theory}, $\bar{S}^d$ is defined in the momentum space with the involution $\textbf{k}\leftrightarrow-\textbf{k}$\cite{kitaev2009periodic}. For the translation symmetry system, their gapped phases are classified by
\begin{equation}\label{kot}
    \widetilde{\textrm{KR}}^{0,q}(\bar{T^d})\cong \bigoplus_{s=0}^{d-1}\left(\begin{array}{c}
                                                              d \\
                                                              s
                                                            \end{array}\right) \widetilde{\textrm{KO}}^{-q}(\bar{S}^s)=\bigoplus_{s=0}^{d-1}\left(\begin{array}{c}
                                                              d \\
                                                              s
                                                            \end{array}\right)\pi_0(R_{q-s}).
\end{equation}
\par
The classification with translation symmetry can be obtained by equation (\ref{kuT}) for $complex ~case$ and (\ref{kot}) for $real~case$, respectively. Such a result has a physical meaning, the classification for higher-dimensional topological phases can be obtained by stacking the lower-dimensional ones. The examples have been discussed in Ref.\cite{wen2012symmetry}, we don't repeat here for simplicity. The gapped fermion system topological phases for general base manifold $M$ can be obtained by
$\widetilde{\textrm{KU}}^{-q}(M)$  for $complex ~case$ or $\widetilde{\textrm{KR}}^{0,q}(M)$ for $real ~case$, respectively. In addition, the 2-period and 8-period pattern also can be established based on Bott periodicity theorem of $\widetilde{\textrm{KU}}$ and $\widetilde{\textrm{KO}}$, respectively.
\par
Now, the classification problem has been solved completely. Following, we will give some more physical interpretation about above classification. The dimension effect on the classification has been revealed according to the discussion above, they effectively cancel the actual symmetries. For example, the classification problem in class $R_0$ and dimension $0$ is equivalent to the one in class $R_1$ and $1$-dimension system since $\pi_0(R_0)=\pi_0(R_{1-1})$. In fact, Kitaev has shown that gapped Hamiltonians in the momentum space are topologically equivalent to nondegenerate mass terms that anticommute with a fixed Dirac operator. Now let us consider the $real~case$ fermion system described by $\hat{H}=\frac{i}{4}\sum_{ij}A_{ij}\hat{\eta}_{i}\hat{\eta}_{j}$, and assuming it has translation symmetry, as well as time-reversal symmetry and fermion number conservation. We also assume that the singe-body energy bands of antisymmetric Hermitian matrix $iA$ have some Dirac points at zero energy and there are no other zero-energy states in the Brillouin zone. The gapless singe-body excitations in the system are described by the continuum limit of $iA$:
\begin{equation}\label{Continuum of A}
    iA=i\sum_{i=1}^{d}\gamma_i\partial_i,
\end{equation}
where we have folded all the Dirac points to the $\textbf{k = 0}$ point.
Without losing generality, we have also assumed that all the Dirac points have the same velocity. Since $\partial_i$ is real and antisymmetric\cite{wen2012symmetry}, $\gamma_i, i = 1,\ldots, d,$ are real symmetric $\gamma$ matrices that satisfy
\begin{equation}\label{gamma d}
    \gamma_i\gamma_j+\gamma_j\gamma_i=2\delta_{ij},~~\gamma_i^*=\gamma_i.
\end{equation}
The addition discrete symmetries imposing on the matrix $A$ require some conditions on the $\gamma$ matrices:
\begin{equation}\label{gamma and rho}
   \gamma_i\rho_j+\rho_j\gamma_i=0 ,
\end{equation}
since they do not affect $\partial_i$, where $q+p$ matrices $\rho_i$ are the symmetry operators that anticommute among themselves with $q$ of them square to 1 and $p$ of them square to -1.
\par
The idea to classify the different topological phases is now equivalent to the  ways to gap the Dirac points through the mass term. The different ways to gap the system (\ref{Continuum of A}) will correspond to various topological gapped phases. If we add a mass term $m$ satisfying
\begin{equation}\label{mass}
    m\gamma_i+\gamma_im=0
\end{equation}
to (\ref{Continuum of A}), and assuming the mass term is nondegenerate, the result Hamiltonian will be fully gapped. For example, if there is no symmetry, we only requires the real antisymmetric mass matrix $m$ to be nondegenerate including the condition (\ref{mass}). Without losing generality, we can choose the mass matrix to also satisfy $m^2=-1$. The space of those mass matrices is given by $R^0_d$. Similarly, if there are some symmetries, the real antisymmetric mass matrix $m$ will satisfy (\ref{mass}) and (\ref{gamma and rho}). Those mass matrices form a space $R_{q+d}^p$. The different disconnected components of $R_{q+d}^p$ represent different "bulk" gapped phases of the free fermions. Therefore, the bulk gapped phases of the free fermions in $d$-dimensions are classified by $\pi_0(R_{q+d}^p) =\pi(R_{p-q-d+2~\textrm{mod}~8})$, with $(q,p)$ depending on the symmetry. So we also obtain the same result as above using K-theory, they effectively cancel the actual symmetries by increasing the dimension of the system.
Because of the same reason, we will get the similar result for $complex ~case$. The different ways to gap the Dirac point using mass term form a space $C_d$, which indicates the gapped phases of the free fermions in $d$-dimensions are classified by $\pi_{0}(C_d)$.
\section{Conclusions and discussions}
\par
In the present paper we study the topological properties in TSMs. We systematically give the topology mathematics  required to read the later sections or subsecions.  In terms of fiber bundle theory, we can give different topological methods, i.e. cohomology, homotopy, K-theory, a unified interpretation. We also add some applications of corresponding topology method now and then during the introduction of these abstract mathematics. The most important tools in our paper are differential forms and fire bundle theory.   Fiber bundles can give one a clear topology picture in question, and the expressions will be very simple and in favor of derivation of complicated formulas if we use the differential forms instead of tensor components.
\par
Based on these mathematics, we give all integral topological invariants expressions with Berry connection and curvature for any Bloch or BdG equation. And we give the prove of these expressions on the sphere model, which can be easily  generalize to other models. A general homotopy discussion about the 1st and 2nd descendants\cite{ryu2010topological,qi2008topological} of topological invariants in same symmetry class and different dimensions. Our approach sets up the one-to-one map between the space $"interpolations"$ used by Qi.el and the loop space of the parent ones.
\par
For TFTs, we show the compact and topology properties of effective action for interacting and non-interacting systems. Following that, we prove the topological invariant $N_{2n+1}[G]$ characterize interacting system can be reduced to Chern-number $\textrm{Ch}_n$ in the non-interacting limit by performing again the integration over the frequency.  The general topological response Eq.(\ref{general current densitiy}) has been obtained, all of the topological physical effect can be read off from $\mathcal {J}_{2n+1}$. We also give the general topological invariants for Dirac model like (\ref{Gauss map number}), which is nothing but Gauss winding number, which generalize the results obtained in Ref.\cite{qi2008topological}. In addition, we show this general winding number can be expressed with Hopf indices and Brouwer degree using topological tensor theory\cite{duan1993topological}. What's more, the Chern-Simons effective action is reformulated in phase-space, and we give the unified generating functional Chern-Simons action in any dimensions. Finally, we review the bulk-boundary correspondence in any dimensions, which can be expressed like a AS index theorem. This index theorem\cite{volovik2003universe} relates the chirality of gapless excitations states of edge to the bulk invariant expressed with $N_D$.
\par
We give the completely classification of all gapped  free fermion systems using homotopy and topoloy K-theory. The classification can be viewed as the classification of corresponding fire bundles in our points. The classification of $d=0$ free fermion systems is equal to the classification of fiber bundles with point base manifold. In this case, we only need to study the topology property of the fiber. We obtain 2-classifying spaces for complex case and 8-classifying spaces for real case from the explicit physical models, respectively. We study the topology properties and their relations by minimal geodesic and loop space\cite{milnor1963morse} discussion. These symmetric classifying spaces can be interpreted as the fiber in fiber bundle theory. In the same logic,  if we choose one specific classifying space, the fiber bundle will have fixed fiber structure. Then, we only need to change the base manifold to respect model for the classification in $d\neq0$ cases. If the base manifold is $M$, this lead to  the classification of all gapped free fermion states using $\widetilde{\textrm{KU}}^{-q}(M)$  for $complex ~case$ or $\widetilde{\textrm{KR}}^{0,q}(M)$ for $real ~case$, respectively. Finally, we give the physical interpretation about the space dimension effect on the classification.
\par
Although we have studied many topological properties about TSM, there are still something interesting topological properties we didn't involve. For example, the famous fractional quantum Hall effect\cite{tsui1982two} (FQHE) is known a fully interacting system, the ground states are degenerate which are depend on the topology of defined space. The general states may be classified by the concept of topological order introduced by Wen\cite{PhysRevB.41.9377}. There are two development direction in this non-trivial states. One route is to study this states from a new mathematic framework called tensor categories\cite{gu2008tensor}.  Another route is to generalize FQHE in 3d spatial dimension \cite{maciejko2010fractional} from topological field theory. The physical realization materials will need to be studied in the future.
Another point we want to mention is that, we  have only studied one of the three universality classes of fermionic vacua\cite{volovik2003universe}, the other two are topological protected fermi surface and topological protected fermi point.  These topological protected systems are still an active study direction in condensed matter physics, such as topological semi-metal\cite{wan2011topological,burkov2011weyl}. Topology has become a useful mathematic tool to predict novel materials, better understand of robust systems of matter, classification principle and so on.

%
\end{document}